\documentclass[prd,%preprint,
superscriptaddress,preprintnumbers,tightenlines,nofootinbib, eqsecnum]{revtex4-2}

\usepackage{amsmath}
\usepackage{amsfonts}
\usepackage{amssymb}
\usepackage{bm}
\usepackage{hyperref}
\usepackage{mathrsfs}
\usepackage{graphicx}
\usepackage{empheq}
\usepackage{ulem}
\usepackage[usenames]{color}
\usepackage{ytableau}
        
\allowdisplaybreaks

\DeclareSymbolFontAlphabet{\mathrsfs}{rsfs}
\DeclareMathAlphabet{\mathcal}{OMS}{cmsy}{m}{n}

\newcommand{\FP}{\mathop{\mathrm{FP}}_{B=0}}

\newcommand{\nn}{\nonumber}
\newcommand\calO{\mathcal{O}}
\newcommand{\dd}{\mathrm{d}}

\newcommand{\dI}{\mathrm{I}}
\newcommand{\dJ}{\mathrm{J}}
\newcommand{\dM}{\mathrm{M}}

\newcommand{\dS}{\mathrm{S}}

\newcommand{\dW}{\mathrm{W}}
\newcommand{\dX}{\mathrm{X}}
\newcommand{\dY}{\mathrm{Y}}
\newcommand{\dZ}{\mathrm{Z}}

\newcommand{\calA}{\mathcal{A}}
\newcommand{\calB}{\mathcal{B}}
\newcommand{\calF}{\mathcal{F}}
\newcommand{\calG}{\mathcal{G}}
\newcommand{\calR}{\mathcal{R}}

\newcommand{\calP}{\mathcal{P}}
\newcommand{\calQ}{\mathcal{Q}}

\newcommand{\ab}{^{\alpha\beta}}
\newcommand{\RR}{_{\mathrm{RR}}}
\newcommand{\BT}{\text{BT}}
\newcommand{\sym}{_{\mathrm{sym}}}
\newcommand{\cons}{_{\mathrm{cons}}}

\newcommand{\integre}{\hbox{$\int$}}

\newcommand{\p}{\partial}
\newcommand{\be}{\begin{equation}}
\newcommand{\ee}{\end{equation}}
\newcommand{\bse}{\begin{subequations}}
\newcommand{\ese}{\end{subequations}}
\newcommand{\Vrr}{V^\mathrm{RR}}
\definecolor{darkgreen}{rgb}{0,0.5,0}

\hypersetup{
 unicode=false,          % non-Latin characters in Acrobat???s bookmarks
 pdftoolbar=true,        % show Acrobat???s toolbar?
 pdfmenubar=true,        % show Acrobat???s menu?
 pdffitwindow=false,     % window fit to page when opened
 pdfstartview={FitH},    % fits the width of the page to the window
 pdftitle={My title},    % title
 pdfauthor={Author},     % author
 pdfsubject={Subject},   % subject of the document
 pdfcreator={Creator},   % creator of the document
 pdfproducer={Producer}, % producer of the document
 pdfkeywords={keyword1} {key2} {key3}, % list of keywords
 pdfnewwindow=true,      % links in new window
 colorlinks=true,       % false: boxed links; true: colored links
 linkcolor=red,          % color of internal links
 citecolor=cyan,        % color of links to bibliography
 filecolor=magenta,      % color of file links
 urlcolor=darkgreen,           % color of external links
 linktocpage=true
}

\makeatletter
\g@addto@macro\bfseries{\boldmath}
\makeatother

\begin{document}
	
\title{Gravitational radiation reaction for compact binary systems \\at the fourth-and-a-half post-Newtonian order}

\author{Luc \textsc{Blanchet}}\email{luc.blanchet@iap.fr}
\affiliation{$\mathcal{G}\mathbb{R}\varepsilon{\mathbb{C}}\calO$, 
	Institut d'Astrophysique de Paris, \\UMR 7095, CNRS, Sorbonne Universit{\'e},
	98\textsuperscript{bis} boulevard Arago, 75014 Paris, France}

\author{Guillaume \textsc{Faye}}\email{faye@iap.fr}
\affiliation{$\mathcal{G}\mathbb{R}\varepsilon{\mathbb{C}}\calO$, 
Institut d'Astrophysique de Paris, \\UMR 7095, CNRS, Sorbonne Universit{\'e},
98\textsuperscript{bis} boulevard Arago, 75014 Paris, France}
%\affiliation{Centre for Strings, Gravitation and Cosmology, Department of Physics,
%Indian Institute of Technology Madras, Chennai 600036, India}

\author{David \textsc{Trestini}}\email{trestini@fzu.cz}
\affiliation{CEICO, Institute of Physics of the Czech Academy of Sciences, Na Slovance 2, 182 21 Praha 8, Czechia}

\date{\today}

\begin{abstract}
We compute the gravitational radiation-reaction force on a compact binary source at the fourth-and-a-half post-Newtonian (4.5PN) order of general relativity, \textit{i.e.}, 2PN order beyond the leading 2.5PN radiation reaction. The calculation is valid for general orbits in a general frame, but in a particular coordinate system which is an extension of the Burke-Thorne coordinate system at the lowest order. With the radiation-reaction acceleration, we derive (from first principles) the flux-balance laws associated with the energy, the angular and  linear momenta,  and  the center-of-mass position, in a general frame and up to 4.5PN order. Restricting our attention to the frame of the center of mass, we point out that the equations of motion acquire a non-local-in-time contribution at the 4.5PN order, made of the integrated flux of linear momentum (responsible for the recoil of the source) together with the instantaneous flux of center-of-mass position. The non-local contribution was overlooked in the past literature, which assumed locality of the radiation-reaction force in the center of mass frame at 4.5PN order. We discuss the consequences of this non-local effect and obtain consistent non-local equations of motion and flux balance laws at 4.5PN order in the center-of-mass frame. 
\end{abstract}


\maketitle

\section{Introduction}\label{sec:introduction}

%\Luc{David fait maintenant à partir de Sec IV}
%\Luc{Guillaume fait maintenant l'App $v^{\mu\nu}$}
%\Luc{Luc fait maintenant les Secs II et III et l'Appendice Gopu}

\subsection{Overview}

Our goal is to compute the gravitational radiation-reaction (RR) force acting on an orbiting compact binary system through the fourth-and-a-half post-Newtonian (4.5PN) approximation of general relativity (GR), \emph{i.e.}, at order $\calO(9)\equiv\calO(c^{-9})$, whereas the dominant, lowest order RR occurs at the 2.5PN $\sim\calO(5)$ order~\cite{CE70, BuTh70, Bu71, Miller74, Ehl80, K80a, K80b, PapaL81, BD84} (see~\cite{Dcargese} for a review). We thus intend to control the RR force through 2PN order beyond the dominant level. More precisely, we shall compute the 2.5PN, 3.5PN and 4.5PN contributions, which have odd-parity powers of $1/c$, and correspond to purely dissipative RR effects. We do not address the 4PN tail term, which contains both conservative and dissipative contributions, because it has been computed elsewhere~\cite{BD88,FStail}.

As is well known~\cite{Schafer81, Schafer82}, the RR force takes different expressions depending on the choice of coordinate system. A very interesting choice is the Burke-Thorne (BT)  coordinate system~\cite{BuTh70, Bu71, MTW}, in which the 2.5PN RR force (density) takes the simple scalar form $F^i\RR = \rho\,\partial_i V\RR + \calO(7)$. Here, $\rho$ is the Newtonian mass density in the source and the scalar RR potential is explicitly given by\footnote{See Sec.~\ref{sec:notation} for the notations.}
\begin{equation}\label{eq:VRR}
V\RR(\mathbf{x},t) = - \frac{G}{5 c^5}\,x^i
x^j\,\frac{\dd^5\dM_{ij}}{\dd t^5} + \calO(7)\,,
\end{equation} 
where $\dM_{ij}$ is the trace-free mass quadrupole moment of the source, reducing at lowest order to the usual Newtonian quadrupole moment of the Newtonian mass density,
\begin{equation}\label{eq:MijN}
\dM_{ij}(t) = \int_\text{source}\dd^3\mathbf{x}\Bigl[x^ix^j-\frac{1}{3}\delta_{ij}\mathbf{x}^2\Bigr] \rho(\mathbf{x},t) + \calO(2)\,.
\end{equation} 
Despite the linear-in-$G$ character of the RR force in BT coordinates, its derivation requires the control of the non-linear effects to justify that they do not play any role in this coordinate system~\cite{BD84} (see also~\cite{WalkW80} for discussions on non-linear effects taking place in other coordinate systems). Of course, Eqs.~\eqref{eq:VRR}--\eqref{eq:MijN} reproduce the known fluxes (in particular the energy flux given by the Einstein quadrupole formula) computed asymptotically far from the source.

At the 3.5PN order (or 1PN relative order) and in the case of compact binaries, the RR force has been derived in an arbitrary coordinate system, but only in  the  center-of-mass (CM) frame~\cite{IW93, IW95}, by assuming the validity of the flux-balance equations for energy and angular momentum at the relative 1PN order. We shall refer to the latter approach for computing the RR as the ``flux-balance'' method. Since it is restricted to the frame of the CM, only the fluxes for energy and angular momentum are needed.\footnote{The flux-balance method in the CM frame has been extended to include spin-orbit RR effects at 3.5PN order~\cite{ZW07}.} By contrast, a ``first-principle'' approach is to perform the direct calculation of the RR at 1PN order by integrating the field equations in the source's near zone, \emph{i.e.}, without \textit{a priori} assuming the validity of the flux-balance laws. The first-principle approach proves, at some given order, that the RR force implies the flux-balance laws, where the fluxes are given by their values as computed at future null infinity. It has been implemented at 1PN order in various coordinate systems~\cite{JaraS97, PW02, KFS03, NB05, itoh3}, with results consistent with the balance laws, which means that the RR corresponds to a unique determination of the Iyer-Will gauge parameters~\cite{IW93, IW95}. On the other hand, some general investigations of RR and the balance laws in the case of isolated systems (not restricted to compact binaries) were performed up to relative 1.5PN order beyond the leading RR force~\cite{B93, B97, BF19}, at which the gravitational wave tails first appear (this corresponds to 4PN beyond the Newtonian acceleration~\cite{BD88, FStail}). In particular, the works~\cite{B93, B97} introduced the extension to 1.5PN order of the leading-order BT scalar RR potential~\eqref{eq:VRR}.

The flux-balance method, still restricted to the CM frame, has been extended to 4.5PN order (namely 2PN relative order) by Gopakumar, Iyer and Iyer in Ref.~\cite{GII97}, henceforth referred to as ``GII''. To date, the only calculation of the RR to relative 2PN order using the first-principle approach (\textit{i.e.}, not relying on balance equations) was done using effective field theory (EFT)~\cite{LPY23}. In Ref.~\cite{LPY23}, the computations were performed in a general frame, but only the CM frame results were published; moreover, Ref.~\cite{LPY23} claimed agreement with GII but only the case of circular orbits was shown to be consistent with them. 

In the present paper, we shall follow the first-principle approach as well, but the results presented are valid in a general frame, and only subsequently restricted to the CM frame. Moreover, they are valid in a different coordinate system, namely a specific extension of the BT coordinates at the 4.5PN order. At that order and in that coordinate system, the RR is described by some specific scalar, vector, and tensor RR potentials,
\begin{align}\label{eq:VRRsvt}
		V\RR\ab \equiv \Bigl(V\RR, V\RR^i, V\RR^{ij}\Bigr)\,,
\end{align}
where $V\RR$ generalizes~\eqref{eq:VRR},  $V\RR^i$ appears only at 1PN relative order as a generalization of the vector RR introduced in~\cite{B93,B97}, and the tensor $V\RR^{ij}$ appears at 2PN order. Since our calculation will not be restricted to the CM frame, it will enable us to check (and, actually, to prove) the flux-balance laws, not only for the energy and the angular momentum, but also for the linear momentum and the CM position. We shall thus recover from the RR force the energy and angular momentum fluxes to 2PN relative order (\textit{i.e.}, 2.5PN + 3.5PN + 4.5PN), as well as the linear momentum and CM fluxes to 1PN relative order (3.5PN + 4.5PN). In particular, we shall recover and confirm to 1PN order the flux associated with the CM position obtained in Refs.~\cite{KQ16, KNQ18, N18, BF19, COS20}. 
%
%Note also that at 1.5PN relative order of 4PN there is a RR contribution coming from radiating tails that we consistently add to our final results. 

Next, we consider the problem of the definition of the CM frame. Here, by CM, we mean the barycenter not only of the matter distribution, but also of the radiation, therefore taking into account the linear momentum kick or recoil of the source by the GW emission. As a result, we find that the relation between the positions of the bodies in the CM frame and the relative positions and velocities acquires a non-local-in-time contribution at the 3.5PN order, which is given by the time integral of the flux of linear momentum. We refer to such a contribution as 
%``hereditary'' or 
``semi-hereditary'', by which we mean that it is given by the integral in time of some local or ``instantaneous'' term~\cite{BD92}, and thus its time-derivative is entirely instantaneous (conversely, ``hereditary'' terms such as tails are non-local-in-time, as well as their iterated time-derivatives). In turn, we find that the RR force itself acquires in the CM frame a non-local (semi-hereditary) contribution at the 4.5PN order, which involves the gravitational recoil of the system as well as the flux of CM. We conclude that at the 4.5PN order in the CM frame, it becomes impossible (in any coordinate system) to express the RR and equations of motion (EOM) as local or instantaneous functionals of the source's parameters. 

On this point, we are in conflict with the flux-balance approach extended to 4.5PN order in~\cite{GII97}, which assumed as a basic hypothesis that the RR and EOM in the CM frame are local, thereby missing the physical consequence of the binary's GW recoil. We prove that our EOM in the CM frame preserve the correct flux-balance laws in the CM frame for the energy and  the angular momentum at 2PN order, despite these quantities not being in the local form postulated in~\cite{GII97}. We shall provide the formulas correcting GII~\cite{GII97} (\textit{i.e.}, incorporating the gauge invariant non-local effect) which will allow us to uniquely determine the GII gauge parameters that correspond to our extended BT gauge at the 4.5PN order. Finally, we discuss the interesting case of quasi-circular orbits.

The plan of this paper is as follows. In Sec.~\ref{sec:NZ}, we recall the general structure of the PN metric in the near zone of a general isolated source and the RR terms therein. In Sec.~\ref{sec:RR}, we introduce the extended BT coordinate system and define the scalar, vector, and tensor RR potentials~\eqref{eq:VRRsvt} up to 4.5PN order. In Sec.~\ref{sec:acceleration}, we obtain the EOM with RR terms included up to 4.5PN order in a general frame and in extended BT coordinates. In Sec.~\ref{sec:flux_balance}, we prove that all the flux balance equations (energy, angular and linear momenta, CM) are satisfied to the required order. In Sec.~\ref{sec:CoM}, we solve the problem of the passage to the CM frame, taking into account the physical effects of GW recoil and displacement of the CM position. In Sec.~\ref{sec:GIIcorrected}, we show how to modify the flux-balance approach at 2PN order~\cite{GII97} to include the recoil and CM displacement, and we compute the set of GII gauge parameters corresponding to our chosen coordinate system. In Sec.~\ref{sec:circular}, we obtain the RR contributions in the case of quasi-circular orbits. The paper ends by a short conclusion in Sec.~\ref{sec:conclusion}. The Appendix~\ref{app:vmunu} deals with the control of certain required terms in the metric, the technical Appendix~\ref{app:id_dim} discusses dimensional identities, and the Appendix~\ref{app:Schott} contains the expressions of Schott-like terms.

\subsection{Notation}
\label{sec:notation}

\begin{enumerate}
	\item The Einstein field equations in a general coordinate system are denoted 
	\begin{align}\label{eq:EE}
		\Box h^{\alpha\beta} - \partial H^{\alpha\beta} = \frac{16\pi G}{c^4} \tau^{\alpha\beta}\,,
	\end{align}
	where $h^{\alpha\beta} \equiv \sqrt{-g}\, g^{\alpha\beta} - \eta^{\alpha\beta}$ is the gothic-type metric deviation; $g \equiv \mathrm{det}(g_{\alpha\beta})$ and $\eta^{\alpha\beta} \equiv \text{diag}(-1,1,1,1)$; $\Box\equiv\Box_\eta$ is the flat d'Alembertian operator; $H^\alpha\equiv\partial_\beta h\ab$ denotes the ``harmonicity'' and we pose $\partial H^{\alpha\beta} \equiv \partial^\alpha H^\beta + \partial^\beta H^\alpha - \eta^{\alpha\beta} \partial_\gamma H^\gamma$, so that $\partial_\beta\partial H\ab = \Box H^\alpha$; $\tau\ab$ is the matter $+$ gravitation pseudo-tensor,
	\begin{align}\label{eq:tauab}
		\tau^{\alpha\beta} = \vert g\vert T^{\alpha\beta} + \frac{c^4}{16\pi G}\Lambda^{\alpha\beta}[h]\,,
	\end{align}
	with $T^{\alpha\beta}$ being the matter tensor and $\Lambda^{\alpha\beta}$ the gravitational non-linear part, at least quadratic in $h^{\alpha\beta}$ and its space-time derivatives; we have $\partial_\beta\tau\ab = 0$ as a consequence of the field equations~\eqref{eq:EE}.
	\item Multi-spatial indices are denoted $L=i_1i_2\cdots i_\ell$, made of $\ell$ (spatial) indices ranging from 1 to 3; it is always understood that contracted multi-indices $L$ involve $\ell$ summations (over the indices they contain); the multi-derivative operator $\partial_L$ is a short-hand for $\partial_{i_1}\cdots\partial_{i_\ell}$; the symmetric-trace-free (STF) operation is denoted with a hat or sometimes by angular brackets surrounding indices, \textit{e.g.}, $\hat{\partial}_L=\partial_{\langle L\rangle}=\text{STF}\,[\partial_L]$; similarly $x_L=x_{i_1}\cdots x_{i_\ell}$ is a multi-spatial vector and $\hat{x}_L=x_{\langle L\rangle}=\text{STF}\,[x_L]$; $n_L=x_L/r^\ell$ with $r=\vert\mathbf{x}\vert$ and $\hat{n}_L=\text{STF}\,[n_L]$; The Levi-Civita antisymmetric symbol is denoted $\varepsilon_{ijk}$ (with $\varepsilon_{123}=1$); parenthesis refer to symmetrization, $T_{(ij)}=\frac{1}{2}(T_{ij}+T_{ji})$.
	\item Superscripts $(q)$ indicate $q$ time differentiations; time derivatives are also sometimes indicated by dots; time antiderivatives are denoted by superscripts $(-q)$ or by an explicit integral sign, \textit{e.g.},
	%$\int\!f(t) = f^{(-1)}(t) = \int^t_{-\infty} \dd t' f(t')$ and $\int\!\!\int f(t) = f^{(-2)}(t) =\int^t_{-\infty} \dd t' \int f(t')$.
	\begin{align}\label{eq:antiderive}
	\hbox{$\int$}\!f(t) \equiv f^{(-1)}(t) = \int^t_{-\infty} \dd t' f(t')\,;\qquad\hbox{$\int\!\!\int$} f(t) \equiv f^{(-2)}(t) = \int^t_{-\infty} \dd t' \hbox{$\int$} f(t')\,.
	\end{align}
	We assume a matter system which is stationary in the remote past~\cite{BD86}, thus all time varying functions are zero for any time before some finite instant $-\mathcal{T}$.
	\item For any smooth function of time $f(t)$ we use the specific notation for the antisymmetric, \textit{i.e.}, half-retarded minus half advanced, combination of homogeneous waves (beware of the factor $\frac{1}{2}$ included)
	\begin{align}\label{eq:antisym}
		\bigl\{f\bigr\}(t,r) \equiv \frac{f(t-r/c) - f(t+r/c)}{2r}\,,
	\end{align}
	where $r=\vert\mathbf{x}\vert$. This corresponds to a monopolar wave, and by applying repeatedly STF spatial derivatives $\hat{\partial}_L$ we obtain corresponding multipolar waves, given in closed form as~\cite{B93}
	\begin{align}\label{eq:antisymL}
		\hat{\partial}_L\bigl\{f\bigr\} = - \frac{\hat{x}_L}{c^{2\ell+1}(2\ell+1)!!}\int_{-1}^{1}\dd z \,\delta_\ell(z) \,f^{(2\ell+1)}(t + z r/c)\,,
	\end{align}
	where we pose $\delta_\ell (z) = \frac{(2\ell+1)!!}{2^{\ell+1} \ell!} (1-z^2)^\ell$, so that $\int_{-1}^{1}\dd z \,\delta_\ell(z)=1$. Performing the PN expansion or equivalently the near-zone expansion ($r/c\to 0$) of such antisymmetric multipolar wave one obtains the regular expansion
	\begin{align}\label{eq:PNantisymL}
		\hat{\partial}_L\bigl\{f\bigr\} = - \frac{\hat{x}_L}{c^{2\ell+1}}\sum_{k=0}^{+\infty}\frac{r^{2k}}{c^{2k}(2k)!!(2k+2\ell+1)!!} f^{(2\ell+2k+1)}(t)\,.
	\end{align}
	The multipolar antisymmetric wave is smaller by a factor $\calO(2\ell+1)=\calO(c^{-2\ell-1})$ than the corresponding symmetric or retarded wave (as seen in the near zone, when $r\to 0$).
	\item The usual small post-Newtonian remainder is denoted $\calO(n)\equiv\calO(c^{-n})$, and $\calO(n,p)\equiv\calO(c^{-n},c^{-p})$ means a PN remainder which is at once of order $\calO(n)$ for conservative even-parity contributions, and $\calO(p)$ for dissipative odd-parity (RR) contributions.
\end{enumerate}

\section{Post-Newtonian iteration in the near zone}\label{sec:NZ}

\subsection{General structure of the PN metric}

We provide some necessary reminders on the construction of the PN metric in the near zone of an isolated system. We shall indicate the formal PN expansion with an overbar. The general expression of the PN metric in the near zone (valid up to any PN order) has been obtained by asymptotic matching to the multipolar-post-Minkowskian~\cite{BD86} metric in the zone exterior to the system. In harmonic coordinates, \textit{i.e.}, $\overline{H}^\alpha=\partial_\beta\mathop{\overline{h}}_{}\!{}\ab=0$, the result is~\cite{B98mult, PB02, BFN05} 
\begin{equation}\label{eq:usymRR}
	\mathop{\overline{h}}_{}\!{}\ab \equiv \mathop{\overline{u}}_{}\!{}\ab + \mathop{\overline{u}}_{}\!{}\ab_{\mathrm{RR}}\,.
\end{equation} 
The first term corresponds to the standard PN iteration using the ``symmetric'' inverse d'Alembertian operator, 
\begin{equation}\label{eq:usym}
	\mathop{\overline{u}}_{}\!{}\ab \equiv \frac{16\pi G}{c^4}
	\,\widetilde{\Box}^{-1}_\text{sym}\big[\mathop{\overline{\tau}}_{}\!{}\ab\big] \,,
\end{equation} 
where $\mathop{\overline{\tau}}_{}\!{}\ab$ is the PN expansion of the stress-energy pseudo-tensor~\eqref{eq:tauab}. The symmetric inverse d'Alembertian $\widetilde{\Box}^{-1}_\text{sym}$ must be endowed with a crucial infrared (IR) regularization based on Hadamard's finite part (FP). The regularization accounts for the divergence of Poisson-like integrals at spatial infinity ($r\to +\infty$), known to occur at high PN orders. In detail, we explicitly have
\begin{equation}\label{eq:propsym}
	\widetilde{\Box}^{-1}\sym \big[\mathop{\overline{\tau}}_{}\!{}\ab\big] \equiv \FP\Box^{-1}_\text{sym}\big[\widetilde{r}^B \mathop{\overline{\tau}}_{}\!{}\ab\big] = \sum_{k=0}^{+\infty}\left(\frac{\partial}{c\partial
		t}\right)^{\!\!2k}\FP\Delta^{-k-1} \big[\widetilde{r}^B \mathop{\overline{\tau}}_{}\!{}\ab\big] \,,
\end{equation} 
where $\widetilde{r}^B=(r/r_0)^B$ is the IR regularization factor ($r_0$ is an arbitrary length scale), and $\FP$ denotes the finite part, \emph{i.e.}, the coefficient of $B^0$ in the Laurent expansion when $B\to 0$. By definition of the Hadamard finite part, any pole in $1/B$ is discarded. The $k$-th iteration of the Poisson operator in~\eqref{eq:propsym} is given by
\begin{equation}\label{eq:genpoissiter}
	\widetilde{\Delta}^{-k-1}\big[\mathop{\overline{\tau}}_{}\!{}\ab\big] \equiv \FP\Delta^{-k-1}\big[\widetilde{r}^B \mathop{\overline{\tau}}_{}\!{}\ab\big] = -\frac{1}{4\pi}\,\FP \,\int
	\dd^3\mathbf{x}'\,\frac{\vert\mathbf{x}-\mathbf{x}'\vert^{2k-1}}{(2k)!}
	\,\widetilde{r}'^B \,\mathop{\overline{\tau}}_{}\!{}\ab(\mathbf{x}',t) \,.
\end{equation}
The first term in~\eqref{eq:usymRR} represents a particular solution of the (PN-expanded) d'Alembertian equation which is sourced by the (PN-expanded) pseudo-tensor, \textit{i.e.}, we have
\begin{align}
	\Box\mathop{\overline{u}}_{}\!{}\ab = \frac{16\pi G}{c^4} \mathop{\overline{\tau}}_{}\!{}\ab\,.
\end{align}

To the particular solution, one must add a specific homogeneous solution, such that the matching equation is satisfied~\cite{PB02}. Furthermore, the homogeneous solution must be regular all over the source's near zone, \textit{i.e.}, be of the ``retarded-minus-advanced'' type. The homogeneous solution is associated with RR effects occuring in the source's near zone. This is the second term in~\eqref{eq:usymRR}, given by
\begin{equation}\label{eq:uRR}
	\mathop{\overline{u}}_{}\!{}\ab_{\mathrm{RR}} = -
	\frac{4G}{c^4}\sum^{+\infty}_{\ell=0}
	\frac{(-)^\ell}{\ell!}\hat{\partial}_L \bigl\{ \calA_L\ab\bigr\}\,.
\end{equation} 
It is parametrized by the multipole-moment functions $\mathcal{A}_L\ab(t)$, which are STF in their $\ell$ indices $L$. We use the notation~\eqref{eq:antisym}--\eqref{eq:antisymL} for antisymmetric type homogeneous solutions (which are regular in the source, when $r\to 0$). As shown in~\cite{PB02} the multipole moments functions in~\eqref{eq:uRR} are actually composed of two distinct contributions: 
\begin{align}\label{eq:AFR}
	\mathcal{A}_L\ab(t) = \mathcal{F}_L\ab(t) +\calR_L\ab(t)\,.
	\end{align}
The first one represents the multipole moments of the source as seen from the exterior region, and is given by
\begin{equation}\label{eq:FL}
	\calF^{\alpha\beta}_L (t) = \FP \int \dd^3 \mathbf{x}
	\, \widetilde r^B \hat{x}_L \,\int^1_{-1} \dd z \,\delta_\ell(z) \,{\overline
		\tau}^{\alpha\beta}(\mathbf{x}, t+z r/c)\,.
\end{equation}
The function $\delta_\ell(z)$ is defined in Sec.~\ref{sec:notation}. Note that the integral in~\eqref{eq:FL} is symmetric under time reversal and the argument $t+ z r/c$ can be indifferently replaced by $t- z r/c$. The second term in~\eqref{eq:AFR} corresponds to the so-called radiation modes due to tails and higher non-linear effects occuring in the GW propagation. Those effects start at the 4PN order in the near-zone metric and the source's dynamics~\cite{BD88, B93}. The general expression is
\begin{equation}\label{eq:RL}
	\calR_L\ab(t) = -2
	\FP \int\,\dd^3\mathbf{x}\,\widetilde r^B \hat{x}_L\int_{1}^{+\infty}\dd
	z\, \delta_\ell(z)\,\mathcal{M}(\tau\ab)\,\left(\mathbf{x},t -z
	r/c\right) \,,
\end{equation}
where $\mathcal{M}(\tau\ab)$ is the multipole expansion of the pseudo-tensor,
%, and $\gamma_\ell(z)=-2\delta_\ell(z)$ is normalized in such a way that $\int_{1}^{+\infty}\dd z\,\gamma_\ell(z)=1$. 
and the integral in~\eqref{eq:RL} is truly retarded. In the following, we shall mostly consider the term $\calF_L\ab$, since the other term $\calR_L\ab$ contributes only at the 4PN order, and we are only interested in the odd parity 4.5PN term. However, although the 4PN tail in~\eqref{eq:RL} has a formal even parity, it corresponds to genuine RR effects, so we shall briefly discuss it following Refs.~\cite{B93, B97}, see Eqs.~\eqref{eq:Ftail}.

\subsection{Explicit nonlinear iteration of the metric}
\label{sec:nonlinear}

We take advantage of the fact that  the PN solution~\eqref{eq:usymRR} satisfies the harmonic coordinate condition $\mathop{\overline{H}}_{}\!{}^\alpha=0$ by construction. Computing the divergence of the first term~\eqref{eq:usym} using the conservation law of the pseudo-tensor $\partial_\beta\overline{\tau}\ab=0$, we obtain
\begin{equation}\label{eq:div1}
	\partial_\beta \!\mathop{\overline{u}}_{}\!{}\ab = \FP\Box^{-1}_\text{sym}\Bigl[B\,\widetilde{r}^{B} \frac{n^i}{r} \mathop{\overline{\Lambda}}_{}\!{}^{i\alpha}\Bigr]\,.
\end{equation} 
We observe that the result comes entirely from the differentiation of the regularization factor $\widetilde{r}^B$, and consequently that there is an explicit factor $B$ in the integrand. This factor $B$ kills the matter contribution to the pseudo-tensor, since the regularization applies only to the portion of the integral at spatial infinity from the matter source. 
%Hence the term~\eqref{eq:div1} is at least second order (say $\propto G^2$) in a non-linearity expansion. 
On the other hand, the harmonicity condition also implies
\begin{equation}\label{eq:div2}
	\partial_\beta \!\mathop{\overline{u}}_{}\!{}\ab = - \partial_\beta \!\mathop{\overline{u}}_{}\!{}\ab_{\mathrm{RR}} = \frac{4G}{c^4}\partial_\beta\biggl[\sum^{+\infty}_{\ell=0}
	\frac{(-)^\ell}{\ell!}\hat{\partial}_L \bigl\{ \calA_L\ab\bigr\}\biggr]\,,
\end{equation} 
which can be transformed following the procedure of Eqs.~(4.4)--(4.7) in~\cite{B98mult}. Defining
\begin{equation}\label{eq:BL}
	\calB_L^\alpha \equiv \frac{1}{c}\dot{\calA}^{0\alpha}_L - \ell \calA^{\alpha\langle i_\ell}_{L-1\rangle} - \frac{1}{(2\ell+3)c^2} \ddot{\calA}^{i\alpha}_{i L}\,,
\end{equation} 
where $\calA^{\alpha\langle i_\ell}_{L-1\rangle}$ denotes the STF part of $\calA^{\alpha i_\ell}_{L-1}$, and where the overdot refers to the time derivative, we obtain 
\begin{equation}\label{eq:harmcond}
	\partial_\beta \!\mathop{\overline{u}}_{}\!{}\ab = \frac{4G}{c^4} \sum^{+\infty}_{\ell=0}
	\frac{(-)^\ell}{\ell!}\hat{\partial}_L \bigl\{ \calB_L^\alpha\bigr\}\,.
\end{equation} 
%
%%%%%%%%%%ù
%%%%%%%%%%ù
Next, we construct from~\eqref{eq:harmcond} a related object $\overline{v}\ab$, which is not only an antisymmetric homogeneous solution of the wave equation ($\Box\mathop{\overline{v}}_{}\!{}\ab=0$ and $\mathop{\overline{v}}_{}\!{}\ab$ is regular when $r\to 0$), but also has the property that its divergence cancels out the divergence of $\mathop{\overline{u}}_{}\!{}\ab$, namely 
\begin{equation}\label{eq:divh}
\partial_\beta\bigl(\mathop{\overline{u}}_{}\!{}\ab+\mathop{\overline{v}}_{}\!{}\ab\bigr) = 0\,.
\end{equation}
%
%
%%%%%%%%%%%%ù
%%%%%%%%%%%ù
%Next, we construct from~\eqref{eq:harmcond} a related object $\overline{v}\ab$, which will be like~\eqref{eq:harmcond} an antisymmetric homogeneous solution of the wave equation ($\Box\mathop{\overline{v}}_{}\!{}\ab=0$ and $\mathop{\overline{v}}_{}\!{}\ab$ is regular when $r\to 0$), and is such that its divergence cancels out the divergence of $\mathop{\overline{u}}_{}\!{}\ab$, namely 
%%
%\begin{equation}\label{eq:divh}
%\partial_\beta\bigl(\mathop{\overline{u}}_{}\!{}\ab+\mathop{\overline{v}}_{}\!{}\ab\bigr) = 0\,.
%\end{equation}
%
The definition of $\mathop{\overline{v}}_{}\!{}\ab$ (and of some alternative forms) is relegated to Appendix~\ref{app:vmunu}, where we shall prove that it does not contribute to RR up to the 4.5PN level.

The above construction allows us to define what will appear to be a ``linearized'' approximation (of order $G$) for the PN near-zone metric. To this end, we notice that by defining $\overline{v}\ab\RR \equiv -\overline{v}\ab$, the quantity
\begin{equation}\label{eq:hRR1}
	G \mathop{\overline{h}}_{}\!{}\ab_{\mathrm{RR}\,1} \equiv \mathop{\overline{u}}_{}\!{}\ab\RR + \mathop{\overline{v}}_{}\!{}\ab\RR
\end{equation}
happens to be a linearized vacuum solution of the field equations in harmonic coordinates, \textit{i.e.}, $\Box\mathop{\overline{h}}_{}\!{}\ab_{\mathrm{RR}\,1}=0$ and $\partial_\beta\!\mathop{\overline{h}}_{}\!{}\ab_{\mathrm{RR}\,1}=0$, which is associated with the RR effects at the linear level --- hence the explicit factor $G$ introduced into the definition~\eqref{eq:hRR1}, which we will use as a formal bookkeeping parameter. Still, we have to add the conservative effects, and this results in the following ``linear'' approximation for both conservative and RR effects,
\begin{equation}\label{eq:h1def}
	\mathop{\overline{h}}_{}\!{}\ab_1 \equiv \frac{16\pi}{c^4}
	\,\Box^{-1}_\text{sym}\big[\mathop{\overline{T}}_{}\!{}\ab\big] + \mathop{\overline{h}}_{}\!{}\ab_{\mathrm{RR}\,1}\,,
\end{equation}
where the first term is the symmetric d'Alembertian operator applied to the PN expansion of the matter tensor $T\ab$. Since the matter part is of compact support, we do not need the FP regularization in this term. Beware that in this set up (following~\cite{B98mult}), we are considering that the components of the matter tensor $\mathop{\overline{T}}_{}\!{}\ab$ are of order $G^0$ in the non-linear $G$-expansion. Similarly the multipole moment functions $\calA_L\ab$ and $\calB_L^\alpha$ are considered to be of order $G^0$. The linear solution~\eqref{eq:h1def} obeys the expected linear-looking field equation
\begin{equation}\label{eq:boxh1}
	\Box \mathop{\overline{h}}_{}\!{}\ab_1 \equiv \frac{16\pi}{c^4}
	\,\mathop{\overline{T}}_{}\!{}\ab\,.
\end{equation}
However, it does not separately satisfy  the harmonic coordinate condition, and the complete solution (which \emph{does} satisfy the harmonic coordinate condition) can be rewritten as the sum of the latter linearized solution and non-linear corrections, in the following form:
\begin{equation}\label{eq:hfull}
	\mathop{\overline{h}}_{}\!{}\ab = G \mathop{\overline{h}}_{}\!{}\ab_1 + \widetilde{\Box}^{-1}_\text{sym}\Bigl[\overline{\Sigma}\ab\Bigr] + \mathop{\overline{v}}_{}\!{}\ab\,.
\end{equation}
The second and third terms represent the non-linear corrections (of order $G^2$, $G^3$, and so on) to be added to the linear solution in order to have the complete metric. In the second term, the gravitational source term is given by
\begin{align}
	\mathop{\overline{\Sigma}}_{}\!{}\ab=\mathop{\overline{\Lambda}}_{}\!{}\ab+\frac{16\pi G}{c^4}\bigl(\vert g\vert-1\bigr)\mathop{\overline{T}}_{}\!{}\ab = \mathop{\overline{\Lambda}}_{}\!{}\ab+ \bigl(\vert g\vert-1\bigr) \Box \bigl(G  \mathop{\overline{h}}_{}\!{}\ab_1\bigr) \,,
\end{align}
and is indeed at least quadratic in $h$, hence of order $G^2$ at least; the third term is also of order $G^2$ at least since it is generated by Eq.~\eqref{eq:div1}. The non-linear iteration starting from the linear solution~\eqref{eq:h1def} can be justified using a reasoning along the lines given around Eqs.~(4.15)--(4.17) in~\cite{B98mult}.\footnote{We have defined the linear approximation with the matter tensor $T\ab$, rather than with $\vert g\vert T\ab$, because the iteration, which is implicit in~\eqref{eq:hfull}, is then very close to the explicit iteration we have performed with \textit{Mathematica} using the potentials in Sec.~\ref{sec:potentials}.}

An important point, which we shall extensively use in this paper, is that once the linear solution $\mathop{\overline{h}}_{}\!{}\ab_1$ is specified, the non-linear iteration ``automatically'' follows through order by order. Suppose that we change the linear solution by adding a linear gauge transformation, \textit{i.e.}, of the type $\partial \xi^{\alpha\beta} \equiv \partial^\alpha \xi^\beta + \partial^\beta \xi^\alpha - \eta^{\alpha\beta} \partial_\gamma \xi^\gamma$ for some vector $\xi^\alpha$ (of order~$G^0$). The non-linear iteration will proceed automatically from the new linear approximation and generate non-linear corrections, such that in the end the new solution will be valid in a different coordinate system, including all non-linear orders beyond the linear gauge transformation. This solution will evidently be as legitimate as the original one. In the next section, we shall play with this freedom to define the RR terms in a particularly interesting coordinate system (coined the extended BT coordinate system), differing from the standard harmonic coordinates by a well chosen linear gauge transformation at the linear level.

\section{The radiation-reaction potentials}\label{sec:RR}

The RR potentials we shall use in the present paper come from the specific multipolar structure of the linearized antisymmetric solution of the Einstein field equations in vacuum defined by Eq.~\eqref{eq:hRR1}. Following previous works~\cite{B96,B98mult,BFIS08}, we can perform the iteration in two equivalent ways: a general way, where the linear metric is parametrized by ``source'' moments $(\dI_L, \dJ_L)$ and ``gauge'' moments $(\dW_L, \dX_L, \dY_L, \dZ_L)$, and a canonical way where it is parametrized by ``canonical'' moments $(\dM_L, \dS_L)$. The relations between these sets of moments have been elucidated in previous papers~\cite{BFIS08,BFL22,BFHLT23b}. Since the two constructions differ at linear level by a linear gauge transformation we are free to adopt either construction (see the argument at the end of Sec.~\ref{sec:nonlinear}). Choosing the canonical version, we parametrize the linear antisymmetric metric~\eqref{eq:hRR1} by the STF mass and current canonical moments as
\bse\label{eq:hRR1canonical}
	\begin{align}
		\mathop{\overline{h}}_{}\!{}^{00}_{\mathrm{RR}\,1} &= -\frac{4}{c^2}\sum_{\ell=0}^{+\infty}
		\frac{(-)^\ell}{\ell !} \partial_L \bigl\{\dM_L\bigr\}\,, \\ \mathop{\overline{h}}_{}\!{}^{0i}_{\mathrm{RR}\,1} &=
		\frac{4}{c^3}\sum_{\ell=1}^{+\infty} \frac{(-)^\ell}{\ell !}
		\left[ \partial_{L-1} \bigl\{
		\dM_{iL-1}^{(1)}\bigr\} + \frac{\ell}{\ell+1}
		\varepsilon_{iab} \partial_{aL-1} \bigl\{ \dS_{bL-1}\bigr\}\right]\,, \\ \mathop{\overline{h}}_{}\!{}^{ij}_{\mathrm{RR}\,1} &=
		-\frac{4}{c^4}\sum_{\ell=2}^{+\infty} \frac{(-)^\ell}{\ell !}
		\left[ \partial_{L-2} \bigl\{ \dM_{ijL-2}^{(2)} \bigr\} + \frac{2\ell}{\ell+1}
		\partial_{aL-2} \bigl\{ \varepsilon_{ab(i}
		\dS_{j)bL-2}^{(1)} \bigr\}\right]\,,
	\end{align}
\ese
following the notation~\eqref{eq:antisym}--\eqref{eq:antisymL} for antisymmetric multipolar waves. Note that, in fact, the expressions~\eqref{eq:hRR1canonical} involve only dynamical (time-varying) moments $\dM_L(t)$ and $\dS_L(t)$ with $\ell\geqslant 2$. On the linear metric~\eqref{eq:hRR1canonical}, we perform a linear gauge transformation with vector $\xi^{\alpha}_{1}$ given by
\bse\label{eq:h1can}
\begin{align}
	\xi^{0}_{1} &= \frac{2}{c}\sum_{\ell=2}^{+\infty}
	\frac{(-)^\ell}{\ell !} \frac{2\ell+1}{\ell(\ell-1)} \partial_L \bigl\{\dM_L^{(-1)}\bigr\}\,, \\ \xi^{i}_{1} &=
	-2\sum_{\ell=2}^{+\infty} \frac{(-)^\ell}{\ell!\ell(\ell-1)} \left[(2\ell+1)(2\ell+3)
	\hat{\partial}_{iL} \bigl\{
	\dM_{L}^{(-2)}\bigr\} - \frac{\ell(2\ell-1)}{c^2}\partial_{L-1}\bigl\{\dM_{iL-1}\bigr\} \right]\\
	&~ + \frac{4}{c^2} \sum_{\ell=2}^{+\infty} \frac{(-)^\ell \ell}{(\ell+1)!} \frac{2\ell+1}{\ell-1}
	\varepsilon_{iab} \partial_{aL-1} \bigl\{ \dS_{bL-1}^{(-1)}\bigr\}\,.
\end{align}
\ese
We have introduced some time anti-derivatives of the multipole moments, \textit{e.g.}, $\dM_L^{(-1)}(t)$, but such anti-derivatives will always disappear when PN expanded following the formula~\eqref{eq:PNantisymL}, as shown in Eqs.~\eqref{eq:PNVRR} below. Under the linear gauge transformation $\partial\xi_1\ab \equiv \partial^\alpha\xi_1^\beta + \partial^\beta\xi_1^\alpha - \eta\ab\partial_\gamma\xi_1^\gamma$, the RR part of the metric $\mathop{\overline{h}}_{}\!{}\ab_{\mathrm{RR}\,1}$ is transformed into
\begin{equation}
	\label{eq:h1RRprime}
	%\mathop{\overline{h}'}_{}\!{}\ab_{\mathrm{RR}\,1} =
	\mathop{\overline{h}}_{}\!{}\ab_{\mathrm{RR}\,1} + \partial\xi_1\ab \equiv 
	\left\{\begin{array}{l} \displaystyle - \frac{4}{G c^2}V\RR \\[0.4cm]
		\displaystyle - \frac{4}{G c^3}V\RR^i \\[0.4cm]
		\displaystyle - \frac{4}{G c^4}V\RR^{ij}
	\end{array}\right. \,.
\end{equation}
This allows us to define some scalar, vector, and tensor RR potentials $V\RR\ab\equiv(V\RR, V\RR^{i}, V\RR^{ij})$ from, respectively, the~$00$, $0i$, and $ij$ components of the new linearized metric, as described in~\eqref{eq:h1RRprime}. The explicit expressions of the RR potentials for any multipolarity $\ell$ is~\cite{B93}
\bse\label{eq:defRRpot}
	\begin{align}
 V\RR &= G \sum_{\ell=2}^{+\infty}
		\frac{(-)^\ell}{\ell!} \frac{(\ell+1)(\ell+2)}{\ell(\ell-1)} \partial_L \bigl\{\dM_L\bigr\}\,, \\
%%%%%%%%%%%%%%%%%%%%%%%%%%%%%%%%%%%%%%%%%%%%%%%%%%%%%%%%%%%%%%%%%%%%%%%%
 V\RR^{i} &= - c^2 G \sum_{\ell=2}^{+\infty} \frac{(-)^\ell}{\ell!} \frac{\ell+2}{\ell-1} \left[ \frac{2\ell+1}{\ell}
		\,\hat{\partial}_{iL} \bigl\{
		\dM_{L}^{(-1)}\bigr\} - \frac{\ell}{(\ell+1)c^2}
		\,\varepsilon_{iab} \partial_{aL-1} \bigl\{ \dS_{bL-1}\bigr\}\right]\,, \\ 
%%%%%%%%%%%%%%%%%%%%%%%%%%%%%%%%%%%%%%%%%%%%%%%%%%%%%%%%%%%%%%%%%%%%%%%%%
 V\RR^{ij} &=
		c^4 G \sum_{\ell=2}^{+\infty} \frac{(-)^\ell}{\ell!}\frac{2\ell+1}{\ell-1}
		\left[ \frac{2\ell+3}{\ell}\hat{\partial}_{ijL} \bigl\{ \dM_{L}^{(-2)} \bigr\} - \frac{2\ell}{(\ell+1)c^2}
		\varepsilon_{ab(i} \hat{\partial}_{j)aL-1} \bigl\{ 
		\dS_{bL-1}^{(-1)} \bigr\}\right]\,.
	\end{align}\ese
The crucial point about choosing the RR potentials is that, when computing the PN expansion using~\eqref{eq:PNantisymL}, one finds that the PN orders of each of the potentials is dominantly $\calO(5)$. This means that, taking into account the factors of~$1/c$ in Eq.~\eqref{eq:h1RRprime},  $V\RR$ starts contributing at 2.5PN order [see the leading 2.5PN term~\eqref{eq:VRR}], while $V\RR^i$ affects the RR dominantly at the 3.5PN order, and $V\RR^{ij}$ starts only at the 4.5PN order. This is to be contrasted with the situation in standard harmonic coordinates, where all components of the corresponding RR potentials (scalar, vector, and tensor) contribute dominantly to the 2.5PN order. Expanding Eqs.~\eqref{eq:defRRpot} up to 2PN relative order, \textit{i.e.}, corresponding to leading 2.5PN, next-to-leading 3.5PN and next-to-next-to-leading 4.5PN terms, one finds
\bse\label{eq:PNVRR}
	\begin{align}
		V\RR &= - \frac{G}{5 c^5} x^{ab} \dM_{ab}^{(5)} + \frac{G}{c^7} \left[ \frac{1}{189} x^{abc} \dM_{abc}^{(7)} - \frac{1}{70} r^2 x^{ab} \dM_{ab}^{(7)}\right] \nn\\ &~ + \frac{G}{c^9} \left[ - \frac{1}{9072} x^{abcd} \dM_{abcd}^{(9)} + \frac{1}{3402} r^2 x^{abc} \dM_{abc}^{(9)} - \frac{1}{2520} r^4 x^{ab} \dM_{ab}^{(9)}\right] + \calO\bigl(11\bigr) \,, \\
		%%%%%%%%%%%%%%%%%%%%%%%%%%%%%%%%%%%%%%%%%%%%%%%%%%%%%%%%%%%%%%%%%%%%%%%%
		V\RR^{i} &= \frac{G}{c^5} \left[\frac{1}{21} \hat{x}^{iab} \dM_{ab}^{(6)} -\frac{4}{45} \varepsilon_{iab} x^{ac} \dS_{bc}^{(5)}\right] \nn\\&~ + \frac{G}{c^7} \left[-\frac{1}{972} \hat{x}^{iabc} \dM_{abc}^{(8)} + \frac{1}{378} r^2\hat{x}^{iab} \dM_{ab}^{(8)} + \frac{1}{336} \varepsilon_{iab} \hat{x}^{acd} \dS_{bcd}^{(7)} - \frac{2}{315} \varepsilon_{iab} r^2 \hat{x}^{ac} \dS_{bc}^{(7)} \right] + \calO\bigl(9\bigr) \,, \\ 
		%%%%%%%%%%%%%%%%%%%%%%%%%%%%%%%%%%%%%%%%%%%%%%%%%%%%%%%%%%%%%%%%%%%%%%%%%
		V\RR^{ij} &= \frac{G}{c^5} \left[ - \frac{1}{108} \hat{x}^{ijab} \dM_{ab}^{(7)} + \frac{2}{63} \varepsilon_{ab(i} \hat{x}^{j)ac} \dS_{bc}^{(6)}\right] + \calO\bigl(7\bigr)\,.
	\end{align}\ese
The coordinate system in which the previous RR potentials hold (after non-linear iteration) generalizes the BT coordinate system~\cite{Bu71, BuTh70} to 2PN relative order and will be referred to as the ``extended BT'' coordinate system.

\section{The radiation-reaction acceleration at 4.5PN order}
\label{sec:acceleration}

\subsection{Metric and equations of motion in terms of symmetric and radiation-reaction potentials}
\label{sec:potentials}

We shall write the metric in extended BT coordinates for the RR contributions, and in the usual harmonic coordinates for the conservative contributions. It is convenient to use the same notation  as in previous works in harmonic coordinates for the PN elementary potentials $V$, $V_i$, $ \hat{W}_{ij}$, $\hat{X}$, and $\hat{R}_i$ (see the full definition in Appendix A of~\cite{MHLMFB20}), but beware that the RR contributions will not have the same meaning as in standard harmonic coordinates. We thus define the metric in extended BT coordinates as 
\bse\label{eq:metric_g}\begin{align}
g_{00} &= -1 + \frac{2}{c^2} V - \frac{2}{c^4} V^2 + \frac{8}{c^6} \Bigl[\hat{X} + V_i V_i + \frac{1}{6} V^3\Bigr] + \calO\bigl( 8, 13\bigr) \,, \\
g_{0i} &= - \frac{4}{c^3} V_i - \frac{8}{c^5} \hat{R}_i + \calO\bigl( 7, 12\bigr) \,, \\
g_{ij} &= \delta_{ij}\Bigl[1+ \frac{2}{c^2} V + \frac{2}{c^4} V^2 \Bigr]
+ \frac{4}{c^4} \hat{W}_{ij} + \calO\bigl(6,11\bigr) \,.
\end{align}\ese
The PN accuracy of the metric is defined for both the conservative (time-even) and dissipative (time-odd, RR) parts: recall the notation $\calO(p,q)$, which means that we neglect even terms $\calO(p)$ and odd terms $\calO(q)$. Thus, Eq.~\eqref{eq:metric_g} is accurate to relative order 2PN for both the conservative and RR terms, as can be seen from the fact that the symbols $\calO(p,q)$ in~\eqref{eq:metric_g} have $q-p=5$, which reflects the fact that $\calO(5)$ is the leading order of RR. In extended BT coordinates, the definition of the potentials is as follows. The leading scalar, vector, and tensor potentials $V$, $V_i$, and $\hat{W}_{ij}$ contain the RR potentials given by~\eqref{eq:PNVRR},
\bse\label{eq:VViWij}\begin{align}
	V &= V\sym + V\RR \,,\\
	V_i &= V\sym^i + V\RR^i \,,\\
	\hat{W}_{ij} &= \hat{W}\sym^{ij} + V\RR^{ij} \,,
\end{align}\ese
while the subleading ones $\hat{R}_i$ and $\hat{X}$ do not contain them explicitly, 
\bse\label{eq:XRi}\begin{align}
	\hat{R}_{i} &= \hat{R}\sym^{i} \,,\\
	\hat{X} &= \hat{X}\sym \,.
\end{align}\ese
In~\eqref{eq:VViWij}--\eqref{eq:XRi}, the subscript ``sym'' refers to the way the potentials are defined, \textit{i.e.}, by means of the (regularized) symmetric propagator given by Eq.~\eqref{eq:propsym}, in contrast with other works in harmonic coordinates where they are defined using the retarded operator (see~\cite{MHLMFB20}). This is because the RR parts of the potentials, which would generally arise from the odd powers in the $1/c$ expansion of the retardations, are now included separately in the form of the potentials~\eqref{eq:PNVRR}. We thus have
\bse\label{eq:sym_potentials_def}
\begin{align}
	V\sym &= \widetilde{\Box}^{-1}\sym \Big[- 4 \pi G \sigma \Big]\,,\\
	V\sym^i &= \widetilde{\Box}^{-1}\sym \Big[- 4 \pi G \sigma_i \Big] \,,\\
	\hat{W}\sym^{ij} &= \widetilde{\Box}^{-1}\sym \Big[- 4 \pi G \bigl(\sigma_{ij} - \delta_{ij} \sigma_{kk}\bigr) - \p_i V \p_j V \Big] \,, \\
	\hat{R}\sym^i &= \widetilde{\Box}^{-1}\sym \Big[- 4 \pi G\big( V \sigma_i - V_i \sigma \big) - 2 \p_k V \p_i V_k - \frac{3}{2} \p_t V \p_i V \Big] \,, \\
	\hat{X}\sym &=  \widetilde{\Box}^{-1}\sym \Big[-4 \pi G V \sigma_{kk} + \hat{W}_{ij} \p_{ij} V + 2 V_i \p_t \p_i V + V \p_t^2 V + \frac{3}{2} (\p_t V)^2 - 2 \p_i V_j \p_j V_i \Big]\,. 
\end{align}\ese
Very importantly in this approach, the ``symmetric'' potentials actually involve some RR contributions, because their sources in the right-hand sides of~\eqref{eq:sym_potentials_def} are defined with the complete potentials~\eqref{eq:VViWij}. Thus, the RR terms in the metric~\eqref{eq:metric_g} are to be computed by iteration (see Sec.~\ref{sec:RRsym}).
The iteration of the RR terms in the metric, with the RR potentials $V\ab\RR$ introduced \textit{linearly} in the scalar, vector, and tensor potentials $V$, $V_i$, and $\hat{W}_{ij}$ [see~\eqref{eq:VViWij}], is justified by the investigation of Sec.~\ref{sec:nonlinear}, which constructs the RR part of the metric by iteration of the linear approximation~\eqref{eq:hRR1} parametrized (linearly) by the potentials $V\ab\RR$.

In the case of compact binary systems of non-spinning point-particles, the matter source takes the form
\bse\label{eq:sigma}
\begin{align}
\sigma &= \tilde{\mu}_1 \delta_1 + (1 \leftrightarrow 2) \,,\\*
\sigma_i &= \mu_1 v_1^i \delta_1 + (1 \leftrightarrow 2) \,,\\*
\sigma_{ij} &= \mu_1 v_1^i  v_1^i \delta_1 + (1 \leftrightarrow 2) \,,
\end{align}\ese
where $\delta_1\equiv\delta^{(3)}(\mathbf{x}-\mathbf{y}_1)$ is the three-dimensional delta-function at the location of the particle 1, $v_1^i\equiv\dd y_1^i/\dd t$ is the coordinate velocity of the particle (with $t\equiv x^0/c$ the coordinate time), and $(1 \leftrightarrow 2)$ refers to the corresponding contribution for the particle 2. The effective masses $\tilde{\mu}_1(t)$ and $\mu_1(t)$ are just functions of time and are defined from the regularization of the stress-energy tensor of point particles as 
\bse\label{eq:mu}\begin{align}
	\tilde{\mu}_1 &= \biggl(1+\frac{v_1^2}{c^2}\biggr) \mu_1 \,,\\ \mu_1 &= \Bigg(\frac{m_1 c}{\sqrt{g \,g_{\alpha\beta} \,v_1^\alpha v_1^\beta}}\Biggr)_1\,,
\end{align}\ese
where $m_1$ and $m_2$ are the constant PN masses (we neglect the black hole absorption), where we pose $v_1^\alpha\equiv(c,v_1^i)$, and where the metric $g_{\alpha\beta}$ and its determinant $g$ are evaluated at the location of the particle 1 following the regularization. In principle, the ultraviolet (UV) dimensional regularization scheme should be systematically applied, but at the relatively low 2PN order, the Hadamard regularization is sufficient in practice. The explicit expression of $\tilde{\mu}_1$ is given by Eq.~(4.2) of~\cite{BFP98},  and reads
\begin{align}\label{eq:muTilde}
	\tilde{\mu}_1 = m_1 \Bigg( 1 + \frac{1}{c^2} \left[-V + \frac{3}{2}
	v_1^2\right] + \frac{1}{c^4} \left[-2\hat{W} + \frac{1}{2} V^2 + 
	\frac{1}{2} V v_1^2
	- 4 V_i v_1^i + \frac{7}{8} v_1^4\right] \Bigg)_1 + \calO(6,11) \,,
\end{align}
where $V$, $V_i$, and $\hat{W}_{ij}$ (as well as $\hat{W}=\hat{W}_{ii}$) are given by~\eqref{eq:VViWij}, and the notation $(\dots)_1$ refers to the UV regularization applied to compute the potentials at the location of the particle 1. We see that the explicit expressions of $\mu_1$ and $\tilde{\mu}_1$ include the contribution of the potentials $V\RR\ab$.

%\subsection{Equations of motion in terms of the potentials}

The EOM follow from the conservation of the matter tensor $\nabla_{\beta} T^{\alpha\beta} = 0$. For non-spinning point particles, they are conveniently written as
\begin{equation}\label{eq:dPdtF}
	\frac{\dd P_1^i}{\dd t} = F_1^i \,,
\end{equation}
where $P_1^i$ and $F_1^i$ result from the geodesic equations on the metric generated by the particles themselves as
\begin{align} 
	P_1^i &= \Bigg(\frac{g_{i\alpha}\,c \,v_1^\alpha}{ \sqrt{- g_{\beta\gamma} \,v_1^\beta
				v_1^\gamma}}\Bigg)_1\,,\\
	F_1^i &= \Bigg(\frac{c}{2}\frac{ \partial_i g_{\alpha\beta} \,v_1^\alpha
		v_1^\beta}{ \sqrt{- g_{\gamma\delta} \,v_1^\gamma v_1^\delta}}\Bigg)_1\,.
\end{align}
% 
%Further denoting $P_1^i=v_1^i+Q_1^i$, the ordinary acceleration $a_1^i=\dd v_1^i/\dd t$ of the particle is $a_1^i = F_1^i - \frac{\dd Q_1^i}{\dd t}$.
%
%\begin{equation}\label{eq:accFQ}
%	a_1^i = F_1^i - \frac{\dd Q_1^i}{\dd t}\,,
%\end{equation}
%
With the metric~\eqref{eq:metric_g} written in terms of potentials in ``standard'' form, we can directly import known formulas for the 2PN-accurate EOM in terms of elementary potentials~\cite{BFP98}. These formulas will be valid for both the conservative and dissipative effects up to the PN remainder $\calO(6,11)$. We have\footnote{We also notice that $F_1^{i}$ can be written in the form of a spatial gradient:
\begin{align*}
	F_1^{i} &= \Bigg(\partial_i \Biggl\{V +\frac{1}{c^2} \left[ -\frac{1}{2}V^2
	+\frac{3}{2} V v_1^2-4 V_j v_1^j \right] +
	\frac{1}{c^4} \left[\frac{7}{8} V v_1^4-2 V_j v_1^j
	v_1^2 + \frac{9}{4} V^2 v_1^2 +2 \hat{W}_{jk} v_1^j
	v_1^k \right.\nn\\& \left.\qquad\qquad
	-4 V_j V v_1^j -8 \hat{R}_j v_1^j+\frac{1}{6} V^3
	+ 4 V_j V_j+4 \hat{X} \right]\Biggr\}\Bigg)_1 + \calO(6,11) \,.
\end{align*}}
\bse\label{eq:QiFi}\begin{align}
	P_1^{i} &= v_1^i + \Bigg(\frac{1}{c^2} \left[\frac{1}{2} v_1^2v_1^i +3 V v_1^i-4 V_i
	\right] \nn\\*&\qquad\quad + \frac{1}{c^4} \left[\frac{3}{8} v_1^4 v_1^i+\frac{7}{2} V
	v_1^2 v_1^i-4 V_j v_1^i v_1^j -2 V_i v_1^2 + \frac{9}{2} V^2 v_1^i-4 V V_i +4
	\hat{W}_{ij} v_1^j-8 \hat{R}_i \right]\Bigg)_1 + \calO\bigl(6,11\bigr)\,,\label{eq:Qi}\\*
	%%%%%%%%%%%%%%%%%%%%%%%%%%%%%%%%%%%%%%%%%%%%%%%%%%%%%%%%%%%%%%%%
	F_1^{i} &= \Bigg(\partial_i V +\frac{1}{c^2} \left[ -V \partial_i V
	+\frac{3}{2} \partial_i V v_1^2-4 \partial_i V_j v_1^j \right]\nn\\* &\qquad +
	\frac{1}{c^4} \left[\frac{7}{8} \partial_i V v_1^4-2 \partial_i V_j v_1^j
	v_1^2 + \frac{9}{2} V \partial_i V v_1^2 +2 \partial_i \hat{W}_{jk} v_1^j
	v_1^k \right.\nn\\*& \left.\qquad\qquad -4 V_j \partial_i V v_1^j-4 V
	\partial_i V_j v_1^j -8 \partial_i \hat{R}_j v_1^j+\frac{1}{2} V^2
	\partial_i V +8 V_j \partial_i V_j+4 \partial_i \hat{X} \right]\Bigg)_1 + \calO(6,11) \,.
\end{align}\ese
Alternatively, the acceleration $a_1^i = F_1^i - \dd Q_1^i/\dd t$ (with $Q_1^{i} \equiv P_1^{i}-v_1^i$) can be written in fully expanded form as
\begin{align}\label{eq:a1i_pot}
	a_1^i &= \Bigg( \p_i V\ + \frac{1}{c^2}\Big[ (v_1^2 - 4 V) \p_i V + 4 \p_t V_i -  8 v_1^j \p_{[i}V_{j]} - 3 v_1^i \p_t V  - 4 v_1^i v_1^j \p_j V  \Big]\nn \\*
	&\qquad\qquad\, +\frac{1}{c^4}\Big[4 v_1^i V_j \p_j V + 4 v_1^i v_1^j v_1^k \p_j V_k + 8 v_1^j V_i \p_j V + 8 \p_t \hat{R}_i + v_1^i v_1^2 \p_t V  + 4 V_i \p_t V \nn\\*
	&\qquad\qquad\qquad - 8 V \p_t V_i - 4 v_1^j \p_t \hat{W}_{ij} + 8 v_1^j \p_j\hat{R}_i - 8 V v_1^j\p_j V_i - 4 \hat{W}_{ij} \p_j V - 4 v_1^j v_1^k \p_k\hat{W}_{ij}\nn\\*
	&\qquad\qquad\qquad   - 8 v_1^j \p_i \hat{R}_j + 8 V^2 \p_i V + 8 V v_1^j  \p_i V_j + 8 V_j \p_i V_j + 2 v_1^j v_1^k  \p_i \hat{W}_{jk} + 4 \p_i \hat{X}   \Big] \Bigg)_1 +\calO(6,11) \,.
\end{align}
We can compute the expression of the acceleration $a_1^i$ in terms of the positions ($y_1^i$, $y_2^i$) and velocities ($v_1^i$, $v_2^i$) in two ways. The first one consists in using the formulation~\eqref{eq:QiFi} and computing the time derivative of $Q_1^{i}$ with the ``convective'' time derivative, namely
%\footnote{This relation can easily be checked for any ``translation-invariant'' function $F(\mathbf{r}_A,\bm{v}_A)$, \textit{i.e.}, depending on the position $\mathbf{x}$ only through the distance to the trajectories $\mathbf{r}_A(t)=\mathbf{x}-\bm{y}_A(t)$ (which is the case of all functions encountered in the present problem). For such a function, we have $\partial^i \!F + \partial_1^i F + \partial_2^i F =0$, where $\partial_A^i \equiv\partial/\partial y_A^i$ is the partial derivative with respect to the source point position.}
%
\begin{align}\label{eq:convective}  
	\frac{\dd}{\dd t}(F)_1 = \Bigl(\p_t F + v_1^i \,\p_i F\Bigr)_1\,.
\end{align}
The second method, used at 1PN order in Ref.~\cite{IW95}, resorts to the expanded form~\eqref{eq:a1i_pot}. This involves the partial time derivative $\partial_t F$ of some potentials, and the equivalence with the first method is guaranteed by the formula~\eqref{eq:convective}. In practice, we prefer following the first method, \textit{i.e.}, computing the quantities~\eqref{eq:QiFi} and applying the convective time derivative~\eqref{eq:convective}. Finally, to avoid possible confusion, we remind the reader that the regularization of a product of (derivatives of) potentials should be equal to the product of the regularizations of these potentials. This is the so-called ``distributivity'' property of the regularization, $(FG)_1=(F)_1(G)_1$, which holds with dimensional regularization but is violated by the Hadamard regularization (this is the main source of ambiguities with the Hadamard regularization, see~\cite{BDE04}). 
In the present work, we use Hadamard's regularization supplemented by the prescription that, e.g., $(V \p_i V)_1$ should be computed as $(V)_1 (\p_i V)_1$. This corresponds to the ``pure Hadamard-Schwartz'' regularization, which is equivalent to dimensional regularization at the low relative PN orders we are working at~\cite{BDE04}.

\subsection{Radiation-reaction contributions arising from symmetric potentials}
\label{sec:RRsym}

In this section, we determine the RR contributions to the acceleration that are hidden in the symmetric potentials defined by Eqs.~\eqref{eq:sym_potentials_def}. In order to compute them, a certain number of tools will be handy. First, most inverse Laplace regularized operators defined by Eq.~\eqref{eq:genpoissiter} reduce to 
\begin{align}\label{eq:invLaplace}
   \widetilde{\Delta}^{-1}\Big[r_1^\lambda \,\hat{n}_1^L\Big] = \FP \Delta^{-1}\Big[r^B r_1^\lambda \,\hat{n}_1^L\Big]\,,
\end{align}
where we have set $r_0=1$ for convenience. When computing~\eqref{eq:invLaplace}, we have to worry about the presence of possible poles $\propto 1/B$. As we can argue that there cannot be double poles~\cite{BDE04}, it is sufficient to expand $r^B$ around $r_1^B$ to first order in $B$, \textit{i.e.}, $r^B=r_1^B\big[1+B\frac{(n_1y_1)}{r_1}+\calO(B^2)\big]$. Hence, the formula used in our calculations is 
\begin{align}
	\widetilde{\Delta}^{-1}\Big[r_1^\lambda \,\hat{n}_1^L\Big] = \FP \biggl\{\Delta^{-1}\Big[r_1^{B+\lambda} \,\hat{n}_1^L\Big] + B y_1^i \Delta^{-1}\Big[r_1^{B+\lambda-1} \,\hat{n}_1^L n_1^i\Big] + \calO(B^2)\biggr\}\,, 
\end{align}
together with the Matthieu formula valid in the absence of poles,
\begin{align}\label{eq:matthieu}
	\Delta^{-1}\Bigl[r_1^{B+\lambda} \hat{n}_1^L\Bigr]\bigg|_{B=0} =  \frac{r_1^{\lambda+2}\hat{n}^L_1}{(\lambda-\ell+2)(\lambda+\ell+3)} \,.
\end{align}
We have explicitly found that no poles $1/B$ occur in our 2PN-accurate calculation; hence we just use~\eqref{eq:matthieu} in all cases. 

In addition to the Matthieu formula, we  also require the solutions of more complicated Poisson equations, in the form of a hierarchy of elementary ``kernel'' functions, the most famous one being the Fock~\cite{Fock} kernel function $g_\text{\,Fock} = \ln (r_1 + r_2 + r_{12})$. The Poisson equations we need to solve for the present purpose are 
\begin{align}
		\Delta g &= \frac{1}{r_1r_2}\,, & \Delta f &= g\,, \nn\\ 
		\Delta f_{12} &= \frac{r_1}{2r_2}\,, & \Delta f_{21} &= \frac{r_2}{2r_1}\,, \nn\\
	     \Delta h_{12} &= \frac{r_1^3}{24\,r_2}\,, & \Delta h_{21} &= \frac{r_2^3}{24\,r_1}\,.\label{eq:kernels_def}
\end{align}
%
%
%\bse\label{eq:kernels_def}
%\begin{align}
%	\Delta g &= \frac{1}{r_1r_2}\,, \\ \Delta f &= g\,, \qquad\quad \Delta f_{12} = \frac{r_1}{2r_2}\,, \qquad\quad \Delta f_{21} = \frac{r_2}{2r_1}\,, \\ \Delta h &= f\,, \qquad\quad \Delta h_{12} = \frac{r_1^3}{24\,r_2}\,, \qquad\quad \Delta h_{21} = \frac{r_2^3}{24\,r_1}\,.
%\end{align}\ese
%
Particular solutions of those equations can be constructed starting from $g_\text{\,Fock}$,  and to these we must add some specific homogeneous solutions, which ensure the matching between the near zone and the exterior zone. The general procedure to obtain the homogeneous solutions is described in Sec.~V C 2 of~\cite{MHLMFB20} (see also~\cite{BFeom, JaraS98, JaraS15}). Posing $S \equiv r_1 + r_2 + r_{12}$, the required ``matched'' solutions to Eq.~\eqref{eq:kernels_def} (valid in the sense of distribution theory) are 
%\Luc{D'où vient le $h12$ est-ce que c'est mon répertoire Noyaux?}
%
\bse\label{eq:kernels}
	\begin{align}
		g &= \ln\left(\frac{S}{2r_0}\right)-1\,, \\
		%%%%%%%%%%%%%%%%%%%%%%%%%%%%%%%%%%%%%%%%%%%%%%%%%%%
		f &= \frac{1}{12}\Biggl[\bigl(r_{1}^2+r_{2}^2-r_{12}^2\bigr)\biggl( \ln\left(\frac{S}{2r_0}\right) + \frac{1}{6}\biggr) + r_{1}r_{12}+r_{2}r_{12}-r_{1}r_{2} +2(xy_1)+2(xy_2)-3r^2\Biggr]\,,\\
		%%%%%%%%%%%%%%%%%%%%%%%%%%%%%%%%%%%%%%%%%%%%%%%%%%%%%%%%%%%%
	f_{12} &=  \frac{1}{12}\Biggl[\bigl(r_{1}^2+r_{12}^2-r_{2}^2\bigr)\biggl( \ln\left(\frac{S}{2r_0}\right) + \frac{1}{6} \biggr) + r_{1}r_{2}+r_{12}r_{2}-r_{1}r_{12} +2(xy_1)+2(y_1y_2)-3y_1^2\Biggr]\,,\\
	%%%%%%%%%%%%%%%%%%%%%%%%%%%%%%%%%%%%%%%%%%%%%%%%%%%%%%%%%%%%%%%%%%%%
	h_{12} &= \frac{1}{360} \Biggl[\frac{9}{8} \biggl( r_{1}^4 + r_{12}^4 - 2 r_{12}^2 r_{2}^2 + r_{2}^4 + \frac{2}{3} r_{1}^2 r_{12}^2 - 2 r_{1}^2 r_{2}^2\biggl) \ln \Bigl(\frac{S}{2 r_{0}}\Bigr)+ \frac{69}{160} r_{1}^4 + \frac{69}{160} r_{12}^4
 + \frac{15}{8} r_{12}^3 r_{2} \nn\\ & \qquad\quad -  \frac{69}{80} r_{12}^2 r_{2}^2
 -  \frac{9}{8} r_{12} r_{2}^3 + \frac{69}{160} r_{2}^4 + r_{1}^3 \Bigl(- \frac{9}{8} r_{12}
 + \frac{15}{8} r_{2}\Bigr) + r_{1}^2 \Bigl(\frac{63}{80} r_{12}^2+ \frac{9}{8} r_{12} r_{2}
 -  \frac{69}{80} r_{2}^2\Bigr)\nn\\ & \qquad\quad + \frac{9}{8} r_{1} \bigl(- r_{12}^3
 + r_{12}^2 r_{2} + r_{12} r_{2}^2 - r_{2}^3\bigr)
 +3 (y_2^i - 2 y_1^i) y_1^j \hat{x}^{ij} - 4 (x y_{1}) (y_{1} y_{2}) + 3 (x y_{2}) (y_{1} y_{2}) \nn\\
& \qquad\quad 
 - 6 (y_{1} y_{2})^2 + \frac{19}{2} (x y_{1}) y_{1}^{2} - \frac{9}{2} (x y_{2}) y_{1}^{2}
 + \frac{19}{2} (y_{1} y_{2}) y_{1}^{2} - \frac{25}{4} y_{1}^{4}
 - (x y_{1}) y_{2}^{2} + 2 y_{1}^{2} y_{2}^{2}\Biggl]\,,
\end{align}\ese
where we denote scalar products by, for instance, $(x y_1) = \mathbf{x}\cdot\bm{y}_1$. From \eqref{eq:kernels}, the expressions for $f_{21}$ and $h_{21}$ are simply obtained by exchanging the particle labels 1 and 2. 
%with the source points, \textit{i.e.}
%
%\bse\label{eq:f12part}
%\begin{align}
%        f_{21} &= f_{12} \big|_{{\bf y}_1 \longleftrightarrow {\bf y}_2}\,, &  h_{21} &=h_{12} \big|_{{\bf y}_1 \longleftrightarrow {\bf y}_2}\,.
%\end{align}
%\ese
%
%For the present application we need the functions $h_{12}$ and $h_{21}$ but not the function $h$. 
For the present application, we have found that the aforementioned homogeneous solutions do not contribute to the end results.

We compute the inverse Laplace operators thanks to~\eqref{eq:matthieu} and the above elementary kernels, and obtain the RR parts in the symmetric type potentials. Those can appear in four ways: (i) through the explicit dependence on the canonical moments $\dM_L$ and $\dS_L$ or on the RR potentials $V\RR\ab$; (ii) from the expressions~\eqref{eq:mu} of $\mu_A$ and $\tilde{\mu}_A$, which explicitly contain $V\RR\ab$; (iii) from the symmetric potentials entering $\mu_A$ and $\tilde{\mu}_A$, which  implicitly depend on the RR terms; (iv) from the expressions of the accelerations arising from the time derivatives hitting on $r_A$, $v_A^i$, $\mu_A$ and $\tilde{\mu}_A$ (only the 3.5PN acceleration is needed, and given by (3.11) of~\cite{IW95}). The results are 
\bse\label{eq:symRR}
\begin{align}
	V\sym\Big|\RR &= G\bigg(\frac{\tilde{\mu}_1}{r_1} + \frac{1}{2 c^2} \frac{\dd^2}{\dd t^2}\left[\tilde{\mu}_1 r_1 \right] + \frac{1}{24 c^4} \frac{\dd^4}{\dd t^4}\left[\tilde{\mu}_1 r_1^3 \right] \bigg)\Big|\RR + (1 \leftrightarrow 2) + \calO(6,11) \,,\\
	%%%%%%%%%%%%%%%%%%%%%%%%%%%%%%%%%%%%%%%%%%%%%%%%%%%%%%%%%%%%%%%%%%%%%%%%%%%%
	V\sym^i\Big|\RR &= G \bigg(\frac{\mu_1 v_1^i}{r_1} + \frac{1}{2 c^2} \frac{\dd^2}{\dd t^2}\left[\mu_1 v_1^i r_1 \right]\bigg)\Big|\RR + (1 \leftrightarrow 2) + \calO(4,9) \,, \\
	%%%%%%%%%%%%%%%%%%%%%%%%%%%%%%%%%%%%%%%%%%%%%%%%%%%%%%%%%%%%%%%%%%%%%%%%%%%%
	\hat{W}\sym^{ij}\Big|\RR &=  \frac{G^2 m_1}{5 c^5} \dM_{a (i}^{(5)}\left( 2 n_1^{j)}y_1^a + r_1 n_1^{j)} n_1^{a} - r_1 \delta_{j) a}\right) + (1 \leftrightarrow 2) + \calO(2,7) \,, \\
	%%%%%%%%%%%%%%%%%%%%%%%%%%%%%%%%%%%%%%%%%%%%%%%%%%%%%%%%%%%%%%%%%%%%%%%%%%%%
	\hat{R}\sym^i\Big|\RR &=  \frac{G m_1}{r_1} \Big(v_1^i V\RR - V\RR^i \Big)_1  \nn\\
	& + \frac{G^2 m_1 }{c^5} \Bigg[\frac{1}{5} n_1^i v_1^j \left(r_1 n_1^{k} +2  y_1^k  \right)\dM_{jk}^{(5)}  - \frac{1}{20}\left(r_1 v_1^j + 3 r_1 n_1^j n_1^k v_1^k + 6 y_1^j n_1^k v_1^k \right)\dM_{ij}^{(5)}\nn\\
	& \qquad\qquad + \frac{1}{1290}\left(59 r_1^2 n_1^i n_1^j n_1^k + 153 r_1 n_1^i n_1^j y_1^k  + 129 n_1^i y_1^j y_1^k + 24 r_1 y_1^i n_1^j n_1^k + 48 y_1^i y_1^j n_1^k \right) \dM_{jk}^{(6)}\nn \\
	&  \qquad\qquad -  \frac{1}{1290}\left( 11 r_1^2 n_1^j + 36 r_1 n_1^j n_1^k y_1^k + 120 y_1^j n_1^k y_1^k  - 3 r_1 y_1^j - 24 n_1^j y_1^k y_1^k \right) \dM_{ij}^{(6)} \nn\\
	& \qquad\qquad  + \frac{2}{45} \varepsilon_{ijk}\left(r_1 n_1^k n_1^l + 2 n_1^k y_1^l \right)\dS_{jl}^{(5)}  + \frac{4}{45}\varepsilon_{jkl} n_1^k y_1^l \dS_{ij}^{(5)}   \Bigg] 
	+ (1 \leftrightarrow 2) + \calO(2,7) \,, \\
	%%%%%%%%%%%%%%%%%%%%%%%%%%%%%%%%%%%%%%%%%%%%%%%%%%%%%%%%%%%%%%%%%%%%%%%%%%%%
	\hat{X}\sym\Big|\RR &= \frac{G m_1 v_1^2}{r_1}\left(V\RR\right)_1 - \frac{G^2 m_1^2 n_1^j a_1^j}{2 r_1} + G^2 m_1 m_2  a_1^j \p_1^j g  \nn\\
	&\quad + \frac{G^3 m_1^2}{30 c^5}\left(n_1^{ij} - 12 \frac{y_1^i n_1^j}{r_1}\right)\dM_{ij}^{(5)} + \frac{4 G^3 m_1 m_2}{5 c^5}\left[  2 \p_1^{ik}\p_2^{jk}  h_{12}- y_1^i \p_1^k \p_2^{jk} f_{12} - \p_2^{ij} f_{12}  \right]\dM_{ij}^{(5)}\nn\\
	& \quad - \frac{G^2 m_1}{5 c^5} \Bigg( r_1 v_1^i v_1^j - \frac{G m_1}{12} n_1^i n_1^j   - 2 G m_2 \p_1^i \p_2^j f \Bigg) \,  \dM_{ij}^{(5)} \nn\\
	& \quad +
	\frac{G^2 m_{1}}{c^5} \Biggl[- \frac{29}{2520}  n_{1}^{a} y_{1}^{i} \dM^{(7)}_{ai} y_{1}^{2}
	-  \frac{1}{378} \frac{n_{1}^{b} y_{1}^{j} \dM^{(7)}_{bj}}{r_{1}} (n_{1} y_{1}) y_{1}^{2}
	+  \dM^{(7)}_{bi} \Bigl(\frac{1}{945} n_{1}^{bi} r_{1}^2 (n_{1} y_{1})\nn\\
	& \qquad \qquad + \frac{23}{1890} n_{1}^{b} r_{1} y_{1}^{i} (n_{1} y_{1})
	+ \frac{1}{36} y_{1}^{bi} (n_{1} y_{1})
	-  \frac{31}{7560} n_{1}^{bi} r_{1} y_{1}^{2}
	-  \frac{1}{1512} \frac{1}{r_{1}} y_{1}^{bi} y_{1}^{2}\Bigr)
	+ \dM^{(7)}_{ij} \Bigl(\frac{1}{7560} n_{1}^{ij} r_{1} (n_{1} y_{1})^2\nn\\
	& \qquad \qquad + \frac{1}{504} n_{1}^{i} y_{1}^{j} (n_{1} y_{1})^2
	+ \frac{1}{216} \frac{1}{r_{1}} y_{1}^{ij} (n_{1} y_{1})^2
	-  \frac{1}{1260} n_{1}^{ij} (n_{1} y_{1}) y_{1}^{2}
	+ \frac{1}{3780} n_{1}^{ij} \frac{1}{r_{1}} y_{1}^{4}\Bigr)\nn\\
	& \qquad \qquad + \dM^{(7)}_{ab} \Bigl(- \frac{11}{315} r_{1}^3 n_{1}^{ab}
	-  \frac{103}{945} n_{1}^{a} r_{1}^2 y_{1}^{b}
	-  \frac{67}{540} r_{1} y_{1}^{ab}\Bigr)
	+ \dM^{(6)}_{bi} \Bigl(\frac{67}{630} n_{1}^{bi} r_{1}^2 (n_{1} v_{1})\nn\\
	& \qquad \qquad + \frac{197}{630} n_{1}^{b} r_{1} y_{1}^{i} (n_{1} v_{1})
	+ \frac{23}{70} y_{1}^{bi} (n_{1} v_{1})
	-  \frac{11}{315} n_{1}^{b} r_{1} v_{1}^{i} (n_{1} y_{1})
	-  \frac{13}{105} v_{1}^{b} y_{1}^{i} (n_{1} y_{1})
	+ \frac{1}{63} n_{1}^{bi} r_{1} (v_{1} y_{1})\nn\\
	& \qquad \qquad -  \frac{1}{21} \frac{1}{r_{1}} y_{1}^{bi} (v_{1} y_{1})\Bigr)
	-  \frac{2}{105} \frac{n_{1}^{b} y_{1}^{j} \dM^{(6)}_{bj}}{r_{1}} (n_{1} v_{1}) y_{1}^{2}
	+ \dM^{(6)}_{ai} \Bigl(\frac{1}{105} n_{1}^{a} y_{1}^{i} (v_{1} y_{1})
	+ \frac{2}{105} \frac{v_{1}^{a} y_{1}^{i}}{r_{1}} y_{1}^{2}\Bigr)\nn\\
	& \qquad \qquad + \dM^{(6)}_{ij} \Bigl(\frac{1}{315} n_{1}^{ij} r_{1} (n_{1} v_{1}) (n_{1} y_{1})
	+ \frac{1}{35} n_{1}^{i} y_{1}^{j} (n_{1} v_{1}) (n_{1} y_{1})
	+ \frac{1}{21} \frac{1}{r_{1}} y_{1}^{ij} (n_{1} v_{1}) (n_{1} y_{1})\nn\\
	& \qquad \qquad -  \frac{1}{105} n_{1}^{ij} (n_{1} v_{1}) y_{1}^{2}\Bigr)
	+ \dM^{(6)}_{ab} \Bigl(- \frac{151}{630} n_{1}^{a} r_{1}^2 v_{1}^{b}
	+ \frac{4}{105} n_{1}^{a} y_{1}^{2} v_{1}^{b}
	-  \frac{319}{630} r_{1} v_{1}^{a} y_{1}^{b}\Bigr)\nn\\
	& \qquad \qquad + \dM^{(5)}_{ij} \Bigl(\frac{1}{30} n_{1}^{ij} r_{1} (n_{1} v_{1})^2
	+ \frac{1}{10} n_{1}^{i} y_{1}^{j} (n_{1} v_{1})^2
	+ \frac{1}{10} \frac{1}{r_{1}} y_{1}^{ij} (n_{1} v_{1})^2\Bigr)
	-  \frac{1}{10} n_{1}^{a} y_{1}^{i} \dM^{(5)}_{ai} v_{1}^{2}\nn\\
	& \qquad \qquad + \dM^{(5)}_{bi} \Bigl(\frac{1}{15} n_{1}^{b} r_{1} v_{1}^{i} (n_{1} v_{1})
	+ \frac{1}{5} v_{1}^{b} y_{1}^{i} (n_{1} v_{1})
	-  \frac{1}{30} n_{1}^{bi} r_{1} v_{1}^{2}
	-  \frac{1}{10} \frac{1}{r_{1}} y_{1}^{bi} v_{1}^{2}\Bigr)
	+ \biggl(- \frac{1}{15} r_{1} v_{1}^{ab}\nn\\
	& \qquad \qquad + \frac{G m_{2}}{r_{12}^2} \biggl [ \frac{1}{30} n_{12}^{a} n_{1}^{b} r_{1}^2
	+ \frac{1}{10} n_{12}^{a} r_{1} y_{1}^{b}
	+ r_{12} \Bigl(\frac{1}{60} r_{1} n_{1}^{ab}
	+ \frac{1}{20} n_{1}^{a} y_{1}^{b}
	+ \frac{1}{20} \frac{1}{r_{1}} y_{1}^{ab}\Bigr)\nn\\
	& \qquad \qquad + \frac{1}{r_{12}} \Bigl(\frac{1}{60} r_{1}^3 n_{1}^{ab}
	-  \frac{1}{60} r_{1} r_{2}^2 n_{1}^{ab}
	+ \frac{1}{20} n_{1}^{a} r_{1}^2 y_{1}^{b}
	-  \frac{1}{20} n_{1}^{a} r_{2}^2 y_{1}^{b}
	+ \frac{1}{20} r_{1} y_{1}^{ab}
	-  \frac{1}{20} \frac{r_{2}^2}{r_{1}} y_{1}^{ab}\Bigr)\biggl]\biggl) \dM^{(5)}_{ab}\nn\\
	& \qquad \qquad + \varepsilon_{aij} \dS^{(6)}_{bj} \Bigl(\frac{2}{105} n_{1}^{bi} r_{1} y_{1}^{a}
	-  \frac{11}{210} y_{1}^{bi} n_{1}^{a}\Bigr)
	+ \varepsilon_{bjp} \dS^{(6)}_{ip} \Bigl(- \frac{1}{63} \frac{n_{1}^{b}}{r_{1}} y_{1}^{ij} (n_{1} y_{1})
	-  \frac{1}{315} n_{1}^{ij} \frac{y_{1}^{b}}{r_{1}} y_{1}^{2}\nn\\
	& \qquad \qquad + \frac{1}{210} n_{1}^{ij} (n_{1} y_{1}) y_{1}^{b}\Bigr)
	+ \varepsilon_{bjp} \dS^{(5)}_{ip} \Bigl(- \frac{4}{45} \frac{n_{1}^{b}}{r_{1}} y_{1}^{ij} (n_{1} v_{1})
	+ \frac{2}{45} n_{1}^{ij} (n_{1} v_{1}) y_{1}^{b}\Bigr)\nn\\
	& \qquad \qquad + \varepsilon_{aij} \dS^{(5)}_{bj} \Bigl(- \frac{2}{45} n_{1}^{bi} r_{1} v_{1}^{a}
	+ \frac{4}{45} \frac{v_{1}^{a}}{r_{1}} y_{1}^{bi}
	+ \frac{2}{15} n_{1}^{a} v_{1}^{b} y_{1}^{i}\Bigr)
	+ \frac{2}{45} \varepsilon_{bij} n_{1}^{a} v_{1}^{b} y_{1}^{i} \dS^{(5)}_{aj}\nn\\
	& \qquad \qquad + \frac{4}{45} \varepsilon_{abj} n_{1}^{a} v_{1}^{b} y_{1}^{i} \dS^{(5)}_{ij}\Biggl] + (1 \leftrightarrow 2) + \calO(2,7) \,.
\end{align}\ese
Here, we have discarded terms that are low order conservative contributions which do not belong to the RR piece. 
To finish, we will now discuss  the origin of the different terms in~\eqref{eq:symRR}. In the potential $\hat{W}^{ij}\sym|\RR$, all terms shown come from the non-compact part of the source of the type ``symmetric~$\times$~RR''. In $\hat{R}^i\sym|\RR$: (i) the first line comes from the compact part of the source; (ii) the other terms come from the non-compact part of the source of type ``symmetric~$\times$~RR''. In $\hat{X}\sym|\RR$: (i) the first term comes from the compact part of the source; (ii) the second and third terms come from the non-compact $V\sym\times\p_t^2 V\sym$ part of the source which involves a replacement of acceleration; (iii) the second line comes from the non-compact $\hat{W}^{ij}\sym\times\p_{ij} V\sym$ part; (iv) the third line comes from the non-compact $\hat{W}^{ij}\sym\times\p_{ij} \Vrr$ part of the source; (v) all the other terms come from the other non-compact source terms of the type ``symmetric~$\times$~RR''.

\subsection{Final expression for the equations of motion}

Once all the contributions are accounted for, we find that the RR part of the acceleration of body 1 through 4.5PN order in extended BT coordinates reads 
\begin{align}\label{eq:a1i_moments}
a_{\mathrm{RR}\,1}^{i} = a_{\text{2.5PN}\,1}^{i} + a_{\text{3.5PN}\,1}^{i} + a_{\text{4.5PN}\,1}^{i} +\calO(11)\,,
\end{align}
where all the PN pieces are given by (see also the Supplemental Material \cite{SuppMaterial})
\bse\label{eq:acc45PN1}\begin{align}
	a_{\text{2.5PN}\,1}^{i} &= - \frac{2G}{5c^5} y_{1}^{a} \dM^{(5)}_{ia}\,,\\
%%%%%%%%%%%%%%%%%%%%%%%%%%%%%%%%%%%%%%%%%%%%%%%%%%%%%%%%%%%%%%%%%%%%%%%%%%%%%%%%%%%
	a_{\text{3.5PN}\,1}^{i} &= \frac{G}{c^7} \biggl\{- \frac{11}{105} y_{1}^{b} \dM^{(7)}_{ib} y_{1}^{2}
	+ \frac{17}{105} y_{1}^{iab} \dM^{(7)}_{ab}
	-  \frac{8}{15} y_{1}^{b} \dM^{(6)}_{ib} (v_{1}{} y_{1}{})
	+ \Bigl(\frac{8}{15} y_{1}^{bi} v_{1}^{a}
	+ \frac{3}{5} v_{1}^{i} y_{1}^{ab}\Bigr) \dM^{(6)}_{ab}
	-  \frac{2}{5} y_{1}^{b} \dM^{(5)}_{ib} v_{1}^{2}\nonumber\\
	& \qquad + \frac{G \dM^{(5)}_{ia}}{r_{12}} \Bigl(\frac{7}{5} m_{2} n_{12}^{a} r_{12}
	+ \frac{1}{5} m_{2} y_{1}^{a}\Bigr)
	+ \biggl [\frac{8}{5} v_{1}^{bi} y_{1}^{a}
	+ \frac{G}{r_{12}} \Bigl(\frac{1}{5} n_{12}^{bi} m_{2} y_{1}^{a}
	-  \frac{m_{2} n_{12}^{i}}{r_{12}} y_{1}^{ab}\Bigr)\biggl] \dM^{(5)}_{ab}
	+ \frac{1}{63} \dM^{(7)}_{iab} y_{1}^{ab}\nonumber\\
	& \qquad -  \frac{16}{45} \varepsilon_{ibj} \dS^{(6)}_{aj} y_{1}^{ab}
	-  \frac{16}{45} \varepsilon_{ibj} v_{1}^{a} y_{1}^{b} \dS^{(5)}_{aj}
	-  \frac{32}{45} \varepsilon_{iaj} v_{1}^{a} y_{1}^{b} \dS^{(5)}_{bj}
	+ \frac{16}{45} \varepsilon_{abj} v_{1}^{a} y_{1}^{b} \dS^{(5)}_{ij}\biggl\}\,.\\
%%%%%%%%%%%%%%%%%%%%%%%%%%%%%%%%%%%%%%%%%%%%%%%%%%%%%%%%%%%%%%%%%%%%%%%%%%%%%%%%%%%%%%%
	a_{\text{4.5PN}\,1}^{i} &= \frac{G}{c^9} \Biggl\{ - \frac{19}{3780} y_{1}^{j} \dM^{(9)}_{ij} y_{1}^{4}
	+ \frac{17}{1890} y_{1}^{ibj} y_{1}^{2} \dM^{(9)}_{bj}
	-  \frac{46}{945} y_{1}^{j} \dM^{(8)}_{ij} (v_{1}{} y_{1}{}) y_{1}^{2}
	+ \frac{2}{945} v_{1}^{a} \dM^{(8)}_{ia} y_{1}^{4}
	+ \frac{71}{1890} v_{1}^{i} \dM^{(8)}_{bj} y_{1}^{2} y_{1}^{bj}\nonumber\\
	& \qquad + \frac{26}{945} y_{1}^{ij} y_{1}^{2} v_{1}^{a} \dM^{(8)}_{aj}
	+ \frac{1}{27} y_{1}^{ibj} (v_{1}{} y_{1}{}) \dM^{(8)}_{bj}
	+ \Bigl(- \frac{23}{756} n_{12}^{abi} m_{2} r_{12}^3
	-  \frac{13}{1890} m_{2} y_{1}^{abi}\Bigr) \frac{G \dM^{(7)}_{ab}}{r_{12}}\nonumber\\
	& \qquad + \frac{G \dM^{(7)}_{jk}}{r_{12}} \Bigl(- \frac{1}{630} n_{12}^{ijk} m_{2} r_{12} (n_{12}{} y_{1}{})^2
	-  \frac{1}{105} m_{2} n_{12}^{ijk} (n_{12}{} y_{1}{}) y_{1}^{2}
	-  \frac{1}{315} n_{12}^{ijk} \frac{m_{2}}{r_{12}} y_{1}^{4}\Bigr)\nonumber\\ 
	& \qquad + \frac{G \dM^{(7)}_{ia}}{r_{12}} \Bigl(\frac{2629}{3780} m_{2} n_{12}^{a} r_{12}^3
	+ \frac{359}{945} m_{2} n_{12}^{a} r_{12} y_{1}^{2}
	-  \frac{2629}{3780} m_{2} r_{12}^2 y_{1}^{a}\Bigr)
	\nonumber\\ 
	& \qquad+ \dM^{(7)}_{ib} \biggl(\frac{8}{945} (v_{1}{} y_{1}{}) y_{1}^{2} v_{1}^{b} + \frac{G}{r_{12}} \biggl [- \frac{47}{189} m_{2} n_{12}^{b} r_{12}^2 (n_{12}{} y_{1}{})
	-  \frac{227}{945} m_{2} r_{12} y_{1}^{b} (n_{12}{} y_{1}{}) \nonumber\\ & \qquad \qquad \qquad \qquad \qquad \qquad \qquad \quad
	+ m_{2} \Bigl(- \frac{4}{945} n_{12}^{b} (n_{12}{} y_{1}{}) y_{1}^{2}
	+ \frac{16}{945} y_{1}^{2} y_{1}^{b}\Bigr)\biggl]\biggl)\nonumber\\
	& \qquad + \dM^{(7)}_{ij} \biggl [\frac{G}{r_{12}} \Bigl(- \frac{13}{1890} m_{2} n_{12}^{j} r_{12} (n_{12}{} y_{1}{})^2
	+ \frac{11}{378} m_{2} y_{1}^{j} (n_{12}{} y_{1}{})^2\Bigr)
	-  \frac{11}{189} (v_{1}{} y_{1}{})^2 y_{1}^{j}
	-  \frac{32}{945} v_{1}^{2} y_{1}^{2} y_{1}^{j}\biggl]\nonumber\\
	& \qquad + \frac{G \dM^{(7)}_{ab}}{r_{12}} \biggl [m_{2} r_{12}^2 \Bigl(\frac{23}{270} y_{1}^{i} n_{12}^{ab}
	-  \frac{5}{756} n_{12}^{bi} y_{1}^{a}\Bigr)
	+ m_{2} r_{12} \Bigl(- \frac{26}{135} y_{1}^{bi} n_{12}^{a}
	+ \frac{17}{135} n_{12}^{i} y_{1}^{ab}\Bigr)\biggl]\nonumber\\
	& \qquad + \frac{G \dM^{(7)}_{jk}}{r_{12}} \biggl [\frac{1}{42} n_{12}^{ik} m_{2} y_{1}^{j} (n_{12}{} y_{1}{})^2
	+ \frac{m_{2}}{r_{12}} \Bigl(\frac{2}{63} n_{12}^{ik} (n_{12}{} y_{1}{}) y_{1}^{2} y_{1}^{j}
	-  \frac{1}{18} n_{12}^{i} (n_{12}{} y_{1}{})^2 y_{1}^{jk}\Bigr)\biggl]
	\nonumber\\
	& \qquad + \biggl(\frac{2}{27} y_{1}^{ij} (v_{1}{} y_{1}{}) v_{1}^{b} -  \frac{2}{135} y_{1}^{i} y_{1}^{2} v_{1}^{bj}
	+ \frac{14}{135} v_{1}^{ij} y_{1}^{2} y_{1}^{b}
	+ \frac{14}{135} v_{1}^{i} (v_{1}{} y_{1}{}) y_{1}^{bj} 
	\nonumber\\
	& \qquad \quad + \frac{G}{r_{12}} \biggl [m_{2} r_{12} \Bigl(\frac{1}{135} y_{1}^{i} n_{12}^{bj} (n_{12}{} y_{1}{})
	-  \frac{61}{945} n_{12}^{ij} (n_{12}{} y_{1}{}) y_{1}^{b}\Bigr) -  \frac{4}{63} \frac{m_{2} n_{12}^{i}}{r_{12}} y_{1}^{2} y_{1}^{bj}\nonumber\\ & \qquad \qquad \quad ~
	+ m_{2} \Bigl(- \frac{1}{27} y_{1}^{ij} n_{12}^{b} (n_{12}{} y_{1}{})
	+ \frac{1}{135} y_{1}^{i} n_{12}^{bj} y_{1}^{2}
	-  \frac{4}{945} n_{12}^{ij} y_{1}^{2} y_{1}^{b}  + \frac{127}{1890} n_{12}^{i} (n_{12}{} y_{1}{}) y_{1}^{bj}\Bigr)\biggl]\biggl) \dM^{(7)}_{bj}\nonumber\\
	& \qquad
	+ \biggl [\frac{13}{945} y_{1}^{bij} v_{1}^{2}
	+ \frac{G}{r_{12}} \Bigl(- \frac{1}{378} n_{12}^{bij} m_{2} r_{12}^2 (n_{12}{} y_{1}{})  + \frac{37}{1890} n_{12}^{bij} m_{2} r_{12} y_{1}^{2}\Bigr)\biggl] \dM^{(7)}_{bj} \nonumber\\
	& \qquad
	+ \frac{G m_{2} \dM^{(6)}_{aj}}{r_{12}^2} \Bigl(\frac{16}{105} n_{12}^{i} v_{1}^{a} y_{1}^{j} y_{1}^{2}
	+ \frac{8}{105} n_{12}^{a} v_{1}^{i} y_{1}^{j} y_{1}^{2}
	-  \frac{8}{105} n_{12}^{a} v_{2}^{i} y_{1}^{j} y_{1}^{2}\Bigr)\nonumber\\
	& \qquad + \frac{G \dM^{(6)}_{jk}}{r_{12}} \Bigl(- \frac{4}{105} n_{12}^{ijk} m_{2} r_{12} (n_{12}{} v_{2}{}) (n_{12}{} y_{1}{})
	-  \frac{4}{35} m_{2} n_{12}^{ijk} (n_{12}{} v_{2}{}) y_{1}^{2}\Bigr)
	\nonumber\\
	&  \qquad + \frac{G \dM^{(6)}_{ia}}{r_{12}} \biggl [m_{2} r_{12} \Bigl(\frac{8}{15} n_{12}^{a} (v_{1}{} y_{1}{})+ \frac{64}{45} n_{12}^{a} (v_{2}{} y_{1}{})\Bigr)
	+ m_{2} r_{12}^2 \Bigl(\frac{4}{5} v_{1}^{a}
	-  \frac{35}{18} v_{2}^{a}\Bigr)\biggl] \nonumber\\
	& \qquad
	+ \frac{G \dM^{(6)}_{ib}}{r_{12}} \biggl [m_{2} r_{12}^2 \Bigl(\frac{4}{3} n_{12}^{b} (n_{12}{} v_{1}{})
	-  \frac{119}{45} n_{12}^{b} (n_{12}{} v_{2}{})\Bigr) \nonumber \\ & \qquad \qquad \qquad + m_{2} r_{12} \Bigl(- \frac{44}{45} (n_{12}{} y_{1}{}) v_{2}^{b}
	-  \frac{12}{5} (n_{12}{} v_{1}{}) y_{1}^{b}
	+ \frac{79}{45} (n_{12}{} v_{2}{}) y_{1}^{b}\Bigr)\biggl]
	\nonumber\\
	& \qquad + \frac{G \dM^{(6)}_{ij}}{r_{12}} \biggl [m_{2} r_{12} \Bigl(- \frac{8}{15} n_{12}^{j} (n_{12}{} v_{1}{}) (n_{12}{} y_{1}{})+ \frac{4}{9} n_{12}^{j} (n_{12}{} v_{2}{}) (n_{12}{} y_{1}{})\Bigr) \nonumber \\ & \qquad \qquad \qquad
	+ m_{2} \Bigl(\frac{16}{15} (n_{12}{} v_{1}{}) (n_{12}{} y_{1}{}) y_{1}^{j}
	-  \frac{4}{5} (n_{12}{} v_{2}{}) (n_{12}{} y_{1}{}) y_{1}^{j}\Bigr)\biggl]\nonumber\\
	& \qquad + \frac{G \dM^{(6)}_{ab}}{r_{12}} \biggl [m_{2} r_{12}^2 \Bigl(\frac{23}{126} v_{2}^{i} n_{12}^{ab}
	-  \frac{8}{15} n_{12}^{bi} v_{1}^{a}
	+ \frac{53}{90} n_{12}^{bi} v_{2}^{a}\Bigr)
	+ m_{2} r_{12} \Bigl(\frac{4}{5} n_{12}^{i} v_{1}^{a} y_{1}^{b}
	+ \frac{3}{5} n_{12}^{a} v_{1}^{i} y_{1}^{b}
	+ \frac{2}{9} n_{12}^{i} v_{2}^{a} y_{1}^{b}\nonumber\\
	& \qquad \qquad \qquad \qquad \qquad \qquad \qquad -  \frac{332}{315} n_{12}^{a} v_{2}^{i} y_{1}^{b}
	-  \frac{8}{15} n_{12}^{a} v_{1}^{b} y_{1}^{i}
	-  \frac{4}{9} n_{12}^{a} v_{2}^{b} y_{1}^{i}\Bigr)
	+ m_{2} \Bigl(- \frac{3}{5} v_{1}^{i} y_{1}^{ab}
	+ \frac{127}{210} v_{2}^{i} y_{1}^{ab}\Bigr)\biggl]\nonumber\\
	& \qquad + \frac{G \dM^{(6)}_{jk}}{r_{12}} \biggl [\frac{12}{35} n_{12}^{ik} m_{2} y_{1}^{j} (n_{12}{} v_{2}{}) (n_{12}{} y_{1}{})
	+ \frac{m_{2}}{r_{12}} \Bigl(\frac{8}{35} n_{12}^{ik} (n_{12}{} v_{2}{}) y_{1}^{2} y_{1}^{j}
	-  \frac{4}{7} n_{12}^{i} (n_{12}{} v_{2}{}) (n_{12}{} y_{1}{}) y_{1}^{jk}\Bigr)\biggl]\nonumber\\
	& \qquad + \biggl(\frac{8}{35} v_{1}^{ij} (v_{1}{} y_{1}{}) y_{1}^{b}
	-  \frac{1}{105} v_{1}^{i} v_{1}^{2} y_{1}^{bj}
	+ \frac{G}{r_{12}} \biggl [m_{2} r_{12} \Bigl(\frac{8}{15} y_{1}^{i} n_{12}^{bj} (n_{12}{} v_{1}{})
	-  \frac{4}{9} y_{1}^{i} n_{12}^{bj} (n_{12}{} v_{2}{})
	+ \frac{4}{315} v_{2}^{i} n_{12}^{bj} (n_{12}{} y_{1}{})\nonumber\\
	& \qquad \qquad \quad + \frac{8}{15} n_{12}^{ij} (n_{12}{} y_{1}{}) v_{1}^{b}
	-  \frac{28}{45} n_{12}^{ij} (n_{12}{} y_{1}{}) v_{2}^{b}
	-  \frac{124}{315} n_{12}^{ij} (n_{12}{} v_{2}{}) y_{1}^{b}\Bigr) \nonumber \\ & 
	\qquad \qquad \quad + m_{2} \Bigl(- \frac{16}{15} y_{1}^{ij} n_{12}^{b} (n_{12}{} v_{1}{})
	+ \frac{4}{5} y_{1}^{ij} n_{12}^{b} (n_{12}{} v_{2}{})-  \frac{16}{15} n_{12}^{i} v_{1}^{b} y_{1}^{j} (n_{12}{} y_{1}{})
	+ \frac{16}{15} n_{12}^{i} v_{2}^{b} y_{1}^{j} (n_{12}{} y_{1}{}) \nonumber \\ &
	\qquad \qquad \qquad \quad -  \frac{4}{35} n_{12}^{b} v_{2}^{i} y_{1}^{j} (n_{12}{} y_{1}{})
	+ \frac{4}{105} v_{2}^{i} n_{12}^{bj} y_{1}^{2}
	+ \frac{16}{15} n_{12}^{ij} (v_{1}{} y_{1}{}) y_{1}^{b} -  \frac{16}{15} n_{12}^{ij} (v_{2}{} y_{1}{}) y_{1}^{b}
	+ \frac{27}{70} n_{12}^{i} (n_{12}{} v_{2}{}) y_{1}^{bj}\Bigr) \nonumber\\
	& \qquad \qquad \quad
	+ \frac{m_{2}}{r_{12}} \Bigl(- \frac{4}{21} v_{1}^{i} (n_{12}{} y_{1}{}) y_{1}^{bj}
	+ \frac{4}{21} v_{2}^{i} (n_{12}{} y_{1}{}) y_{1}^{bj}-  \frac{8}{21} n_{12}^{i} (v_{1}{} y_{1}{}) y_{1}^{bj}\Bigr)\biggl]\biggl) \dM^{(6)}_{bj} \nonumber \\ & \qquad
	+ \biggl(\frac{G}{r_{12}} \biggl [\frac{1}{126} n_{12}^{bij} m_{2} r_{12}^2 (n_{12}{} v_{2}{})
	+ m_{2} r_{12} \Bigl(- \frac{8}{15} n_{12}^{bij} (v_{1}{} y_{1}{})
	+ \frac{28}{45} n_{12}^{bij} (v_{2}{} y_{1}{})\Bigr)\biggl]-  \frac{8}{105} v_{1}^{bij} y_{1}^{2}\biggl) \dM^{(6)}_{bj}
	\nonumber\\
	& \qquad + \Bigl(\frac{47}{30} m_{1} m_{2} n_{12}^{abi}
	- 2 m_{2}^2 n_{12}^{abi}\Bigr) \frac{G^2 \dM^{(5)}_{ab}}{r_{12}}
	-  n_{12}^{ijk} G m_{2} \dM^{(5)}_{jk} (n_{12}{} v_{2}{})^2  + G m_{2} \dM^{(5)}_{bj} \Bigl(- \frac{8}{5} n_{12}^{bij} (v_{1}{} v_{2}{})
	+ \frac{26}{15} n_{12}^{bij} v_{2}^{2}\Bigr)\nonumber\\
	& \qquad
	+ \dM^{(5)}_{ia} \biggl(G m_{2} \Bigl(- \frac{8}{5} n_{12}^{a} (v_{1}{} v_{2}{})
	+ \frac{1}{5} n_{12}^{a} v_{1}^{2}
	+ \frac{44}{15} n_{12}^{a} v_{2}^{2}\Bigr)\nonumber\\
	& \qquad \qquad \quad + \frac{G^2}{r_{12}^2} \biggl [r_{12} \Bigl(- \frac{97}{15} m_{1} m_{2} n_{12}^{a}
	-  \frac{14}{5} m_{2}^2 n_{12}^{a}\Bigr)
	+ \frac{3}{5} m_{1} m_{2} y_{1}^{a}
	-  \frac{4}{5} m_{2}^2 y_{1}^{a}\biggl]\biggl)
	\nonumber\\
	& \qquad + \frac{G \dM^{(5)}_{ib}}{r_{12}} \biggl [m_{2} r_{12} \Bigl(\frac{4}{5} (n_{12}{} v_{1}{}) v_{1}^{b}-  \frac{4}{5} (n_{12}{} v_{2}{}) v_{1}^{b}
	-  \frac{16}{5} (n_{12}{} v_{1}{}) v_{2}^{b}
	+ \frac{5}{3} (n_{12}{} v_{2}{}) v_{2}^{b}\Bigr)
	\nonumber \\ & \qquad \qquad \qquad + m_{2} \Bigl(\frac{4}{5} (v_{1}{} v_{2}{}) y_{1}^{b}
	-  \frac{3}{5} v_{1}^{2} y_{1}^{b}
	-  \frac{2}{5} v_{2}^{2} y_{1}^{b}\Bigr)\biggl]\nonumber\\
	& \qquad + \frac{G \dM^{(5)}_{ij}}{r_{12}} \biggl [m_{2} r_{12} \Bigl(- \frac{2}{5} n_{12}^{j} (n_{12}{} v_{1}{})^2
	+ \frac{4}{5} n_{12}^{j} (n_{12}{} v_{1}{}) (n_{12}{} v_{2}{})
	-  \frac{31}{30} n_{12}^{j} (n_{12}{} v_{2}{})^2\Bigr)
	\nonumber\\
	& \qquad \qquad \qquad + m_{2} \Bigl(\frac{4}{5} (n_{12}{} v_{1}{})^2 y_{1}^{j} -  \frac{7}{10} (n_{12}{} v_{2}{})^2 y_{1}^{j}\Bigr)\biggl] \nonumber \\ & \qquad 
	+ \frac{G \dM^{(5)}_{bj}}{r_{12}} \biggl [m_{2} r_{12} \Bigl(\frac{8}{5} v_{2}^{i} n_{12}^{bj} (n_{12}{} v_{1}{})
	+ \frac{6}{5} v_{1}^{i} n_{12}^{bj} (n_{12}{} v_{2}{})
	-  \frac{38}{15} v_{2}^{i} n_{12}^{bj} (n_{12}{} v_{2}{})-  \frac{4}{15} n_{12}^{ij} (n_{12}{} v_{2}{}) v_{2}^{b}\Bigr) \nonumber\\
	& \qquad \qquad \qquad
	+ m_{2} \Bigl(- \frac{4}{5} n_{12}^{b} v_{1}^{i} y_{1}^{j} (n_{12}{} v_{1}{})
	-  \frac{12}{5} n_{12}^{b} v_{2}^{i} y_{1}^{j} (n_{12}{} v_{1}{})
	-  \frac{8}{5} n_{12}^{i} v_{1}^{b} y_{1}^{j} (n_{12}{} v_{2}{}) -  \frac{3}{5} n_{12}^{b} v_{1}^{i} y_{1}^{j} (n_{12}{} v_{2}{})
	\nonumber\\
	& \qquad \qquad \qquad \qquad ~+ \frac{9}{5} n_{12}^{i} v_{2}^{b} y_{1}^{j} (n_{12}{} v_{2}{})
	+ 3 n_{12}^{b} v_{2}^{i} y_{1}^{j} (n_{12}{} v_{2}{})
	+ \frac{12}{5} n_{12}^{ij} (v_{1}{} v_{2}{}) y_{1}^{b}
	+ \frac{1}{5} n_{12}^{ij} v_{1}^{2} y_{1}^{b}-  \frac{14}{5} n_{12}^{ij} v_{2}^{2} y_{1}^{b}\Bigr) \nonumber\\
	& \qquad \qquad \qquad
	+ \frac{m_{2}}{r_{12}} \Bigl(\frac{4}{5} v_{1}^{i} (n_{12}{} v_{1}{}) y_{1}^{bj}
	-  \frac{4}{5} v_{2}^{i} (n_{12}{} v_{1}{}) y_{1}^{bj}
	-  \frac{3}{5} v_{1}^{i} (n_{12}{} v_{2}{}) y_{1}^{bj}
	+ \frac{3}{5} v_{2}^{i} (n_{12}{} v_{2}{}) y_{1}^{bj}\nonumber\\
	& \qquad \qquad \qquad \qquad ~+ \frac{4}{5} n_{12}^{i} (v_{1}{} v_{2}{}) y_{1}^{bj}
	-  \frac{1}{5} n_{12}^{i} v_{1}^{2} y_{1}^{bj}
	-  \frac{2}{5} n_{12}^{i} v_{2}^{2} y_{1}^{bj}\Bigr)\biggl]
	\nonumber \\ & \qquad + \frac{G \dM^{(5)}_{jk}}{r_{12}} \Bigl(\frac{3}{2} n_{12}^{ik} m_{2} y_{1}^{j} (n_{12}{} v_{2}{})^2
	+ \frac{3}{10} \frac{m_{2} n_{12}^{i}}{r_{12}} (n_{12}{} v_{2}{})^2 y_{1}^{jk}\Bigr)\nonumber\\
	& \qquad + \biggl(\frac{G}{r_{12}} \biggl [m_{2} r_{12} \Bigl(\frac{4}{5} v_{1}^{bi} n_{12}^{a}
	+ \frac{4}{15} v_{2}^{bi} n_{12}^{a}
	-  \frac{4}{5} n_{12}^{i} v_{1}^{ab}
	+ \frac{12}{5} n_{12}^{i} v_{1}^{a} v_{2}^{b}
	-  \frac{8}{5} n_{12}^{a} v_{1}^{i} v_{2}^{b}
	-  \frac{12}{5} n_{12}^{a} v_{1}^{b} v_{2}^{i}
	-  \frac{1}{15} n_{12}^{i} v_{2}^{ab}\Bigr)\nonumber\\
	& \qquad \qquad \quad + m_{2} \Bigl(- \frac{4}{5} v_{1}^{bi} y_{1}^{a}
	+ \frac{3}{5} v_{2}^{bi} y_{1}^{a}
	-  \frac{3}{5} v_{1}^{i} v_{2}^{a} y_{1}^{b}
	+ \frac{4}{5} v_{1}^{a} v_{2}^{i} y_{1}^{b}\Bigr)\biggl] \nonumber \\ &
	\qquad \quad + \frac{G^2}{r_{12}^2} \biggl [- \frac{22}{5} n_{12}^{bi} m_{1} m_{2} y_{1}^{a}
	+ \frac{4}{5} n_{12}^{bi} m_{2}^2 y_{1}^{a}
	+ \frac{1}{r_{12}} \Bigl(6 m_{1} m_{2} n_{12}^{i} y_{1}^{ab}
	+ \frac{24}{5} m_{2}^2 n_{12}^{i} y_{1}^{ab}\Bigr)\biggl]\biggl) \dM^{(5)}_{ab}
	+ \frac{1}{378} \dM^{(9)}_{ibj} y_{1}^{2} y_{1}^{bj} \nonumber \\ &
	\qquad -  \frac{2}{567} y_{1}^{iabj} \dM^{(9)}_{abj}
	+ \frac{1}{63} \dM^{(8)}_{ibj} (v_{1}{} y_{1}{}) y_{1}^{bj} + \Bigl(- \frac{1}{63} y_{1}^{bij} v_{1}^{a}
	-  \frac{1}{63} v_{1}^{i} y_{1}^{abj}\Bigr) \dM^{(8)}_{abj}
	+ \frac{1}{378} n_{12}^{iabj} G m_{2} r_{12} \dM^{(7)}_{abj}
	\nonumber\\
	& \qquad + \frac{G \dM^{(7)}_{iab}}{r_{12}} \Bigl(\frac{1}{18} m_{2} r_{12}^2 n_{12}^{ab}
	-  \frac{1}{9} m_{2} n_{12}^{a} r_{12} y_{1}^{b} -  \frac{1}{126} m_{2} y_{1}^{ab}\Bigr)
	+ \frac{1}{63} \dM^{(7)}_{ibj} v_{1}^{2} y_{1}^{bj} \nonumber \\ & \qquad
	+ \biggl [- \frac{4}{63} v_{1}^{ai} y_{1}^{bj}
	+ \frac{G}{r_{12}} \Bigl(- \frac{1}{126} m_{2} n_{12}^{ai} y_{1}^{bj}
	+ \frac{5}{189} \frac{m_{2} n_{12}^{i}}{r_{12}} y_{1}^{abj}\Bigr)\biggl] \dM^{(7)}_{abj}\nonumber\\
	& \qquad -  \frac{1}{2268} \dM^{(9)}_{iabj} y_{1}^{abj}
	-  \frac{8}{315} \varepsilon_{ijk} \dS^{(8)}_{bk} y_{1}^{2} y_{1}^{bj}
	-  \frac{4}{315} \varepsilon_{ijk} v_{1}^{a} y_{1}^{j} \dS^{(7)}_{ak} y_{1}^{2}
	+ \frac{4}{105} \varepsilon_{ajk} v_{1}^{a} y_{1}^{j} \dS^{(7)}_{ik} y_{1}^{2}
	-  \frac{16}{315} \varepsilon_{iak} v_{1}^{a} y_{1}^{j} \dS^{(7)}_{jk} y_{1}^{2}\nonumber\\
	& \qquad -  \frac{4}{35} \varepsilon_{ijk} \dS^{(7)}_{bk} (v_{1}{} y_{1}{}) y_{1}^{bj}
	+ \frac{4}{315} \varepsilon_{abk} y_{1}^{ijk} v_{1}^{a} \dS^{(7)}_{bj}
	+ \frac{2}{21} n_{12}^{ijk} G m_{2} \varepsilon_{abk} y_{1}^{a} \dS^{(6)}_{bj}\nonumber\\
	& \qquad + \frac{G \varepsilon_{bjl} \dS^{(6)}_{jk}}{r_{12}} \Bigl(- \frac{2}{35} n_{12}^{ikl} m_{2} y_{1}^{b} (n_{12}{} y_{1}{})
	-  \frac{4}{105} n_{12}^{ikl} \frac{m_{2} y_{1}^{b}}{r_{12}} y_{1}^{2}\Bigr)
	\nonumber \\ & \qquad + \dS^{(6)}_{ik} \Bigl(- \frac{2}{45} \frac{G m_{2} \varepsilon_{kbj} n_{12}^{b} y_{1}^{j}}{r_{12}} (n_{12}{} y_{1}{}) + \frac{4}{45} \varepsilon_{kbj} (v_{1}{} y_{1}{}) v_{1}^{b} y_{1}^{j}\Bigr) \nonumber\\
	& 
	\qquad + \frac{G \varepsilon_{ibj} \dS^{(6)}_{aj}}{r_{12}} \Bigl(- \frac{88}{315} m_{2} r_{12}^2 n_{12}^{ab}
	+ \frac{40}{63} m_{2} n_{12}^{a} r_{12} y_{1}^{b}
	+ \frac{2}{63} m_{2} y_{1}^{ab}\Bigr)\nonumber\\
	& \qquad + \frac{4}{21} n_{12}^{ib} \frac{G m_{2} \varepsilon_{bkl} \dS^{(6)}_{jl}}{r_{12}^2} (n_{12}{} y_{1}{}) y_{1}^{jk}
	+ \Bigl(\frac{2}{45} \frac{G m_{2} \varepsilon_{kbj}}{r_{12}} y_{1}^{ij} n_{12}^{ab}
	-  \frac{4}{45} \varepsilon_{kbj} y_{1}^{ij} v_{1}^{ab}\Bigr) \dS^{(6)}_{ak}
	-  \frac{2}{63} G m_{2} \varepsilon_{iaj} n_{12}^{a} y_{1}^{b} \dS^{(6)}_{bj}\nonumber\\
	& \qquad + \Biggl[\frac{G}{r_{12}} \biggl(\varepsilon_{ijk} \biggl [- \frac{2}{105} m_{2} r_{12} n_{12}^{bj} (n_{12}{} y_{1}{})
	+ m_{2} \Bigl(- \frac{4}{315} n_{12}^{bj} y_{1}^{2}
	+ \frac{2}{105} n_{12}^{b} (n_{12}{} y_{1}{}) y_{1}^{j}\Bigr)\biggl]
	-  \frac{2}{21} m_{2} \varepsilon_{kaj} n_{12}^{ia} y_{1}^{bj}\biggl)\nonumber\\
	& \qquad \quad + \varepsilon_{ijk} \Bigl(\frac{8}{315} y_{1}^{2} v_{1}^{bj}
	-  \frac{4}{105} (v_{1}{} y_{1}{}) v_{1}^{b} y_{1}^{j}
	-  \frac{4}{63} v_{1}^{2} y_{1}^{bj}\Bigr)\Biggl] \dS^{(6)}_{bk}
	+ \frac{2}{3} G m_{2} \varepsilon_{abj} n_{12}^{a} y_{1}^{b} \dS^{(6)}_{ij}\nonumber\\
	& \qquad + \Bigl(- \frac{4}{63} \frac{G m_{2} \varepsilon_{ikb} n_{12}^{b} y_{1}^{j}}{r_{12}} (n_{12}{} y_{1}{})
	+ \frac{8}{63} \varepsilon_{ikb} v_{1}^{b} y_{1}^{j} (v_{1}{} y_{1}{})\Bigr) \dS^{(6)}_{jk}
	-  \frac{8}{15} n_{12}^{ikl} \frac{G m_{2} \varepsilon_{bjl} y_{1}^{b} \dS^{(5)}_{jk}}{r_{12}} (n_{12}{} v_{2}{})\nonumber\\
	& \qquad + G m_{2} \varepsilon_{iaj} \dS^{(5)}_{bj} \Bigl(\frac{16}{45} n_{12}^{a} v_{1}^{b}
	-  \frac{16}{45} n_{12}^{a} v_{2}^{b}\Bigr)
	+ G m_{2} \varepsilon_{ibj} \dS^{(5)}_{aj} \Bigl(\frac{32}{45} n_{12}^{a} v_{1}^{b}
	+ \frac{56}{45} n_{12}^{a} v_{2}^{b}\Bigr)
	\nonumber\\
	& \qquad + G m_{2} \varepsilon_{abj} \dS^{(5)}_{ij} \Bigl(\frac{16}{45} n_{12}^{a} v_{1}^{b}+ \frac{8}{5} n_{12}^{a} v_{2}^{b}\Bigr)
	+ \frac{G m_{2} \varepsilon_{bjk} \dS^{(5)}_{ak}}{r_{12}} \Bigl(- \frac{32}{45} n_{12}^{ij} v_{1}^{a} y_{1}^{b}
	+ \frac{32}{45} n_{12}^{ij} v_{2}^{a} y_{1}^{b}\Bigr) \nonumber \\ & \qquad
	+ \frac{G m_{2} \varepsilon_{bjk} \dS^{(5)}_{ik}}{r_{12}} \Bigl(- \frac{32}{45} n_{12}^{b} (n_{12}{} v_{1}{}) y_{1}^{j}+ \frac{8}{15} n_{12}^{b} (n_{12}{} v_{2}{}) y_{1}^{j}\Bigr) \nonumber\\
	& \qquad
	+ \frac{G \dS^{(5)}_{jk}}{r_{12}} \biggl [m_{2} \varepsilon_{abk} \Bigl(- \frac{32}{45} n_{12}^{ij} v_{1}^{a} y_{1}^{b}
	+ \frac{32}{45} n_{12}^{ij} v_{2}^{a} y_{1}^{b}\Bigr)
	+ m_{2} \varepsilon_{ibk} \Bigl(\frac{64}{45} n_{12}^{b} (n_{12}{} v_{1}{}) y_{1}^{j}-  \frac{16}{15} n_{12}^{b} (n_{12}{} v_{2}{}) y_{1}^{j}\Bigr)\biggl] \nonumber\\
	& \qquad 
	+ \frac{G \dS^{(5)}_{bk}}{r_{12}} \biggl(\varepsilon_{ijk} \biggl [m_{2} r_{12} \Bigl(- \frac{16}{15} n_{12}^{bj} (n_{12}{} v_{1}{})
	+ \frac{8}{9} n_{12}^{bj} (n_{12}{} v_{2}{})\Bigr)
	+ m_{2} \Bigl(\frac{32}{45} n_{12}^{b} (n_{12}{} v_{1}{}) y_{1}^{j} -  \frac{8}{15} n_{12}^{b} (n_{12}{} v_{2}{}) y_{1}^{j}\Bigr)\biggl]
	\nonumber\\
	& \qquad \qquad \qquad + \varepsilon_{ajk} \biggl [m_{2} \Bigl(- \frac{8}{45} v_{2}^{i} y_{1}^{a} n_{12}^{bj}
	-  \frac{64}{45} n_{12}^{ij} v_{1}^{a} y_{1}^{b}
	+ \frac{64}{45} n_{12}^{ij} v_{2}^{a} y_{1}^{b}\Bigr) \nonumber \\ & \qquad \qquad \qquad \qquad ~\,
	+ \frac{m_{2}}{r_{12}} \Bigl(\frac{32}{45} n_{12}^{i} v_{1}^{a} y_{1}^{bj}+ \frac{16}{45} n_{12}^{a} v_{1}^{i} y_{1}^{bj}
	-  \frac{16}{45} n_{12}^{a} v_{2}^{i} y_{1}^{bj}\Bigr)\biggl]\biggl)
+ \frac{16}{15} n_{12}^{ib} \frac{G m_{2} \varepsilon_{bkl} \dS^{(5)}_{jl}}{r_{12}^2} (n_{12}{} v_{2}{}) y_{1}^{jk} 	\nonumber \\ & \qquad 
	+ \biggl [G m_{2} \varepsilon_{abk} \Bigl(- \frac{16}{15} n_{12}^{ijk} v_{1}^{a}+ \frac{56}{45} n_{12}^{ijk} v_{2}^{a}\Bigr)
	+ \frac{16}{45} \varepsilon_{bak} v_{1}^{ijk} y_{1}^{a}\biggl] \dS^{(5)}_{bj}
	+ \frac{1}{84} \varepsilon_{iab} \dS^{(8)}_{bjk} y_{1}^{ajk}
	+ \frac{1}{42} \varepsilon_{ibk} v_{1}^{a} \dS^{(7)}_{ajk} y_{1}^{bj}
	\nonumber\\
	& \qquad + \frac{1}{42} \varepsilon_{iak} v_{1}^{a} \dS^{(7)}_{bjk} y_{1}^{bj}-  \frac{1}{42} \varepsilon_{abk} v_{1}^{a} \dS^{(7)}_{ijk} y_{1}^{bj}\Biggl\}\,.
\end{align}\ese
We deliberately keep the canonical moments $\dM_L$ and $\dS_L$ of the compact binary sources in ``unexpanded'' form, but of course they contain all their PN corrections consistent with the approximation. In particular, the mass quadrupole moment~$\dM_{ij}$ is accurate through 2PN order (and its time derivatives are computed consistently with the 2PN conservative EOM), the mass octupole and current quadrupole $\dM_{ijk}$ and $\dS_{ij}$ are accurate through 1PN order, while the other moments are Newtonian. The expression of the acceleration after replacing the moments by their explicit expressions in terms of the particles is extremely lengthy, and relegated to the Supplemental Material~\cite{SuppMaterial}.

\section{Flux balance laws}\label{sec:flux_balance}

In this section, we shall prove that the RR part of the acceleration obtained in~\eqref{eq:a1i_moments}--\eqref{eq:acc45PN1} enables us to recover the flux-balance equations for the energy, the angular and linear momenta, and the CM position, up to the 2PN relative (next-to-next-to-leading) order. The calculation constitutes therefore a proof of the flux-balance laws, since they are derived from the RR force in the near-zone EOM, and the fluxes are found to agree with the fluxes computed at future null infinity from the source. Thus, we shall prove that there exists some local energy $E$, angular momentum $J^i$, linear momentum $P^i$ and CM position $G^i$ (which can be attributed to the matter system) such that
\bse\label{eq:fluxbalance}\begin{align}
	\frac{\dd E}{\dd t} &= - \calF_{E} \,,\\
	\frac{\dd J^i}{\dd t} &= - \calF^i_{\bm{J}} \,,\\
	\frac{\dd P^i}{\dd t} &= - \calF^i_{\bm{P}} \,,\\
	\frac{\dd G^i}{\dd t} &= P^i - \calF^i_{\bm{G}} \,,
\end{align}\ese
where the fluxes are given up to 2PN relative order (corresponding to 4.5PN RR order) by~\cite{Th80,BF19}
\bse\label{eq:fluxes}\begin{align}
\label{seq:fluxE} \calF_{E} &= \frac{G}{c^5} \left( \frac{1}{5}
		\dM^{(3)}_{ij}\dM^{(3)}_{ij} +
	\frac{1}{c^2}\left[\frac{1}{189}
		\dM^{(4)}_{ijk}\dM^{(4)}_{ijk} +
		\frac{16}{45} \dS^{(3)}_{ij}\dS^{(3)}_{ij}
		\right]  +
	\frac{1}{c^4}\left[\frac{1}{9072}
		\dM^{(5)}_{ijkl}\dM^{(5)}_{ijkl} +
		\frac{1}{84} \dS^{(4)}_{ijk}\dS^{(4)}_{ijk}
		\right] \right) + \calO(10)
		\,,\\
%%%%%%%%%%%%%%%%%%%%%%%%%%%%%%%%%%%%%%%%%%%%%%%%%%%%%%%%%%%%%%%%
\label{seq:fluxJ} \calF_{\bm{J}}^i &= \frac{G}{c^5}
		\varepsilon_{ijk} \left( \frac{2}{5} \dM^{(2)}_{jl} \dM^{(3)}_{kl} +
		\frac{1}{c^2}\left[ \frac{1}{63} \dM^{(3)}_{jlm} \dM^{(4)}_{klm} +
		\frac{32}{45} \dS^{(2)}_{jl} \dS^{(3)}_{kl} \right] +
		\frac{1}{c^4}\left[ \frac{1}{2268} \dM^{(4)}_{jlmn} \dM^{(5)}_{klmn} +
		\frac{1}{28} \dS^{(3)}_{jlm} \dS^{(4)}_{klm} \right] \right) + \calO(10) \,, \\
%%%%%%%%%%%%%%%%%%%%%%%%%%%%%%%%%%%%%%%%%%%%%%%%%%%%%%%%%%%%%%%%
\label{seq:fluxP} \calF_{\bm{P}}^i &= \frac{G}{c^7}\left( \frac{2}{63}
	\dM^{(4)}_{ijk} \dM^{(3)}_{jk} +
	\frac{16}{45} \varepsilon_{ijk}
	\dM^{(3)}_{jl} \dS^{(3)}_{kl}  + \frac{1}{c^2}\left[ \frac{1}{1134}
	\dM^{(5)}_{ijkm} \dM^{(4)}_{jkm} + \frac{4}{63}
	\dS^{(4)}_{ijk} \dS^{(3)}_{jk} +
	\frac{1}{126} \varepsilon_{ijk}
	\dM^{(4)}_{jlm} \dS^{(4)}_{klm} \right] \right) +
	\calO(10) \,,\\
%%%%%%%%%%%%%%%%%%%%%%%%%%%%%%%%%%%%%%%%%%%%%%%%%%%%%%%%%%%%%%%%
	\label{seq:fluxG} \calF_{\bm{G}}^i &= \frac{G}{c^7}\left( \frac{2}{21}
	\dM^{(3)}_{ijk} \dM^{(3)}_{jk} + \frac{1}{c^2}\left[ \frac{2}{567}
	\dM^{(4)}_{ijkl} \dM^{(4)}_{jkl} + \frac{4}{21}
	\dS^{(3)}_{ijk} \dS^{(3)}_{jk} \right] \right) +
	\calO(10)\,.
\end{align}\ese
Again, the multipole moments are left in unexpanded form, but they include all relevant PN corrections.

The expressions~\eqref{eq:fluxes} concern the local-in-time effects, but there are also some non-local-in-time terms arising at the 4PN order due to gravitational-wave tails. To 4PN order, only the tails in the energy and angular momentum fluxes have to be taken into account, and we have
\bse\label{eq:Ftail}
\begin{align}
	\calF_{E}\Big|_\text{tail} &= \frac{2G^2 \dM}{5c^8} \dM^{(1)}_{ij} \!\int^{+\infty}_0
	\!\dd\tau \ln \left(\frac{\tau}{P}\right) \left[ \dM^{(7)}_{ij}(t-\tau) + \dM^{(7)}_{ij}(t+\tau) \right] + \calO(10) \,,\label{eq:Ftail_E}\\
	\calF^i_{\bm{J}}\Big|_\text{tail} &= \frac{4G^2 \dM}{5c^8} \varepsilon_{ijk}\dM_{jl} \!\int^{+\infty}_0
	\!\dd\tau \ln \left(\frac{\tau}{P}\right) \left[ \dM^{(7)}_{kl}(t-\tau) + \dM^{(7)}_{kl}(t+\tau) \right] + \calO(10) \,,
\end{align}\ese
where $\dM$ is the constant (ADM) mass monopole, and the time scale $P$ in the logarithm actually drops from the result. The non-local tail terms come from the dissipative part of the tail integral in the acceleration at the 4PN order, as given by Eq.~(6.2) in~\cite{BBFM17}. In addition to the dissipative part, there is also a conservative part of the tail integral in the acceleration, which contributes to the left-hand sides of the balance laws; both dissipative and conservative non-local 4PN tail terms are discussed in Ref.~\cite{BBFM17}. 

In Eqs.~\eqref{eq:fluxes}, we have written the fluxes in ``canonical'' form (see \textit{e.g.}~\cite{Th80}), but of course one is free to define the fluxes differently, \textit{i.e.}, by using the Leibniz rule to transfer some total time derivatives to the left-hand sides of the balance laws~\eqref{eq:fluxbalance}. For instance, the flux of CM position was derived in Ref.~\cite{BF19} in the canonical form~\eqref{seq:fluxG}, but Refs.~\cite{KNQ18, COS20} obtained the alternative form
\begin{align}\label{eq:fluxGalternatif}
	\widehat{\calF}_{\bm{G}}^i &= \frac{G}{c^7}\left( \frac{1}{21}
	\Bigl(\dM^{(3)}_{ijk} \dM^{(3)}_{jk}-\dM^{(4)}_{ijk} \dM^{(2)}_{jk}\Bigr) + \frac{1}{c^2}\left[ \frac{1}{567}
	\Bigl(\dM^{(4)}_{ijkl} \dM^{(4)}_{jkl}-\dM^{(5)}_{ijkl} \dM^{(3)}_{jkl}\Bigr) + \frac{2}{21}
	\Bigl(\dS^{(3)}_{ijk} \dS^{(3)}_{jk}-\dS^{(4)}_{ijk} \dS^{(2)}_{jk}\Bigr) \right] \right) +
	\calO(10)\,.
\end{align}
Since the two forms differ only by a total time derivative, the balance law stays physically the same, with just a redefinition of the CM position $G^i$ into some $\widehat{G}^i$, and a different split between what we interpret as the contribution of the matter system versus what we attribute to the radiation field.

%\begin{align}
%	\calF_{E} &= \frac{G}{c^5} \left( \frac{1}{5}
%	M^{(3)}_{ij}M^{(3)}_{ij} +
%	\frac{1}{c^2}\left[\frac{1}{189}
%	M^{(4)}_{ijk}M^{(4)}_{ijk} +
%	\frac{16}{45} S^{(3)}_{ij}S^{(3)}_{ij}
%	\right] \right.\nn\\& \left.
%	\qquad\qquad + \frac{2G M}{c^3} M^{(3)}_{ij} \!\int^{+\infty}_0
%	\!\dd\tau \ln \tau \left[ M^{(5)}_{ij}(t-\tau) + M^{(5)}_{ij}(t+\tau) \right]
%	\right.\nn\\& \left.
%	\qquad\qquad\qquad +\frac{1}{c^4}\left[\frac{1}{9072}
%	M^{(5)}_{ijkl}M^{(5)}_{ijkl} +
%	\frac{1}{84} S^{(4)}_{ijk}S^{(4)}_{ijk}
%	\right] \right) + \calO\left( \frac{1}{c^{10}}\right) \,,
%\end{align}

%\subsection{Ad hoc method}

%\David{Luc décrit sa méthode avec un exemple.}
%\subsection{Systematic method}

Let us now show how we derived the flux balance laws~\eqref{eq:fluxbalance} from the 4.5PN RR acceleration, the latter being expressed in terms of the unexpanded multipole moments by Eqs.~\eqref{eq:a1i_moments}--\eqref{eq:acc45PN1}. In our approach, it is important not to replace the moments by their expressions in terms of the binary's parameters, so as to have a better control over the structure of the acceleration and avoid handling extremely long expressions. See, however, Appendix~\ref{app:id_dim} for an alternative method manageable at 3.5PN order.
 
%%%%%%%%%%%%%%%%%%%%%
%%%%%%%%%%%%%%%%%%%%%%
%First, for illustration purposes, we take the example of the balance equation for the energy $E$.
%We form the combination
%
%\begin{align}\label{eq:combE}
%	\frac{\dd}{\dd t}\left(\frac{1}{2}m_1 v_1^2 + \frac{1}{2}m_2 v_2^2\right) &\equiv m_1 v_1^i a_{1}^{i} + m_2 v_2^i a_{2}^{i} \nn\\&= m_1 v_1^i a_{\mathrm{cons}\,1}^{i} + m_1 v_1^i a_{\mathrm{RR}\,1}^{i} + (1 \leftrightarrow 2)\,,
%\end{align}
%
%
%
%%
%where we have decomposed the total acceleration into a conservative part (accurate to 2PN order) and the RR part given by~\eqref{eq:a1i_moments}. The conservation piece in~\eqref{eq:combE} can be written as a total time derivative up to 2PN order (since there is a conserved energy at that order), but because of the replacement of accelerations this will imply some extra RR contributions; calling these terms $q\RR$ we have transformed Eq.~\eqref{eq:combE} into
%%
%\begin{equation}\label{eq:combE2}
%	\frac{\dd E\cons}{\dd t} = q\RR + m_1 v_1^i a_{\mathrm{RR}\,1}^{i} + m_2 v_2^i a_{\mathrm{RR}\,2}^{i}\,.
%\end{equation}
%%
%Replacing the RR acceleration by the explicit results~\eqref{eq:acc45PN1}, and \textit{idem} for the terms $q\RR$, we have to prove that the right-hand side of Eq.~\eqref{eq:combE2} can be written as a total time derivative minus the contribution of the flux. 

%%%%%%%%%%%%%%%%
%%%%%%%%%%%%%%%%%%%

First, for illustration purposes, we take the simple example of the balance equation for the energy $E$ at lowest order~$\calO(0,5)$. We form the combination $m_1 v_1^i a_{1}^{i} + m_2 v_2^i a_{2}^{i}$ and readily obtain
\begin{align}\label{eq:combE}
	\frac{\dd}{\dd t}\left(\frac{1}{2}m_1 v_1^2 + \frac{1}{2}m_2 v_2^2\right) &\equiv m_1 v_1^i a_{1}^{i} + m_2 v_2^i a_{2}^{i} = \frac{\dd}{\dd t}\left(\frac{G m_1 m_2}{r_{12}}\right) - \frac{G}{5 c^5}\dM_{ij}^{(1)}\dM_{ij}^{(5)}+ \calO(2,7)\,.
\end{align}
where we have recognized in the right-hand side the Newtonian expression for $\dM_{ij}^{(1)} = 2m_1 y_1^{\langle i} v_1^{j \rangle} + 2m_2 y_2^{\langle i} v_2^{j \rangle} + \calO(2)$. It is then trivial to transform the latter equation into the canonical flux balance law
\begin{equation} 
	\frac{\dd (E\cons + E\RR)}{\dd t} = - \mathcal{F}_E \,,
\end{equation}
where we recover the usual quadrupole flux $\mathcal{F}_E = \frac{G}{5 c^5}\dM_{ij}^{(3)}\dM_{ij}^{(3)} + \calO(7)$, where the Newtonian conservative energy is $E_\text{cons} = \frac{1}{2}m_1 v_1^2 + \frac{1}{2}m_2 v_2^2 - \frac{G m_1 m_2}{r_{12}} + \calO(2)$, and where $E\RR$ represents some Schott-like terms~\cite{Schott} given by
\begin{equation} 
	E_\text{RR} =  \frac{G}{5 c^5} \Big(\dM_{ij}^{(2)}\dM_{ij}^{(3)}  - \dM_{ij}^{(1)}\dM_{ij}^{(4)} \Big) + \calO(7) \,. 
\end{equation}

In this approach, note that we neglected many higher order terms in the following operations: replacement of the accelerations by their expressions, identification of total derivatives, and recognition of the expression of $\dM_{ij}^{(1)}$. These terms must crucially be controlled and accounted for when deriving the flux-balance laws at the relative 1PN and 2PN orders. Moreover, due to these terms, it will be increasingly difficult in higher order calculations to identify total time derivatives ``by eye''. Thus, in order to treat the full 2PN problem, a systematic approach is needed.

To this end, the first step is to express the flux in an adequate form that is linear in the \textit{unexpanded} multipole moments (rather than quadratic), but nonetheless preserves some symmetry between multipole moments. We constructed such a flux by replacing in Eqs.~\eqref{eq:fluxes} all quadratic combinations of moments by the symmetric product of one unexpanded moment times the other in expanded form. That is,\footnote{It would have been possible to assume a more general, \textit{a priori} non-symmetric split of terms, but it turns out that~\eqref{eq:split} does work.}
\begin{equation}\label{eq:split}
	\mathrm{A}_L^{(n)} \mathrm{B}_{L'}^{(m)} \quad\longrightarrow\quad \frac{1}{2}\Bigl(\mathrm{A}_L^{(n)}\widetilde{\mathrm{B}}_{L'}^{(m)} + \widetilde{\mathrm{A}}_L^{(n)}\mathrm{B}_{L'}^{(m)}\Bigr)\,,
\end{equation}
where $\mathrm{A}_L$ and $\mathrm{B}_L$ can stand for either $\dM_L$ or $\dS_L$, and where the tilde means that we have replaced the multipole moment by its explicit expression in terms of the particles (\textit{i.e.}, expanded form in terms of $m_1$, $y_1^i$, $v_1^i$, \textit{etc.}), while the moments without a tilde are not replaced (\textit{i.e.}, unexpanded form). We then have the generic structure 
\begin{align}\label{eq:balance_systematic_1} 
\frac{\dd H\cons}{\dd t} + \mathcal{F}_H = \sum_{\ell = 2}^{+\infty} \Biggl[ \sum_{p= p_\mathrm{min}(\ell)}^{p_\mathrm{max}(\ell)} \mathcal{C}_L \dM_L^{(p)}+  \sum_{q= q_\mathrm{min}(\ell)}^{q_\mathrm{max}(\ell)} \mathcal{D}_L \dS_L^{(q)}  \Biggr]\,,
\end{align} 
where $H$ stands for either $E$, $J_i$, $P_i$, or $G_i$, and $\mathcal{C}_L$ and $\mathcal{D}_L$ are explicitly given in terms of the particles. Note that in the previous expression, the time derivative $\dd / \dd t$ is taken using the full acceleration (conservative and dissipative); therefore the conservative sector vanishes by construction, and $\dd H\cons/\dd t$ is purely composed of terms arising from the RR sector of the acceleration, which gives rise to terms that are explicitly linear in the multipole moments \textit{via} the expressions~\eqref{eq:acc45PN1}. It is then straightforward to integrate by parts (up to a controlled total time derivative) in a way that reduces the number of time derivatives on the multipole moments and increases it on the coefficients $\mathcal{C}_L$ and $\mathcal{D}_L$ (the conservative 2PN EOM are needed in this operation). Then, it is easy to express the previous equation with some new coefficients $\mathcal{C}'_L$ and $\mathcal{D}'_L$ as
\begin{align} \label{eq:balance_systematic_2} 
\frac{\dd H\cons}{\dd t} + \mathcal{F}_H = \sum_{\ell = 2}^{+\infty} \Biggl[  \mathcal{C}'_L \dM_L^{(p_\mathrm{min})} +  \mathcal{D}'_L  \dS_L^{(q_\mathrm{min})} \Biggr] - \frac{\dd H\RR}{\dd t}\,,
\end{align} 
where the Schott term $H\RR$ is explicitly constructed through the integration by parts. Remarkably, we have found that the new coefficients in fact all identically vanish at this order:\footnote{Two subtleties can arise here. First, it can happen that $\mathcal{C}'_L$ and $\mathcal{D}'_L$ are not zero, but for example constants of motion. It is then sufficient to take a lower value of $p_\mathrm{min}$ to reach the same conclusions. Secondly, when dealing with balance laws for the vectorial quantities $J^i$, $P^i$, and $G^i$, it can happen that these coefficients are not manifestly zero, but can be proven to vanish by computing the Hodge duals of the vectors $\mathcal{C}'_L \dM_L^{(p_\text{min})}$ and $\mathcal{D}'_L \dS_L^{(q_\text{min})}$; see~\eqref{eq:HodgeDual} for details.} 
\begin{equation} \label{eq:CD_vanishing}
	\mathcal{C}'_L \equiv \mathcal{D}'_L \equiv 0\,,
\end{equation}
which means that the left-hand side of~\eqref{eq:balance_systematic_2} is explicitly expressed as a total time derivative. Thus, the balance equation is proven through 4.5PN order, once the Schott term has been incorporated into the definition of $H$, namely $H = H\cons + H\RR$. The previous method has been successfully applied to all the balance equations for $H=\{E, J_i, P_i, G_i\}$ at 4.5PN order. The conservative parts are required at 2PN order and are given in harmonic coordinates in Sec.~IV A of~\cite{ABF01}. We have therefore obtained the Schott terms for all these quantities, such that
%
%\subsection{Results}
%
%The proof of the flux-balance equations~\eqref{eq:fluxbalance}, where the fluxes~\eqref{eq:fluxes} agree with those computed at future null infinity, consists in proving that there exist RR contributions to the energy, angular momentum, linear momentum and CM such that the balance equations~\eqref{eq:fluxbalance} are satisfied with the fluxes given by~\eqref{eq:fluxes}.
%
\bse\label{eq:consRR}\begin{align}
\label{seq:consRRE}E &= E\cons + E\RR \,, \\
\label{seq:consRRJ}J_i &= J\cons^i + J\RR^i \,, \\
\label{seq:consRRP}P_i &= P\cons^i + P\RR^i \,, \\
\label{seq:consRRG}G_i &= G\cons^i + G\RR^i \,, 
\end{align}\ese
and proved that the flux-balance equations~\eqref{eq:fluxbalance} are satisfied to the required order. The RR Schott terms are composed of 2.5PN, 3.5PN and 4.5PN terms, namely
\bse\label{eq:Schott}\begin{align}
	E\RR &= E_{\text{2.5PN}} + E_{\text{3.5PN}} + E_{\text{4.5PN}} + \calO(11)\,,\\
	J\RR^i &= J^i_{\text{2.5PN}} + J^i_{\text{3.5PN}} + J^i_{\text{4.5PN}} + \calO(11)\,,\\
	P\RR^i &= P^i_{\text{3.5PN}} + P^i_{\text{4.5PN}} + \calO(11)\,,\\
	G\RR^i &= G^i_{\text{3.5PN}} + G^i_{\text{4.5PN}} + \calO(11)\,.
\end{align}\ese
The 2.5PN and 3.5PN terms are given in Appendix~\ref{app:Schott}, but the explicit expressions of the 4.5PN terms are very lengthy (even when the moments are left unexpanded), so they are relegated to the Supplemental Material~\cite{SuppMaterial}. Moreover, we provide the expanded form of the RR Schott terms in the Supplemental Material~\cite{SuppMaterial}, and we have verified that the balance equations also hold directly at that level, once the dimensional identities are correctly identified (see App.~\ref{app:id_dim} for details). 

\section{Reduction to the frame of the center of mass}
\label{sec:CoM}

\subsection{Definition of the center-of-mass frame}

We have established in a general frame at 4.5PN order the flux balance laws for linear momentum and CM position~\eqref{eq:fluxbalance}, which we repeat here:
\bse\label{eq:fluxbalancePG}
\begin{align}
	\frac{\dd P^i}{\dd t} &= - \calF^i_{\bm{P}} \,,\\
	\frac{\dd G^i}{\dd t} &= P^i - \calF^i_{\bm{G}} \,,
\end{align}\ese
where $P^i$ and $G^i$ are made of a conservative part and a RR part following~\eqref{eq:consRR}--\eqref{eq:Schott}. We emphasize that $P^i$ and $G^i$ correspond to the linear momentum and CM position of the \textit{matter} system (the compact binary) while the right-hand sides of Eqs.~\eqref{eq:fluxbalancePG} are attribuable to the \textit{radiation}. Evidently, as we have already discussed, this split between matter and radiation is somewhat arbitrary, since we can rewrite the flux in non-canonical form and transfer a total time derivative to the left-hand side [see for instance Eq.~\eqref{eq:fluxGalternatif}].

We integrate the equations~\eqref{eq:fluxbalancePG} in the case where the source is stationary before some initial instant $t_0$, and emits gravitational waves starting from $t_0$. The solution of the balance laws at any time $t\geqslant t_0$ reads
\bse\label{eq:solbalancePG}
	\begin{align}
		P^i(t) &= P_0^i - \int_{t_0}^{t} \dd t' \,\calF^i_{\bm{P}}(t')\,,\\ 
		G^i(t) &= P_0^i\bigl(t-t_0\bigr) + G_0^i
		- \int_{t_0}^{t} \dd t' \int_{t_0}^{t'} \dd t'' \calF^i_{\bm{P}}(t'') - \int_{t_0}^{t} \dd t' \calF^i_{\bm{G}}(t') \,,
	\end{align}
\ese
where $P_0^i$ and $G_0^i$ denote two integration constants, namely the initial values of $P^i$ and $G^i$ at the initial time $t_0$. 

We can always apply a constant Lorentz boost so that we are initially (at $t=t_0$) in the rest frame of the system, for which $P_0^i=0$. Furthermore, we can always apply a constant spatial translation to choose the origin of the coordinate system to coincide with the CM, hence $G_0^i=0$. These choices define the frame of the CM, and by CM we mean the one of the total matter plus radiation system. From now on, we assume that $t_0=-\infty$ (recall that the matter system is stationary in the remote past), and we introduce the special notation for the non-local-in-time (or semi-hereditary\footnote{Following the terminology of Ref.~\cite{BD92}, we may refer to these terms as ``semi-hereditary'', as they are given by some time anti-derivative of local terms; thus applying multiple time derivatives to a semi-hereditary term reduces it to be local.}) integrated fluxes in~\eqref{eq:solbalancePG}:
\bse\label{eq:non-local}
\begin{align}
	\Pi^i(t) &\equiv \int_{-\infty}^{t} \dd t' \,\calF^i_{\bm{P}}(t')\,,\\ 
	\Gamma^i(t) &\equiv \int_{-\infty}^{t} \dd t' \int_{-\infty}^{t'} \dd t'' \calF^i_{\bm{P}}(t'') + \int_{-\infty}^{t} \dd t' \calF^i_{\bm{G}}(t') = \int_{-\infty}^{t} \dd t' \,\Bigl[ \bigl(t-t'\bigr)\,\calF^i_{\bm{P}}(t') + \calF^i_{\bm{G}}(t')\Bigr] \,,
\end{align}
\ese
which verify the identities $\dd \Pi^i / \dd t = \calF_{\bm{P}}^i$ and $\dd \Gamma^i / \dd t = \Pi^i + \calF_{\bm{G}}^i$. Thus, the CM frame is defined by  
%(which immediately implies the other equation) 
%
\begin{subequations}\label{eq:defCM}
\begin{align}
	G^i + \Gamma^i &= 0\,,
\end{align}
which implies
\begin{equation}P^i + \Pi^i = 0 \,.\end{equation}\end{subequations}
Using the expression for $G^i$ in~\eqref{seq:consRRG}, made of conservative and RR terms, where the Schott terms $G\RR^i$ are given in Appendix~\ref{app:Schott}, it is straightforward to obtain the formulas for the passage to the CM frame by solving iteratively the equation $G^i + \Gamma^i = 0$. In this way, we find that the CM positions (extending the usual notation, see \textit{e.g.}~\cite{BBFM17}) read
\bse\label{eq:passage_CoM}\begin{align}
	y_1^i &= x^i \Bigl( X_2 + \nu \Delta \,\calP \Bigr) +  \nu \Delta \,\calQ \,v^i + \calR^i\,, \\
	y_2^i &= x^i \Bigl( -X_1 + \nu \Delta \,\calP \Bigr) +  \nu \Delta \,\calQ \,v^i + \calR^i\,, 
\end{align}\ese
%
%\Luc{mettre aussi les $v1^i$ $v2^i$}
where $x^i=y_1^i-y_2^i$ and $v^i=v_1^i-v_2^i=\dd x^i/\dd t$ are the relative position and velocity, and we pose $m=m_1+m_2$, $X_1=m_1/m$, $X_2=m_2/m$, along with the symmetric mass ratio $\nu=X_1 X_2$ and the mass difference $\Delta=X_1-X_2$. 

Here, the quantities $\calP$ and $\calQ$ are local and can be computed up to 4.5PN order, with the time-odd contributions at 2.5PN, 3.5PN and 4.5PN orders corresponding to the extended BT coordinates [see~\eqref{eq:PQRR}]. In our convention, $\calP$ and $\calQ$ correspond to the contribution of matter, by which we mean that they are obtained by solving $G^i = 0$. The novel feature is the supplementary term $\calR^i$ introduced in~\eqref{eq:passage_CoM}, which corresponds to the contribution of radiation. This radiation term is defined in such a way that the full equations for the passage to the CM frame \eqref{eq:passage_CoM} solve the full equation $G^i + \Gamma^i = 0$, rather than $G^i = 0$. This term is non-local-in-time and directly linked to the integrated fluxes~\eqref{eq:non-local}. It arises at the 3.5PN order followed by a 4.5PN term which can also be controlled with the present accuracy. Taking into account the 1PN corrections in $G^i$ as well as those in $\calP$ and $\calQ$ (actually the latter is zero as $\calQ$ starts at 2PN order) we obtain
\begin{align}\label{eq:Ri}
	\calR^i &= - \frac{\Gamma^i}{m} + \frac{\nu}{m c^2}\left[ \left( \frac{v^2}{2} - \frac{G m}{r}\right)\Gamma^i + v^j\Bigl(\Pi^j+\calF_{\bm{G}}^j\Bigr) x^i \right] + \calO(11)\,,
\end{align}
where we recall that $\Gamma^i$ is of order 3.5PN with the next-to-leading 4.5PN correction included. Therefore, we already see that the definition of the CM frame is altered at the 3.5PN order by a non-local (semi-hereditary) contribution which accounts for the recoil of the  source and the displacement of the CM due to the GW emission.

Now, let us give the local-in-time contributions $\calP$ and $\calQ$ to the formulas for the passage~\eqref{eq:passage_CoM} to the CM frame. Actually, the even parity contributions (\textit{i.e.}, 1PN, 2PN, 3PN and 4PN) have been already given elsewhere, see Eqs.~(B5) and (B6) of~\cite{BBFM17}. Furthermore, as proven in Sec. IV of~\cite{BBFM17}, the tails at 4PN order do not affect the definition of the CM frame. Thus, we only display the dissipative half-integers contributions, \textit{i.e.}, the RR terms corresponding to the extended BT coordinates. We obtain
\bse\label{eq:PQRR}\begin{align}
	\calP\RR &= \frac{G^2 m^2 \nu \dot{r}}{r^2} \Biggl\{\frac{1}{c^{7}} \Biggl[\frac{212 G m}{105 r}
	-  \frac{26}{5} \dot{r}^2
	+ \frac{78}{7} v^{2}\Biggr] \nn\\
	&\qquad\qquad\ \ 	+ \frac{1}{c^{9}} \Biggl[ \frac{G^2 m^2}{r^2}\Bigl(- \frac{18538}{945}+ \frac{88}{21} \nu \Bigr)
	+  \dot{r}^4 \Bigl( -  \frac{1474}{63} + \frac{1172}{63} \nu \Bigr)+  v^{2} \dot{r}^2 \Bigl(\frac{34093}{2205} 
	+ \frac{29314}{2205} \nu\Bigr) \nn\\
	& \qquad\qquad\qquad \quad 
	+ \frac{G m}{r}\dot{r}^2 \Bigl(\frac{569552}{6615} 
	+ \frac{98926}{6615} \nu   \Bigr) +  \frac{G m}{r}v^{2} \Bigl(- \frac{91513}{735}
	+ \frac{5448}{245} \nu \Bigr) + v^{4} \Bigl(\frac{10321}{2205} -  \frac{8402}{147} \nu \Bigr) \Biggr]\Biggr\} + \calO(11) \,,\\
%%%%%%%%%%%%%%%%%%%%%%%%%%%%%%%%%%%%%%%%%%%%%%%%%%%%%%%%%%%%%%%%%%%%%%%%%%%%%%%%%%%%%%%%%%
	\calQ\RR &= G m \nu \Biggl\{\frac{1}{c^{7}} \Biggl[- \frac{48}{35}\frac{ G^2 m^2}{ r^2}
	-  \frac{398}{105} \frac{G m}{r} \dot{r}^2
	-  \frac{58}{105} \frac{G m}{r}  v^{2}
	-  \frac{8}{35} v^{4}\Biggl] \nn\\
	& \qquad \quad
	+ \frac{1}{c^{9}} \Biggl[\frac{G^3 m^3}{r^3} \Bigl(\frac{25114}{2835} -  \frac{3484}{945} \nu \Bigr) 
	+ \frac{G^2 m^2}{r^2}\dot{r}^2  \Bigl(\frac{9518}{441} 	-  \frac{56978}{2205} \nu\Bigr)
	+ \frac{G^2 m^2}{r^2}v^{2}\Bigl( 	\frac{751}{105}+ \frac{278}{315} \nu \Bigr) \nn\\
	& \qquad \qquad\quad
	+  \frac{G m}{r}\dot{r}^4\Bigl(\frac{22712}{2205}+ \frac{9116}{2205} \nu \Bigr) 
	+ \frac{G m}{r} v^{2} \dot{r}^2 \Bigl(- \frac{16453}{2205}  + \frac{5218}{735} \nu\Bigr) 
	+  \frac{G m}{r} v^{4} \Bigl(\frac{1469}{2205}	+ \frac{122}{147} \nu \Bigr) \nn\\
	& \qquad \qquad\quad
	+ v^{6} \Bigl(- \frac{872}{2205}	+ \frac{356}{245} \nu \Bigr) \Biggl]\Biggr\} + \calO(11) \,.
\end{align}\ese
Note that in extended BT coordinates, the 2.5PN contribution vanishes, unlike in harmonic coordinates.
For explicitness, we also provide the expressions of the fluxes of linear and angular momentum to 4.5PN order in the CM frame for two-particle systems, which enter the semi-hereditary quantities $\Pi^i$ and $\Gamma^i$: 
%\David{à vérifier}\Luc{Je peux le faire}\Guillaume{OK pour $\calF_{\bm{G}}^i$}
%	
\bse \label{eq:FP_FG_COM} \begin{align}
	\label{seq:FP_COM} \calF_{\bm{P}}^i&= \frac{G^3 m^4 \Delta \nu^2}{c^7 r^4} \Bigg\{ \dot{r} n^i \Bigg(\frac{32 G m}{35 r} -  \frac{24}{7} \dot{r}^2 + \frac{88}{21} v^2\Bigg) + v^i\Bigg(- \frac{64 G m}{105 r} + \frac{304}{105} \dot{r}^2 -  \frac{80}{21} v^2\Bigg)  \\
	%%%
	&\qquad + \frac{1}{c^2}\Bigg[ \dot{r} n^i \Bigg( \dot{r}^4 \Big(\frac{14744}{315} -  \frac{1184}{45} \nu \Big) + \dot{r}^2 v^2\Big( -  \frac{22672}{315} + \frac{15016}{315} \nu\Big) +  v^{4}  \Big(\frac{6808}{315}-  \frac{6232}{315} \nu\Big) \nn\\
	& \qquad\qquad\qquad+ \frac{G m \dot{r}^2}{r}\Big(\frac{49204}{945} -  \frac{4672}{945} \nu \Big) + \frac{G m v^2}{r} \Big(-  \frac{3508}{63}  + \frac{3824}{315} \nu \Big) + \frac{G^2 m^2}{r^2}\Big(- \frac{472}{63}  + \frac{16}{315} \nu \Big) \Bigg) \nn\\
	&\qquad \quad + v^i \Bigg( \dot{r}^4 \Big(-  \frac{10652}{315} + \frac{5576}{315} \nu \Big) +  \dot{r}^2 v^2\Big( \frac{448}{9} -  \frac{2056}{63} \nu \Big) + v^4\Big( -  \frac{740}{63} + \frac{592}{45} \nu \Big)\nn\\
	& \qquad\qquad\qquad + \frac{G m \dot{r}^2}{r}\Big(- \frac{10796}{315}  - \frac{8}{63} \nu \Big)  +  \frac{G m v^2}{r}\Big( \frac{3628}{105}  - \frac{216}{35} \nu \Big)+  \frac{G^2 m^2 }{r^2}\Big(\frac{32}{5} + \frac{544}{945} \nu\Big)\Bigg) \Bigg] \Bigg\} + \calO(11)\,,\nn\\
	%%%%%%%%%%%%%%%%%%%%%
	%%%%%%%%%%%%%%%%%%%%%
	\label{seq:FG_COM}\calF_{\bm{G}}^i&= \frac{G^2 m^3 \Delta \nu^2}{c^7 r^2} \Bigg\{n^i \Bigg(\frac{272G m}{105r} \dot{r}^2 -  \frac{64G m}{15r} v^2 + \frac{48}{35} \dot{r}^2 v^2 -  \frac{32}{35} v^{4}\Bigg) +  \dot{r} v^i \Bigg(\frac{16 G m}{21 r} -  \frac{24}{7} \dot{r}^2 + \frac{24}{7} v^2\Bigg)  \\
	& \qquad + \frac{1}{c^2}\Bigg[n^i \Bigg( \dot{r}^4 v^2\Big( -  \frac{96}{49}  + \frac{288}{49} \nu \Big) +\dot{r}^2 v^{4}\Big( \frac{344}{105}  -  \frac{376}{35} \nu \Big) +  v^{6}\Big( -  \frac{632}{315} + \frac{144}{35} \nu\Big) \nn\\
	& \qquad\qquad\qquad  + \frac{G m  \dot{r}^4}{r} \Big(- \frac{3908}{245} + \frac{3716}{245} \nu \Big) + \frac{G m \dot{r}^2 v^2}{r} \Big(\frac{50116}{2205}  -  \frac{19148}{735} \nu \Big) + \frac{G m v^{4}}{r} \Big( -  \frac{3272}{441}  + \frac{3008}{245} \nu \Big) \nn\\
	&  \qquad\qquad\qquad + \frac{G^2 m^2  \dot{r}^2}{r^2} \Big(- \frac{166504}{6615} + \frac{752}{735} \nu \Big)  +  \frac{G^2 m^2 v^2}{r^2} \Big(\frac{241616}{6615}  -  \frac{4864}{735} \nu \Big)+ \frac{G^3 m^3 }{r^3}\bigg(- \frac{256}{945} + \frac{256}{315} \nu \bigg) \Bigg)\nn\\
	& \qquad \quad +  \dot{r} v^i\Bigg( \dot{r}^4 \Big(\frac{680}{147} -  \frac{680}{49} \nu \Big)  + \dot{r}^2 v^2 \Big(-  \frac{832}{105} + \frac{1044}{35} \nu \Big) + v^{4}\Big(\frac{2504}{735} -  \frac{3988}{245} \nu \Big)\nn\\
	&\qquad\qquad\qquad + \frac{G m \dot{r}^2}{r}  \Big(\frac{6884}{441}  -  \frac{1396}{147} \nu\Big) + \frac{G m v^2}{r}  \Big( -  \frac{7556}{735}   + \frac{1588}{147} \nu \Big) + \frac{G^2 m^2}{r^2} \Big(- \frac{76352}{6615} + \frac{64}{49} \nu \Big) \Bigg) \Bigg] \Bigg\} + \calO(11) \,.\nn
\end{align}\ese
The 2PN-accurate fluxes of energy and angular momentum in the CM frame are given by Eqs.~(2.14) in~\cite{GII97}.
Finally, note that all these expressions are also provided in the Supplemental Material~\cite{SuppMaterial}.
\subsection{4.5PN acceleration in the center-of-mass frame}

First, we construct the relative acceleration $a^i \equiv a_1^i - a_2^i$, which contains both the conservative terms (up to 2PN order is sufficient here) and the RR contributions given by Eqs.~\eqref{eq:a1i_moments}--\eqref{eq:acc45PN1}. Our task is to reduce this relative acceleration in the CM frame by replacing $(\bm{y}_1, \bm{y}_2,\bm{v}_1,\bm{v}_2)$ with the help of Eqs.~\eqref{eq:passage_CoM}, which contain in particular the time-odd non-local term~\eqref{eq:Ri}. We find that two contributions arise. The first, \emph{direct}, contribution comes from considering the dissipative terms~\eqref{eq:a1i_moments} in the acceleration, to which one straightforwardly applies the formulas for the passage to the CM at 2PN order. The second, \emph{indirect}, contribution arises when considering the conservative terms in the relative acceleration, to which one applies the dissipative sector of the passage to the CM described by Eqs.~\eqref{eq:Ri}--\eqref{eq:PQRR}. 

As is well known,  the Newtonian piece of the relative acceleration is not modified when going to the CM frame. Moreover, there is no 2.5PN term in the formula for the passage to the CM in extended BT coordinates, so the 2PN term in the relative acceleration contributes only beyond the current 4.5PN accuracy. Thus, only the 1PN piece of the relative acceleration yields an indirect (in the previous sense) contribution. In particular, we find that the 3.5PN non-local term $\mathcal{R}^i$ [given by~\eqref{eq:Ri}] in the passage to the CM frame produces a non-local term in the acceleration reduced to the CM at the 4.5PN order. Thus, at 4.5PN order, the CM acceleration cannot be local in time, because it acquires a contribution due to the integrated linear momentum flux (or recoil) of the source. However, this extra contribution does not affect the 3.5PN-accurate expression for the mass-type quadrupole moment in the CM frame [see (A1) of~\cite{FMBI12}], since only the 2.5PN accurate formula for the passage to the CM frame is needed, thanks to the particular structure of the Newtonian term in the quadrupole moment.

We now split the RR piece of the relative acceleration into PN contributions,
\begin{align}\label{eq:aiCM_moments}
	a_{\mathrm{RR}}^{i} = a_{\text{2.5PN}}^{i} + a_{\text{3.5PN}}^{i} + a_{\text{4.5PN}}^{i}+\calO(11)\,.
\end{align}
The 2.5PN and 3.5PN pieces are readily obtained as 
%\Luc{remplacer par les expressions toutes remplacées}
%
\bse\label{eq:accCMPN}
\begin{align}
	a_{\text{2.5PN}}^{i} &= \frac{8 G^2 m^2 \nu}{c^5 r^3} \biggl [v^{i} \Bigl(\frac{2 G m}{5 r} + 3 \dot{r}^2 -  \frac{6}{5} v^{2}\Bigr) + n^{i} \dot{r} \Bigl(\frac{2 G m}{15 r} - 5 \dot{r}^2 + \frac{18}{5} v^{2}\Bigr)\biggl]\,,\\	%%%%%%%%%%%%%%%%%%%%%%%%%%%%%%%%%%%%%%%%%%%%%%%%%%%%%%%%%%%%%%%%%%%%%%%%%%%%%%%%%%%%%%%%%%%%%
	a_{\text{3.5PN}}^{i} &=\frac{8 G^2 m^2 \nu}{c^7 r^3} \Biggl[v^{i} \biggl(\Bigl(- \frac{776}{105}
 -  \frac{11}{3} \nu \Bigr) \frac{G^2 m^2}{r^2}
 + \Bigl(\frac{5}{2}
 -  \frac{35}{2} \nu \Bigr) \dot{r}^4
 + \Bigl(- \frac{39}{10}
 + \frac{111}{10} \nu \Bigr) \dot{r}^2 v^{2}
\nonumber\\ & \qquad \qquad \qquad + \frac{G m}{r} \biggl [\Bigl(- \frac{2591}{60}
 -  \frac{97}{5} \nu \Bigr) \dot{r}^2
 + \Bigl(\frac{4861}{420}
 + \frac{58}{15} \nu \Bigr) v^{2}\biggl]
 + \frac{27}{70} v^{4}\biggl) \nonumber\\
& \qquad \qquad \quad + n^{i} \dot{r} \biggl(\Bigl(\frac{32}{7}
 + \frac{11}{3} \nu \Bigr) \frac{G^2 m^2}{r^2}
 + \Bigl(- \frac{7}{2}
 + \frac{7}{2} \nu \Bigr) \dot{r}^4
 + \Bigl(\frac{5}{2} + \frac{25}{2} \nu \Bigr) \dot{r}^2 v^{2}\nonumber\\
& \qquad \qquad \qquad \quad ~
 + \frac{G m}{r} \biggl [\Bigl(\frac{1353}{20}
 + \frac{133}{5} \nu \Bigr) \dot{r}^2
 + \Bigl(- \frac{5379}{140}
 -  \frac{136}{15} \nu \Bigr) v^{2}\biggl]
 + \Bigl(\frac{87}{70}
 -  \frac{48}{5} \nu \Bigr) v^{4}\biggl)\Biggl]\,.
\end{align}\ese
The 4.5PN piece can be naturally decomposed into a ``matter'' and a ``radiation'' part, namely 
\begin{equation}
a_{\text{4.5PN}}^i = a_{\text{4.5PN}}^i \Big|_{\text{mat}}+a_{\text{4.5PN}}^i \Big|_{\text{rad}} \,.
\end{equation}
Formally, the matter piece of this split is defined as the acceleration one would obtain if we were going to set $\mathcal{R}^i = 0$, \textit{i.e.}, if we were just solving $G^i = 0$. The radiation part is thus a correction to be added thereto. A straightforward calculation yields
\begin{align}\label{eq:accCMrad}
	a_{\text{4.5PN}}^i\bigg|_\text{rad} = \frac{G \Delta}{r^2c^2}\left(2 n^iv^j+n^jv^i\right)\frac{\dd \Gamma^j}{\dd t} = \frac{G \Delta}{r^2c^2}\left(2 n^iv^j+n^jv^i\right)\Bigl[\Pi^j + \calF_{\bm{G}}^j\Bigr]\,.
\end{align}
This piece of the acceleration contains the non-local piece $\Pi^i$, \textit{i.e.}, the integral of the linear momentum flux [recall our definitions~\eqref{eq:non-local}], which is a consequence of the GW recoil of the source. Previous results at 3.5PN order~\cite{IW93, IW95, JaraS97, PW02, KFS03, NB05, itoh3} are unaffected, but more recent results at 4.5PN order~\cite{GII97, LPY23} have neglected this effect and need to be corrected. In particular, Ref.~\cite{GII97} (GII) assumed from the start that the 4.5PN term in the CM acceleration is local in time. Using this hypothesis, they could derive the RR contribution at 4.5PN order using the balance laws for energy and angular momentum (since they work in the CM frame, they do not account for the balance laws for linear momentum and CM position). However, by the previous reasoning, the assumption regarding the local structure of the CM EOM at 4.5PN order ignores the recoil of the source and the displacement of the CM, so the expression for the 4.5PN acceleration in~\cite{GII97} cannot be correct. The correct CM acceleration at 4.5PN order can only be obtained starting from a general frame and then using the full set of flux-balance laws, \textit{i.e.}, not only for energy and angular momentum, but also for linear momentum and CM position. The approach~\cite{GII97} was also adopted in Ref.~\cite{LPY23}, but they defined the CM frame as the solution to $G^i = 0$, whilst we have here shown that it should instead be defined as the solution to $G^i + \Gamma^i = 0$. Thus, Ref.~\cite{LPY23} also neglects the recoil of the source and does not find any non-local contribution, in contrast with our analysis; moreover, they claim agreement with~\cite{GII97} for the limiting case of  circular orbits, but in the case of the full acceleration in the CM frame, they do not give the values of the GII parameters corresponding to their choice of coordinates. We shall show in Sec.~\ref{sec:GIIcorrected} how to correct for the flux-balance approach of Ref.~\cite{GII97} so as to  finally compute the GII parameters corresponding to our coordinate system.

Finally, we report that  the matter piece of the CM acceleration reads
%\David{vérifier que la partie 4.5PN est bien définie selon notre décomposition matière+radiation}\Luc{à corriger pour enlever le terme de flux $F_G$} \Luc{C'est fait}
%\Luc{A remplacer par le version avec moments remplacés}
%
\begin{align}\label{eq:acc45PNCMdev}
	a_{\text{4.5PN}}^{i}\Big|_{\text{mat}} &= \frac{8 G^2 m^2 \nu}{c^9 r^3} \Biggl\{ n^{i} \dot{r} \biggl(\Bigl(- \frac{226063}{3780}
	-  \frac{30973}{315} \nu
	-  \frac{22}{105} \nu^2\Bigr) \frac{G^3 m^3}{r^3}
	+ \Bigl(\frac{229}{2}
	-  \frac{1879}{4} \nu
	+ \frac{1763}{8} \nu^2\Bigr) \dot{r}^6
	\nonumber\\
	& \qquad \qquad + \Bigl(- \frac{811}{4}+ \frac{7401}{8} \nu
	-  \frac{1931}{4} \nu^2\Bigr) \dot{r}^4 v^{2}
	+ \frac{G^2 m^2}{r^2} \biggl [\Bigl(- \frac{267887}{756}
	-  \frac{519199}{945} \nu
	+ \frac{60653}{630} \nu^2\Bigr) \dot{r}^2
	\nonumber \\ & \qquad \qquad \qquad + \Bigl(\frac{319091}{1260}
	+ \frac{18409}{315} \nu
	-  \frac{6773}{630} \nu^2\Bigr) v^{2}\biggl]
	+ \Bigl(\frac{18365}{168}
	-  \frac{22807}{42} \nu
	+ \frac{47095}{168} \nu^2\Bigr) \dot{r}^2 v^{4} \nonumber \\ & \qquad \qquad
	+ \frac{G m}{r} \biggl [\Bigl(- \frac{17099}{126}
	+ \frac{387179}{2520} \nu
	+ \frac{277031}{1260} \nu^2\Bigr) \dot{r}^4 + \Bigl(\frac{911627}{2520}
	-  \frac{1588387}{2520} \nu
	-  \frac{138263}{630} \nu^2\Bigr) \dot{r}^2 v^{2} \nonumber \\ & \qquad \qquad \qquad \quad
	+ \Bigl(- \frac{116827}{840}
	+ \frac{34823}{140} \nu
	+ \frac{2593}{42} \nu^2\Bigr) v^{4}\biggl]
	+ \Bigl(- \frac{1091}{70}
	+ \frac{13757}{168} \nu
	-  \frac{3461}{105} \nu^2\Bigr) v^{6}\biggl)
	\nonumber \\ & \qquad \qquad + v^{i} \biggl(\Bigl(\frac{74867}{1620}
	+ \frac{52019}{630} \nu
	-  \frac{74}{21} \nu^2\Bigr) \frac{G^3 m^3}{r^3}
	+ \Bigl(- \frac{991}{18}
	+ \frac{5815}{36} \nu
	-  \frac{1361}{72} \nu^2\Bigr) \dot{r}^6\nonumber\\
	& \qquad \qquad \qquad + \Bigl(\frac{6109}{84}
	-  \frac{44995}{168} \nu
	+ \frac{6917}{84} \nu^2\Bigr) \dot{r}^4 v^{2}
	+ \frac{G^2 m^2}{r^2} \biggl [\Bigl(\frac{182369}{1260}
	+ \frac{101011}{210} \nu
	-  \frac{57569}{630} \nu^2\Bigr) \dot{r}^2 \nonumber \\ & \qquad \qquad \qquad \qquad
	+ \Bigl(- \frac{23893}{420}
	-  \frac{1549}{30} \nu
	-  \frac{607}{630} \nu^2\Bigr) v^{2}\biggl]
	+ \Bigl(- \frac{21367}{840}
	+ \frac{12919}{105} \nu
	-  \frac{43241}{840} \nu^2\Bigr) \dot{r}^2 v^{4} \nonumber \\ & \qquad \qquad \qquad
	+ \frac{G m}{r} \biggl [\Bigl(- \frac{17512}{315}
	+ \frac{698491}{2520} \nu
	-  \frac{175403}{1260} \nu^2\Bigr) \dot{r}^4
	+ \Bigl(- \frac{116903}{2520}
	-  \frac{33119}{840} \nu
	+ \frac{49703}{630} \nu^2\Bigr) \dot{r}^2 v^{2} \nonumber \\ & \qquad \qquad \qquad \qquad \quad
	+ \Bigl(\frac{46649}{2520}
	-  \frac{271}{14} \nu
	-  \frac{2753}{630} \nu^2\Bigr) v^{4}\biggl]
	+ \Bigl(\frac{173}{126}
	-  \frac{7711}{840} \nu
	+ \frac{293}{105} \nu^2\Bigr) v^{6}\biggl)\Biggl\}\,.
\end{align}
Note that all these expressions are also provided in the Supplemental Material~\cite{SuppMaterial}.
\subsection{Flux-balance laws for energy and angular momentum in the center-of-masss frame}

An important check is that our CM acceleration (derived by CM reduction of the general acceleration) still implies the energy and angular momentum balance equations in the CM frame. For this, we first compute the energy $E = E\cons + E\RR$ and angular momentum $J_i = J\cons^i + J\RR^i$ in the CM frame by applying the formulas~\eqref{eq:passage_CoM} for the passage to the CM to the full expressions of the energy and angular momentum including the RR Schott terms~\eqref{eq:Schott} [as given explicitly in Appendix~\ref{app:Schott}]. The conservative parts of these conserved quantities of course agree with the known conservative results (see \textit{e.g.}~\cite{BBFM17}, which also contains the discussion concerning the fate of dissipative and conservative 4PN tail effects). Moreover, the odd dissipative terms have a direct contribution, that arises by passing $E\RR$ and $J\RR^i$ to the CM frame, as well as indirect contributions arising from applying~\eqref{eq:passage_CoM} to the even terms in $E\cons$ and $J\cons^i$. As previously, we split $E$ and $J^i$ into ``matter'' and ``radiation'' contributions:
\begin{equation}
E = E \Big|_\text{mat} + E \Big|_\text{rad} \  , \qquad\qquad J^i = J^i \Big|_\text{mat} + J^i  \Big|_\text{rad} \,.
\end{equation}
The radiation contributions in the 4.5PN terms of the energy and angular momentum in the CM frame are given by
\bse\label{eq:EJCMrad}\begin{align}
	E_{\text{4.5PN}}\Big|_\text{rad} &= \frac{\nu \Delta}{c^2} v^2 v^i \Bigl[\Pi^i + \calF_{\bm{G}}^i\Bigr] \,,\\
	J_{\text{4.5PN}}^i\Big|_\text{rad} &= \frac{\nu \Delta}{c^2} \varepsilon_{ijk} x^j v^k v^l \Bigl[\Pi^l + \calF_{\bm{G}}^l\Bigr]\,.
\end{align}\ese
The fluxes $\mathcal{F}_{\bm{P}}^i$ and $\mathcal{F}_{\bm{G}}^i$ are needed here at 3.5PN order, and can be found at 4.5PN order in Eqs.~\eqref{eq:FP_FG_COM}. Again, these terms are interesting because they contain non-local contributions starting at 4.5PN order. We then find that the full expressions of the ``matter'' RR contributions to the energy and angular momentum in the CM frame read
%\David{les expressions 4.5PN ne sont pas correctes au regard de notre subdivision matière+rayonnement}\Luc{à vérifier}
%
%
\bse\label{eq:EJCM}
\begin{align}
	E\RR\Big|_\text{mat} &= \frac{8 G^2 m^3 \nu^2 \dot{r}}{c^5 r^2} \Biggl\{ - \dot{r}^2
	+ \frac{6}{5} v^{2}
	+ \frac{1}{c^{2}} \Biggl[\Bigl(\frac{4}{105}
	-  \frac{16}{105} \nu \Bigr) \frac{G^2 m^2}{r^2}
	+ \Bigl(- \frac{1}{2}
	+ \frac{1}{2} \nu \Bigr) \dot{r}^4
	+ \Bigl(-1
	+ 4 \nu \Bigr) \dot{r}^2 v^{2} \nn\\
	& \qquad \qquad \qquad \qquad  \qquad \qquad \quad\ 
	+ \Bigl(\frac{292}{35} + \frac{57}{35} \nu \Bigr) \frac{G m}{r} \dot{r}^2
	+ \Bigl(- \frac{274}{35}
	-  \frac{67}{105} \nu \Bigr) \frac{G m}{r} v^{2}
	+ \Bigl(\frac{99}{70}
	-  \frac{27}{5} \nu \Bigr) v^{4}\Biggr] \nn \\ 
	& \qquad\qquad\qquad
	+ \frac{1}{c^{4}} \Biggl[\Bigl(- \frac{362}{945}
	+ \frac{172}{105} \nu
	-  \frac{80}{189} \nu^2\Bigr) \frac{G^3 m^3}{r^3}
	+ \Bigl(\frac{229}{18} -  \frac{1879}{36} \nu
	+ \frac{1763}{72} \nu^2\Bigr) \dot{r}^6
	 \nn \\ 
	& \qquad\qquad\qquad\quad\ \
	+ \Bigl(- \frac{174}{7}
	+ \frac{2545}{24} \nu
	-  \frac{346}{7} \nu^2\Bigr) \dot{r}^4 v^{2} 
	+  \Bigl(- \frac{9101}{630}
	-  \frac{1831}{60} \nu
	+ \frac{9103}{945} \nu^2\Bigr) \frac{G^2 m^2}{r^2}\dot{r}^2  \nn \\ 
	& \qquad\qquad\qquad\quad\ \
	+ \Bigl(\frac{541}{70}
	+ \frac{977}{42} \nu
	-  \frac{803}{105} \nu^2\Bigr) \frac{G^2 m^2}{r^2} v^{2}	+ \Bigl(\frac{2347}{210}
	-  \frac{13649}{280} \nu
	+ \frac{185}{84} \nu^2\Bigr) \dot{r}^2 v^{4}
	 \nn \\ 
	& \qquad\qquad\qquad\quad\ \
	+\Bigl(- \frac{7856}{315}
	+ \frac{11605}{252} \nu+ \frac{947}{45} \nu^2\Bigr) \frac{G m}{r} \dot{r}^4	
	+ \Bigl(\frac{50263}{945}
	-  \frac{110122}{945} \nu
	-  \frac{18832}{945} \nu^2\Bigr) \frac{G m}{r} \dot{r}^2 v^{2}
	 \nn \\ 
	& \qquad\qquad\qquad\quad\ \
	+ \Bigl(- \frac{2746}{105}
	+ \frac{80723}{1260} \nu
	-  \frac{148}{15} \nu^2\Bigr) \frac{G m}{r} v^{4}
	+ \Bigl(\frac{94}{315}-  \frac{865}{168} \nu
	+ \frac{1663}{60} \nu^2\Bigr) v^{6}\Biggr]\Biggr\} + \calO(11)\,,\\
	%%%%%%%%%%%%%%%%%%%%%%%%%%%%%%%%%%%%%%%%%%%%%%%%%%%%%%%%%%%%%%%%%
	J^{i}\RR\Big|_\text{mat} &= \frac{32}{5} \frac{G^2 m^3 \nu^2 \dot{r}}{c^5 r^2} \varepsilon^{ijk}x^j v^k \Biggl\{ 1
	+ \frac{1}{c^{2}} \Biggl [\Bigl(- \frac{1373}{168}
	-  \frac{29}{14} \nu \Bigr) \frac{G m}{r}
	+ \Bigl(- \frac{51}{112}
	-  \frac{71}{56} \nu \Bigr) \dot{r}^2
	+ \Bigl(\frac{41}{48}
	-  \frac{215}{42} \nu \Bigr) v^{2}\Biggr] \nn\\
	& \qquad\qquad\qquad
	+ \frac{1}{c^{4}} \Biggl[\Bigl(\frac{8963}{1008} + \frac{12259}{336} \nu
	-  \frac{6289}{504} \nu^2\Bigr) \frac{G^2 m^2}{r^2}
	+ \Bigl(- \frac{5465}{504}
	+ \frac{11075}{336} \nu
	-  \frac{4175}{672} \nu^2\Bigr) \dot{r}^4 \nn\\
	& \qquad\qquad\qquad\quad\ \
	+ \Bigl(\frac{16309}{2016}
	-  \frac{11315}{336} \nu
	+ \frac{827}{224} \nu^2\Bigr) \dot{r}^2 v^{2} + \Bigl(\frac{191}{3024}
	+ \frac{5167}{1512} \nu
	-  \frac{36499}{1008} \nu^2\Bigr)  \frac{G m}{r}  \dot{r}^2 \nn \\ 
	&  \qquad\qquad\qquad\quad\ \
	+ \Bigl(- \frac{1994}{63}
	+ \frac{1083}{16} \nu
	-  \frac{121}{24} \nu^2\Bigr)  \frac{G m}{r}  v^{2}
	+ \Bigl(- \frac{1913}{2016}
	-  \frac{155}{504} \nu+ \frac{113}{4} \nu^2\Bigr) v^{4}\Biggr] \Biggr\} + \calO(11)\,.
\end{align}\ese
Note that all these expressions are also provided in the Supplemental Material~\cite{SuppMaterial}.

We now take the time derivative of the total energy and angular momentum in the CM frame, using the relative acceleration in the CM frame given by~\eqref{eq:aiCM_moments}--\eqref{eq:accCMPN}. The non-local contributions from the energy, angular momentum and EOM cancel out perfectly, and we exactly recover the corresponding instantaneous fluxes in the CM frame. Therefore, we have constructed an acceleration at 4.5PN order, in the CM frame, which satisfies the energy and angular momentum flux-balance laws, although it does not fit into the general framework of~\cite{GII97}. 
%As discussed previously, this is because~\cite{GII97} imposed the balance laws already at the level of the CM equations, assuming that the acceleration, energy and angular momentum are local functionals of the CM variables. 
In the next section, we provide the modification of this framework that correctly accounts for the recoil-induced non-local terms.

\section{The flux-balance approach to radiation reaction}
\label{sec:GIIcorrected}

The framework of Ref.~\cite{GII97} (referred to as ``GII''), building on previous work by Iyer and Will~\cite{IW93,IW95}, applies the flux-balance method, restricted to the frame of the CM, to determine the dissipative RR contributions in the EOM of compact binaries. The end result of GII is the CM relative acceleration at orders 2.5PN, 3.5PN and 4.5PN which is of the form
\begin{align}\label{eq:accGII}
	a\RR^{i\,\text{GII}} = - \frac{8}{5}\frac{G^2m^2\nu}{c^3 r^3}\Bigl[-\bigl(A_\text{2.5PN}+A_\text{3.5PN}+A_\text{4.5PN}\bigr) \dot{r} n^i + \left(B_\text{2.5PN}+B_\text{3.5PN}+B_\text{4.5PN}\right) v^i\Bigr] + \calO(11)\,,
\end{align}
where the coefficients $A_\text{$n$PN}$ and $B_\text{$n$PN}$ are given by Eqs.~(2.8), (2.9), (2.11) and (2.16) in~\cite{GII97}. Remembering that RR is intrinsically gauge-dependent, the coefficients depend on a number of arbitrary gauge parameters reflecting the arbitrariness in the choice of a coordinate system, and denoted
\begin{align}\label{eq:p}
	\bigl\{\alpha_3, \beta_2, \xi_1, \xi_2, \xi_3, \xi_4, \xi_5, \rho_5, \psi_1, \psi_2, \psi_3, \psi_4, \psi_5, \psi_6, \psi_7, \psi_8, \psi_9, \chi_6, \chi_8, \chi_9\bigr\}\,.
\end{align}
At 2.5PN order, the two parameters $\alpha_3$ and $\beta_2$ are sufficient, while at 3.5PN order, one must add six more parameters $\xi_{i=1,\cdots,5}$ and $\rho_5$. At  4.5PN order, there are twelve additional parameters $\psi_{i=1,\cdots,9}$ and $\chi_{i=6,8,9}$, thus totalizing twenty gauge parameters for the most general coordinate transformation up to 4.5PN order~\cite{GII97}.

With the RR acceleration~\eqref{eq:accGII}, the flux-balance laws for energy and angular momentum restricted to the frame of the CM are satisfied [for any set of the gauge parameters~\eqref{eq:p}], namely 
\begin{align}\label{eq:fluxbalanceCM}
	\frac{\dd E_\text{GII}}{\dd t} = - \calF_{E} \,,\qquad \frac{\dd J^i_\text{GII}}{\dd t} = - \calF^i_{\bm{J}} \,,
\end{align}
where $\calF_{E}$ and $\calF^i_{\bm{J}}$ are the gauge invariant (relative 2PN-accurate) energy and angular momentum fluxes in the CM frame, and the time derivatives are performed with consistent order reduction of the accelerations using Eq.~\eqref{eq:accGII}. The energy and angular momentum in the left-hand sides take the form
\bse\label{eq:SchottCMstruct}\begin{align}
	E_\text{GII} &= E\cons + E\RR^\text{GII} \,, \\
	J^i_\text{GII} &= J\cons^i + J\RR^{i\,\text{GII}} \,, 
\end{align}\ese
where $E\cons$ and $J\cons^i$ denote the standard conservative pieces at the 2PN order (say, in harmonic coordinates), which are conserved in absence of RR, and augmented by the RR contributions in the form of some 2.5PN, 3.5PN and 4.5PN Schott-like correction terms,
\bse\label{eq:SchottCM}\begin{align}
	E\RR^\text{GII} &= (E_{\text{2.5PN}} + E_{\text{3.5PN}} + E_{\text{4.5PN}})_\text{GII} + \calO(11) \,, \\
	J\RR^{i\,\text{GII}} &= (J_{\text{2.5PN}}^i + J_{\text{3.5PN}}^i + J_{\text{4.5PN}}^i)_\text{GII} + \calO(11)\,.
\end{align}\ese
In contrast to the right-hand sides of~\eqref{eq:fluxbalanceCM}, the Schott terms are gauge-dependent and parametrized up to 4.5PN order by the gauge parameters~\eqref{eq:p}; see~(2.12) and (2.17) in~\cite{GII97}.

The computations of GII are perfectly consistent, but they are based on a fundamental ansatz that the 4.5PN CM acceleration~\eqref{eq:accGII} is local-in-time, \textit{i.e.}, all the variables $r$, $\dot{r}$, $n^i$, $v^i$ in~\eqref{eq:accGII} depend only on the current time $t$, without hereditary or semi-hereditary dependence on earlier times $t'<t$. Yet, we have proved in Sec.~\ref{sec:CoM} that this cannot be correct, because the definition of the CM frame must take into account not only the matter contribution but also the radiation contribution, and this implies the presence at the 4.5PN order of the non-local integral of the flux of linear momentum, which contains the important physical effect of gravitational recoil. Let us show how to correct the end result of GII in order to consistently include the radiation contribution and the non-local effect. We shall then be able to prove that our chosen extended BT coordinate system for RR indeed corresponds to a unique set of the GII gauge parameters~\eqref{eq:p}.

First of all, we establish the necessary and sufficient conditions under which the flux-balance  laws~\eqref{eq:fluxbalanceCM} are preserved despite a modification of the CM acceleration~\eqref{eq:accGII}, 
\bse\label{eq:corrections}
\begin{align}\label{seq:corrections-acc}
\widetilde{a}\RR^i &= a\RR^{i\,\text{GII}} + \delta a\RR^i \,,
\end{align}
together with the corresponding modifications of the RR Schott terms~\eqref{eq:SchottCM}, 
\begin{align}
	%\widetilde{a}\RR^i &= a\RR^{i\,\text{GII}} + \delta a\RR^i \,,\\	
	\widetilde{E}\RR &= E\RR^\text{GII} + \delta E\RR\,, \\	
	\widetilde{J}\RR^i &= J\RR^{i\,\text{GII}} + \delta J\RR^i\,.
\end{align}\ese
 If the modifications $\delta a\RR^i$, $\delta E\RR$ and $\delta J\RR^i$ are local, the answer is easy because they should simply correspond to a redefinition of the gauge parameters~\eqref{eq:p}. However, we want to consider the case where these modifications involve a non-local part. We readily obtain, in the case where $\delta a\RR^i$, $\delta E\RR$ and $\delta J\RR^i$ are at the highest 4.5PN order, the two equations
\bse\label{eq:eqscorr}\begin{align}
	& \frac{\dd \delta E\RR}{\dd t} + m \nu v^i \delta a\RR^i = 0 \,,\\
	& \frac{\dd \delta J\RR^i}{\dd t} + m \nu \varepsilon_{ijk} x^j \delta a\RR^k = 0 \,.
\end{align}\ese
From Sec.~\ref{sec:CoM}, we know that in order to be physically acceptable, the correction terms $\delta a\RR^i$, $\delta E\RR$ and $\delta J\RR^i$ must include the non-local radiation contributions. Furthermore, we are allowed to add some purely local contribution to the CM acceleration, which is to be determined by imposing that the Eqs.~\eqref{eq:eqscorr} are satisfied. Using Eqs.~\eqref{eq:accCMrad} and~\eqref{eq:EJCMrad}, we thus pose
\bse\label{eq:corrections2}\begin{align}
	\delta a\RR^i &= \frac{G \Delta}{r^2c^2}\left(2 n^iv^j+n^jv^i\right)\Bigl[\Pi^j + \calF_{\bm{G}}^j\Bigr] + \delta^\text{loc} \!a\RR^i\,,\\
	\delta E\RR &= \frac{\nu \Delta}{c^2} v^2 v^i \Bigl[\Pi^i + \calF_{\bm{G}}^i\Bigr]\,, \\	
	\delta J\RR^i &= \frac{\nu \Delta}{c^2} \varepsilon_{ijk} x^j v^k v^l \Bigl[\Pi^l + \calF_{\bm{G}}^l\Bigr]\,, 
\end{align}\ese
where $\delta^\text{loc} \!a\RR^i$ is a local modification of the acceleration that is  yet to be determined, and we have used the fact that, up to a redefinition of the parameters~\eqref{eq:p}, we can always set potential local terms $\delta^\text{loc} E\RR$ and $\delta^\text{loc} J\RR^i$  to zero. Plugging Eqs.~\eqref{eq:corrections2} into~\eqref{eq:eqscorr}, we find that the supplementary local acceleration $\delta^\text{loc} a\RR^i$ must satisfy the two equations (recall $\dot{\Pi}^i=\calF_{\bm{P}}^i$),
\bse\label{eq:eqsdeltaa}\begin{align}
	v^i\delta^\text{loc} \!a\RR^i &= - \frac{\Delta}{m c^2} v^2 v^i \Bigl[\calF_{\bm{P}}^i + \dot{\calF}_{\bm{G}}^i\Bigr] \,,\\
	\varepsilon_{ijk} x^j \delta^\text{loc} \!a\RR^k &= - \frac{\Delta}{m c^2} \varepsilon_{ijk} x^j v^k v^l \Bigl[\calF_{\bm{P}}^l + \dot{\calF}_{\bm{G}}^l\Bigr] \,.
\end{align}\ese
The non-local terms have cancelled out, consistently with the fact that GII could obtain the balance laws~\eqref{eq:fluxbalanceCM} with purely local acceleration and Schott terms. The unique solution of the two equations~\eqref{eq:eqsdeltaa} is
\begin{align}\label{eq:soldeltaa}
	\delta^\text{loc} \!a\RR^i &= - \frac{\Delta}{m c^2} v^i v^j \Bigl[\calF_{\bm{P}}^j + \dot{\calF}_{\bm{G}}^j\Bigr] \,.
\end{align}
Finally, we have obtained a new CM acceleration with related quantities at 4.5PN order, given by Eqs.~\eqref{eq:corrections} and~\eqref{eq:corrections2}, together with the explicit determination of $\delta a\RR^i$ in~\eqref{eq:soldeltaa}, such that it (i) satisfies the flux balance laws for energy and angular momentum; (ii) includes the radiation contributions (with the non-local terms therein) in the definition of the CM; and (iii) is as general as the GII acceleration as it depends on the complete set of gauge parameters~\eqref{eq:p}.

The local term~\eqref{eq:soldeltaa} plays a crucial role in ensuring the correctness of the result. Using the expressions of the linear momentum and CM fluxes given by~\eqref{eq:FP_FG_COM} (and restricted to the dominant 3.5PN order) we obtain 
%\Luc{à vérifier}\David{je confirme}
%
\begin{align}\label{eq:deltaalocal}
	\delta^\text{loc} \!a\RR^i &\!=\! -\frac{8}{105}\frac{G^2 m^2 \Delta^2 \nu^2}{c^9 r^3} \!\Bigg[ \!\frac{2G^2 m^2}{r^2}\bigl(23\dot{r}^2-9v^2\bigr) +\! \frac{G m}{r}\bigl(-240\dot{r}^4+391\dot{r}^2v^2-141v^4\bigr) + 3v^2\bigl(45\dot{r}^4-60\dot{r}^2v^2+11v^4\bigr)\!\Bigg] v^i \,.
\end{align}
We notice that $\delta a\RR^i$ is proportional to the velocity $v^i$. Therefore, it is to be combined with the coefficient $B_\text{4.5PN}$ in Eq.~\eqref{eq:accGII}, where it causes some extra contributions to all the coefficients $k_i$ in Eq.~(2.11b) of~\cite{GII97} (except $k_7$ and $k_{10}$, which do not receive extra contributions). 

Finally, by comparing our CM acceleration obtained in Eqs.~\eqref{eq:aiCM_moments}--\eqref{eq:accCMPN} to the modified general acceleration $\widetilde{a}\RR^i$ in an arbitrary coordinate system, as defined in Eqs.~\eqref{eq:corrections}--\eqref{eq:deltaalocal}, we obtain a unique set of gauge parameters~\eqref{eq:p} which correspond to our chosen extended BT coordinates, namely
\begin{equation}\label{eq:paramGII-BT}
	\begin{array}{ll} \displaystyle \alpha_3 = 5\,,& \displaystyle \qquad\qquad \beta_2 = 4\,,\\[0.3cm]
		%%%%%%%%%%%%%%%%%%%%%%%%%%%%%%%%%%%%%%%%%%%%%%%%%%%%%%%%%%%%%%%%%%%%%%%%%%%
		\displaystyle \xi_1 = -\frac{99}{14} + 27\nu\,,& \displaystyle \qquad\qquad \xi_2 = 5 - 20\nu\,,\\[0.4cm]
		%%%%%%%%%%%%%%%%%%%%%%%%%%%%%%%%%%%%%%%%%%%%%%%%%%%%%%%%%%%%%%%%%%%%%%%%%%%
		\displaystyle \xi_3 = \frac{274}{7} + \frac{67}{21}\nu\,,& \displaystyle \qquad\qquad \xi_4 = \frac{5}{2} - \frac{5}{2}\nu\,,\\[0.4cm]
		%%%%%%%%%%%%%%%%%%%%%%%%%%%%%%%%%%%%%%%%%%%%%%%%%%%%%%%%%%%%%%%%%%%%%%%%%%%
		\displaystyle \xi_5 = - \frac{292}{7} - \frac{57}{7}\nu\,,& \displaystyle \qquad\qquad \rho_5 = \frac{51}{28} + \frac{71}{14}\nu\,,\\[0.4cm]
		%%%%%%%%%%%%%%%%%%%%%%%%%%%%%%%%%%%%%%%%%%%%%%%%%%%%%%%%%%%%%%%%%%%%%%%%%%%
		\displaystyle \psi_1 = - \frac{94}{63} + \frac{4325}{168}\nu - \frac{1663}{12}\nu^2\,,& \displaystyle \qquad\qquad \psi_2 = - \frac{2347}{42} + \frac{13649}{56}\nu - \frac{925}{84}\nu^2\,,\\[0.4cm]
		%%%%%%%%%%%%%%%%%%%%%%%%%%%%%%%%%%%%%%%%%%%%%%%%%%%%%%%%%%%%%%%%%%%%%%%%%%%
		\displaystyle \psi_3 = \frac{2746}{21} - \frac{80723}{252}\nu + \frac{148}{3}\nu^2\,,& \displaystyle \qquad\qquad \psi_4 = \frac{870}{7} - \frac{12725}{24}\nu + \frac{1730}{7}\nu^2\,,\\[0.4cm]
		%%%%%%%%%%%%%%%%%%%%%%%%%%%%%%%%%%%%%%%%%%%%%%%%%%%%%%%%%%%%%%%%%%%%%%%%%%%
		\displaystyle \psi_5 = - \frac{541}{14} - \frac{4885}{42}\nu + \frac{803}{21}\nu^2\,,& \displaystyle \qquad\qquad \psi_6 = - \frac{50263}{189} + \frac{110122}{189}\nu + \frac{18832}{189}\nu^2\,,\\[0.4cm]
		%%%%%%%%%%%%%%%%%%%%%%%%%%%%%%%%%%%%%%%%%%%%%%%%%%%%%%%%%%%%%%%%%%%%%%%%%%%
		\displaystyle \psi_7 = - \frac{1145}{18} + \frac{9395}{36}\nu - \frac{8815}{72}\nu^2\,,& \displaystyle \qquad\qquad \psi_8 = \frac{7856}{63} - \frac{58025}{252}\nu - \frac{947}{9}\nu^2\,,\\[0.4cm]
		%%%%%%%%%%%%%%%%%%%%%%%%%%%%%%%%%%%%%%%%%%%%%%%%%%%%%%%%%%%%%%%%%%%%%%%%%%%
		\displaystyle \psi_9 = \frac{9101}{126} + \frac{1831}{12}\nu - \frac{9103}{189}\nu^2\,,& \displaystyle \qquad\qquad \chi_6 = - \frac{16309}{504} + \frac{11315}{84}\nu - \frac{827}{56}\nu^2\,,\\[0.4cm]
		%%%%%%%%%%%%%%%%%%%%%%%%%%%%%%%%%%%%%%%%%%%%%%%%%%%%%%%%%%%%%%%%%%%%%%%%%%%
		\displaystyle \chi_8 = \frac{5465}{126} - \frac{11075}{84}\nu + \frac{4175}{168}\nu^2\,,& \displaystyle \qquad\qquad \chi_9 = - \frac{191}{756} - \frac{5167}{378}\nu + \frac{36499}{252}\nu^2\,.
	\end{array}
\end{equation}
At the 3.5PN level, we recover the parameters corresponding to the extended BT coordinates computed in~\cite{IW95, B97}. Our unique determination of the parameters at the 4.5PN level constitues a non-trivial check on the physical soundness of the RR potentials~\eqref{eq:defRRpot}--\eqref{eq:PNVRR} and the perfect consistency with the flux balance method~\cite{IW93, IW95, GII97}. Finally, note that we provide the corrected parametrized acceleration \eqref{seq:corrections-acc} in the Supplemental Material~\cite{SuppMaterial}, as well as the values of the parameters \eqref{eq:paramGII-BT} corresponding to our choice of coordinate system.

\section{Circular orbits}
\label{sec:circular}

The expression we have found for the 4.5PN acceleration in the CM frame features a novel non-local term $\Pi^i$, which introduces a practical difficulty because it \textit{a priori} depends on the whole past history of the binary. In this section, we show that specifying the orbit allows us to ``localize'' these terms, in a way similar to other non-local effects like tails or memory. We consider the case of circular orbits and focus on the dominant 3.5PN contribution to the fluxes. For circular orbits, the fluxes of linear momentum and CM position read (see Sec.~VI of~\cite{BF19})
\bse\label{eq:fluxcirc}\begin{align}
	\calF_{\bm{P}}^i &= -
	\frac{464}{105}\,\frac{G^4m^5}{c^7r^5}\,\Delta\,\nu^2\,
	v^i + \calO(9)\,,\\
	\calF_{\bm{G}}^i &= - \frac{544}{105}\,\frac{G^4m^5}{c^7r^4}\,\Delta\,\nu^2\,n^i + \calO(9)\,,
\end{align}\ese
where the orbital separation is denoted by $r$ and the unit vector along the binary's separation is $n^i$. When performing the integrations in Eqs.~\eqref{eq:non-local} for $\Pi^i$ and $\Gamma^i$, we find that there are no DC terms --- only oscillatory AC terms. Since the radial velocity $\dot{r}=O(5)$ is negligible at 2PN, we can thus perform the integration by
assuming that the orbit was perfectly circular in the past~\cite{ABIQ04}. This entails integrating using $\dd n^i/\dd t = v^i/r$ and $\dd v^i / \dd t = - \omega^2 x^i$ in the circular case [with orbital frequency $\omega=\sqrt{G m/r^3} + \calO(2)$], all other quantities staying constant. We find\footnote{These results provides reassurance about the following concern. The 4.5PN piece of the energy flux was obtained in the limiting case of circular orbits in Eq.~(5.11) of~\cite{MBF16}, and is due to the hereditary tail-of-tail-of-tails propagating in the wave zone. The argument that was used to compute it was that any \textit{local} term vanishes in the flux at 4.5PN order for circular orbits, and only the non-local terms survive. But this argument does not \textit{a priori} rule out a non-vanishing contribution from the non-local terms in the formulas~\eqref{eq:passage_CoM}--\eqref{eq:Ri}. 
%The latter results permit to solve a concern, whether the 4.5PN piece of the energy flux, Eq.~(5.11) of~\cite{MBF16}, which is due to tail-of-tail-of-tails propagating in the wave zone and was derived in the limiting case of circular orbits, could be affected by the non-local passage to the CM at 4.5PN order. Indeed the argument used in~\cite{MBF16}, namely that any \textit{local} term gives zero in the flux at 4.5PN order for circular orbits, does not \textit{a priori} apply to the non-local terms in the formulas~\eqref{eq:passage_CoM}--\eqref{eq:Ri}. 
Indeed, when inserting~\eqref{eq:passage_CoM}--\eqref{eq:Ri} into the energy flux at 1PN order in a general frame, we do find some contributions at the 4.5PN level containing the non-local term $\calR^i$. Fortunately, now that  $\calR^i=-\frac{1}{m}\Gamma^i+\calO(9)$ is explicitly given by~\eqref{eq:Gamcirc} in the case of circular orbits, we readily check that the  argument of Ref.~\cite{MBF16} still applies to this case and that these terms vanish for circular orbits in the 4.5PN flux.
}
%(discarding the bound at $t'' = - \infty$), namely that $\lambda^i(t') = \sin\Big(\omega(t-t')\Big) n^i(t) + \cos\Big(\omega(t-t')\Big) \lambda^i(t) $, where $\omega$ is the orbital frequency. We find that \David{give more details ore references, e.g. the memory paper by Arun and Blanchet } 
%\David{la contribution non-locale \emph{ne disparait pas} dans le cas circulaire, mais en incluant sa contribution, je suis quand même en accord avec Gopu dans ce cas. Pourquoi ?}
%
\bse\label{eq:PiGamcirc}\begin{align}
	\Pi^i &= - \frac{464}{105}\,\frac{G^4m^5}{c^7r^4}\,\Delta\,\nu^2\,n^i + \calO(9)\,,\\*
	\Gamma^i &= 
	\frac{48}{5}\,\frac{G^3 m^4}{c^7r^2}\,\Delta\,\nu^2\,v^i + \calO(9)
	\,.\label{eq:Gamcirc}
\end{align}\ese
The main difficulty having been treated with~\eqref{eq:fluxcirc}--\eqref{eq:PiGamcirc}, we readily obtain the radiation contribution~\eqref{eq:accCMrad} to the acceleration in the circular case as
\begin{align}\label{eq:accCMradcirc}
	a_{\text{4.5PN}}^i\bigg|_\text{rad} = - \frac{32 G^3 m^3 \nu}{5 c^5 r^4}v^i\bigg[  \gamma^2 \left(\frac{3}{2}\nu - 6 \nu^2\right) + \calO(\gamma^{3}) \bigg]\,.
\end{align}
Then, we can take the full expression~\eqref{eq:acc45PNCMdev} of the acceleration, and reduce it for circular orbits using $v^2 = r^2\omega^2 + \dot{r}^2$ and $\dot{r} = \calO(5)$; this is the direct contribution, and we obtain, after including also the radiation contribution~\eqref{eq:accCMradcirc},
\begin{equation}\label{eq:circdirect}
	a_{\text{circ}}^i \Big|_\text{direct}^\text{RR} = - \frac{32 G^3 m^3 \nu}{5 c^5 r^4}v^i\Bigg[ 1 + \gamma\left(- \frac{3431}{336} + \frac{5}{4} \nu \right) + \gamma^2 \left(\frac{794369}{18144} + \frac{26095}{2016}\nu - \frac{7}{4} \nu^2\right) + \calO(\gamma^{3})   \Bigg]\,.
\end{equation}
This result is in perfect agreement with Eq.~(5.2) of~\cite{GII97}, which points out that it is actually ``gauge invariant'', since any gauge ambiguity is proportional to $\dot{r}$ and plays no role in this case. We notice that the agreement with~\cite{GII97} occurs because the local correction term~\eqref{eq:soldeltaa} happens to exactly cancel the non-local radiation term~\eqref{eq:accCMradcirc} in the circular orbit case. This can immedialely be verified with Eq.~\eqref{eq:deltaalocal}.

On top of the direct expression, we need to account for an indirect expression, arising from replacing $\dot{r}$ in the conservative piece of the acceleration by its expression given by Eq.~(372a) of~\cite{BlanchetLR_2024}. It reads
\begin{equation}\label{eq:circindirect}
	a_{\text{circ}}^i \Big|_\text{indirect}^\text{RR} = - \frac{32 G^3 m^3 \nu}{5 c^5 r^4}v^i\Bigg[  \gamma\left(8-4 \nu \right) + \gamma^2 \left(-\frac{1919}{42} - \frac{1609}{84}\nu + 3 \nu^2\right)+ \calO(\gamma^{3})  \Bigg] \,.
\end{equation}
Finally, the total RR acceleration for circular orbits is the sum of~\eqref{eq:circdirect} and~\eqref{eq:circindirect}. Adding to it the hitherto neglected dissipative 4PN tail term [which originates from Eq.~\eqref{eq:Ftail_E}], we find that the full acceleration at 4.5PN order for circular orbits reads 
%\David{vérifier avec le flux 2PN} \Luc{à vérifier}
%
\begin{equation}
	a_{\text{circ}}^i = -\omega^2 x^i - \frac{32 G^3 m^3 \nu}{5 c^5 r^4} v^i\Bigg[ 1 + \gamma\left(- \frac{743}{336} - \frac{11}{4} \nu \right) + 4\pi\gamma^{3/2} + \gamma^2 \left(-\frac{34639}{18144} - \frac{12521}{2016}\nu + \frac{5}{4} \nu^2\right) + \calO(\gamma^{5/2}) \Bigg]\,,
\end{equation}
where the orbital frequency $\omega$ encodes the conservative PN terms (and is given at 4PN order in~\cite{BBFM17}), and where the dissipative RR terms to 4.5PN order correspond to the extended BT coordinate system. This expression is in agreement with (371) of~\cite{BlanchetLR_2024} up to 4PN relative order, and the 4.5PN term is new.

\section{Conclusion}\label{sec:conclusion}

%\David{Mettre dans les fichiers ancillaires les formules non-développées canoniques (linéaires ds les moments non développés) et les formules toutes développées} 

In this paper, we have computed the RR force on the orbit of non-spinning compact binary systems  up to 2PN relative order, \textit{i.e.}, including the leading 2.5PN effect followed by the next-to-leading 3.5PN and next-to-next-to-leading 4.5PN corrections. Previous calculations at 1PN order were performed in~\cite{JaraS97, PW02, KFS03, NB05, itoh3}. Our calculation is valid for general binary orbits and in a general frame. The coordinate system is an extension to 2PN order of the Burke-Thorne~\cite{BuTh70,Bu71} coordinate system, and greatly simplifies the calculations. In this coordinate system, the RR is described by specific scalar, vector, and tensor potentials parametrized by multipole moments and defined in Sec.~\ref{sec:RR}. The end result for the 4.5PN RR terms is given by Eqs.~\eqref{eq:acc45PN1}. The EOM of compact binary systems are now completely known up to 4.5PN order, since the conservative part up to 4PN order was derived previously, in both ADM~\cite{DJS14,JaraS15} and harmonic~\cite{BBBFMc,BBFM17,FS19, BlumMMS20a} coordinates.

With the expression of the RR force in a general frame, we prove (in Sec.~\ref{sec:flux_balance}) the flux balance laws for the energy, the angular and linear momenta,  and the CM position to 2PN (next-to-next-to-leading) order. Besides the extended BT coordinate system, a simplifying feature of our derivation is that we keep as much as possible the multipole moments of the compact binary source in ``unexpanded'' form, \textit{i.e.}, without replacing the moments (and their time derivatives) by explicit expressions in terms of the particles' positions and velocities. Such a strategy saves a lot of calculations and is crucial in our derivation of the flux balance laws. The RR Schott-like terms~\cite{Schott} that we find in the left-hand sides of the balance laws, which depend on the chosen coordinate system and whose existence constitutes our proof of these laws, are very lengthy and could be presented in Appendix~\ref{app:Schott} only up to 3.5PN order; at the full 4.5PN order, we provide them in the Supplemental Material~\cite{SuppMaterial}.

Next, we have addressed the problem of the definition of the frame of the CM and the reduction of the EOM to this frame. In GR, for an isolated gravitating system, the position of the CM (and likewise the other Poincar\'e invariants) should be defined as the sum of the contribution due to the matter and the one due to the radiation. At 3.5PN order, the total linear momentum is thus given by the one of the matter system plus the correction from the radiation, the latter being the time integral of the linear momentum flux, \textit{i.e.}, the GW recoil. Therefore, at the 3.5PN order, the definition of the CM frame fundamentally involves a non-local-in-time contribution. In turn, such a  contribution implies that the EOM at the 4.5PN order, when reduced to the CM frame, involve a non-local term linked with the GW recoil, as detailed in Sec.~\ref{sec:CoM}.

An important approach to the problem of RR for compact binaries is the flux-balance method pioneered in~\cite{IW93, IW95}. The ansatz used in this method is that the RR acceleration is local in the frame of the CM, and the RR is then determined by imposing the flux-balance laws for energy and angular momentum. The result is not unique but depends on a set of arbitrary gauge parameters reflecting precisely the arbitrariness in the choice of a coordinate system. The point we make in Sec.~\ref{sec:GIIcorrected} is that, when the flux-balance method is extended to 2PN relative order~\cite{GII97}, the ansatz on the locality of the RR acceleration breaks down due to the non-local contributions of the source's recoil and CM displacement by GWs. Nevertheless, we find that a careful modification of the ansatz (adding the contribution of radiation to the CM definition) allows us to  reconcile the flux-balance method with the present ``first-principle'' calculation, and to uniquely determine the gauge parameters corresponding to our extended BT coordinate system. 

\acknowledgments

We thank Laura Bernard for an early investigation on this topic (together with L.B.) and Bala Iyer for a discussion. Most computations as well as the formatting of lengthy equations were performed using the \textit{xAct} library~\cite{xtensor} for \emph{Mathematica}. This work has also made use of the Infinity Cluster hosted by Institut d'Astrophysique de Paris. D.T.~thanks the Institut d'Astrophysique de Paris for its hospitality, and received support from the Czech Academy of Sciences under the Grant No. LQ100102101.

\appendix

\section{Controlling the $v^{\mu\nu}$ term in the harmonicity algorithm}

\label{app:vmunu}
%\section{Contributions from the harmonic}
%\section{Contributions from the harmonicity algorithmComputation of $\overline{v}^{\ab}$}

\subsection{Different forms of the harmonicity algorithm}
%\subsection{Standard construction in harmonic coordinates}
%\subsection{Construction of $\overline{v}\ab$ with the harmonicity algorithm}

In Sec.~\ref{sec:NZ}, we have introduced the object $\overline{v}\ab$ from the so-called ``harmonicity'' algorithm, say $\overline{v}\ab=\mathcal{H}\ab[\overline{w}]$, acting on the components of the divergence $\overline{w}^\alpha\equiv\partial_\beta\overline{u}\ab$ of the object, and which is given by~\eqref{eq:div1}, but also admits the form~\eqref{eq:harmcond}. In this Appendix, we give more details on the definition of this object, and also  present  new alternative expressions, which differ from the former only by a linear gauge transformation.

We decompose the components of the multipole moments $\calB^\mu_L$, see Eqs.~\eqref{eq:BL}--\eqref{eq:harmcond}, into irreducible moments\footnote{Not to be confused with $\mathcal{P}$ and $\mathcal{Q}$ defined in Eqs.~\eqref{eq:PQRR}.} as
\bse\label{eq:irred1}\begin{align}
	\calB^{0}_{L} &= c \,\mathcal{P}_L\,,\\*
	\calB^{i}_{L} &= \calQ^{(+)}_{iL} + \varepsilon_{ai\langle i_\ell} \calQ^{(0)}_{L-1\rangle a} + \delta_{i\langle i_\ell} \calQ^{(-)}_{L-1\rangle} \,,
\end{align}\ese
where the irreducible STF tensors are defined by the inverse formulas as
\bse\label{eq:irred2}\begin{align}
	\mathcal{P}_L &\equiv \frac{1}{c}\,\calB_L^0\,,\\
	\calQ^{(+)}_{L+1} &\equiv \calB^{\langle i_{\ell +1}}_{L\rangle}\,,\\
	\calQ^{(0)}_{L} &\equiv \frac{\ell}{\ell+1}\, \calB^a_{b\langle L-1}\varepsilon_{i_\ell\rangle ab}\,,\\
	\calQ^{(-)}_{L-1} &\equiv \frac{2\ell-1}{2\ell+1}\, \calB^a_{aL-1}\,.
\end{align}\ese
The moments $\calP_L$, $\calQ^{(+)}_L$, $\calQ^{(0)}_L$, and $\calQ^{(-)}_L$ (in which the tail piece is discarded, see below) coincide with the corresponding moments
%$P_L$,  ${}^{(+)}\!Q_L$, ${}^{(0)}\!Q_L$, and ${}^{(-)}\!Q_L$, respectively, 
in~\cite{B98mult}, except for a rescaling of $\calP_L$ by a factor $1/c$ aimed at making this quantity ``Newtonian'' (in the sense that its expression in terms of the sources has a non-zero finite limit when $c\to +\infty$). In terms of  this irreducible decomposition, the components of the divergence $\overline{w}^\alpha$ become
\bse\label{eq:wexpr}
\begin{align}
	\mathop{\overline{w}}_{}\!{}^{0} &= \frac{4G}{c^3} \sum_{\ell \geqslant 0} \frac{(-)^\ell}{\ell!} \partial_{L} \bigl\{\calP_{L}\bigr\} \,,\\*
%%%%%%%%%%%%%%%%%%%%%%%%%%%%%%%%%%%%%%%%%%%%%%%%%%%%%%%%%%%%%%%%%%%%%%%%%
\mathop{\overline{w}}_{}\!{}^{i} &= \frac{4G}{c^4} \biggl\{ \sum_{\ell \geqslant 0} \frac{(-)^\ell}{\ell!} \partial_{L} \bigl\{\calQ^{(+)}_{iL}\bigr\} + \sum_{\ell \geqslant 1}\frac{(-)^\ell}{\ell!} \left(\hat{\partial}_{iL-1} \bigl\{\calQ^{(-)}_{L-1} \bigr\} + \varepsilon_{iab} \partial_{aL-1}\bigl\{\calQ^{(0)}_{bL-1}\bigr\} \right)\biggr\} \,. 
\end{align}\ese
Then, the precise definition of $\mathop{\overline{v}}_{}\!{}\ab$ following the ``standard'' harmonicity algorithm of Refs.~\cite{BD86, B98quad} is
\bse\label{eq:vab}
\begin{align}
	\mathop{\overline{v}}_{}\!{}^{00} &= \frac{4G}{c^2}\biggl\{-\bigl\{\integre\calP \bigr\} +\partial_a \left( \bigl\{\integre \calP_a\bigr\}
	+ \bigl\{ \integre\!\!\integre \calQ^{(+)}_a \bigr\}- \frac{5}{3 \, c^2} \bigl\{\calQ^{(-)}_a\bigr\}
	\right) \biggr\}\,,\\ 
	%%%%%%%%%%%%%%%%%%%%%%%%%%%%%%%%%%%%%%%%%%%%%%%%%%%%%%%%%%%%%%%%%%%%%%%
	\mathop{\overline{v}}_{}\!{}^{0i} &= \frac{4G}{c^3}
	\biggl\{ - \bigl\{\integre\calQ^{(+)}_i \bigr\} + \frac{5}{3 \, c^2} \dot{\calQ}^{(-)}_i + \partial_a \Bigl( \varepsilon_{iab}  \bigl\{\integre \calQ^{(0)}_b\bigr\} \Bigr) - \sum_{\ell \geqslant 2} \frac{(-)^\ell}{\ell!} \partial_{L-1}
	\bigl\{\calP_{iL-1}\bigr\} \biggr\} \,,\\
	%%%%%%%%%%%%%%%%%%%%%%%%%%%%%%%%%%%%%%%%%%%%%%%%%%%%%%%%%%%%%%%%%%%%%%%
	\mathop{\overline{v}}_{}\!{}^{ij} &= \frac{4G}{c^4} \biggl\{ \delta_{ij} \bigl\{\calQ^{(-)}\bigr\} 
	+ \sum_{\ell \geqslant 2} \frac{(-)^\ell}{\ell!} \biggl(2\delta_{ij}\partial_{L-1} \bigl\{\calQ^{(-)}_{L-1}\bigr\} -6 \partial_{L-2(i} \bigl\{\calQ^{(-)}_{j)L-2} \bigr\} - 2 \partial_{aL-2} \left(\varepsilon_{ab(i} \calQ^{(0)}_{j)bL-2} \right)
	\nn\\ &\qquad\qquad    + \partial_{L-2} \left[ \bigl\{\dot\calP_{ijL-2} \bigr\}
	- \frac{(7\ell+3)}{(\ell+1) (2\ell+1) \, c^2} \bigl\{ \ddot{\calQ}^{(-)}_{ijL-2} \bigr\} + \ell \bigl\{ \calQ^{(+)}_{ijL-2} \bigr\} \right] \biggr) \biggr\}\,. 
\end{align}\ese
Recall the notation~\eqref{eq:antisym} for antisymmetric waves, and~\eqref{eq:antiderive} for time anti-derivatives. The main property of this object is that $\partial_\beta\mathop{\overline{u}}_{}\!{}\ab=-\mathop{\overline{w}}_{}\!{}^\alpha$, easily checked from the expressions~\eqref{eq:vab}, and of course that $\mathop{\overline{u}}_{}\!{}\ab$ is an antisymmetric homogeneous solution of the wave equation (hence, regular when $r\to 0$).

Next, we present two forms which are alternative to Eqs.~\eqref{eq:vab}, \textit{i.e.}, which differ from it by some linear gauge transformations. Such forms are interesting because they enable us to simplify the RR calculations at 4.5PN order. The idea is to choose the gauge transformation in such a way that it reduces the number of multipole moments to be computed at the considered approximation level. Because $\ell$-th order STF spatial derivatives of the type $\hat{\partial}_L\{f\}$ are of order $\mathcal{O}(2\ell+1)$ by virtue of Eq.~\eqref{eq:PNantisymL}, we see that high-$\ell$ multipole moments, which also come with high-order STF differentiation, constitute small PN corrections --- and generally can be discarded.

A first strategy thus consists in maximizing the number of spatial STF derivatives entering the analogue of the object~\eqref{eq:vab} after a linear gauge transformation. With this choice we obtain
\bse\label{eq:vBT1}
\begin{align}
  (\overline{v}^{00})' &= \frac{4G}{c^2} \sum_{\ell=2}^{+\infty} \frac{(-)^\ell}{\ell (\ell-1) \ell!} \hat{\partial}_L \left( 2 (2\ell+1) \bigl\{\integre \calP_L\bigr\} + \ell (\ell+1) (2\ell+1) \bigl\{\integre\!\!\integre \calQ^{(+)}_L \bigr\} -\frac{2 (2\ell+3)}{c^2} \bigl\{ \calQ^{(-)}_L \bigr\} \right)\,, \\
%%%%%%%%%%%%%%%%%%%%%%%%%%%%%%%%%%%%%%%%%%%%%%%%%%%%%%%%%%%%%%%%%%%%%%%%%%%%%%%%%%
  (\overline{v}^{0i})' &= \frac{4G}{c^3} \biggl\{ - c^2 \hat{\partial}_{i} \bigl\{ \integre\!\!\integre \calP \bigr\} + \frac{3\, c^2}{2} \hat{\partial}_{ia} \bigl\{\integre\!\!\integre \calP_a \bigr\} + \sum_{\ell = 2}^{+\infty} \frac{(-)^\ell (2\ell+1)}{\ell (\ell-1) (\ell + 1)!} \left[ \hat{\partial}_{iL} \Big( (\ell+1) (\ell + 2) \, c^2 \bigl\{\integre\!\!\integre \calP_L \bigr\}  \right.  \nn \\ & \left.\left. \qquad \qquad \quad + \ell (\ell+1) (2\ell+1) \, c^2 \bigl\{ \integre\!\!\integre\!\!\integre \calQ^{(+)}_L \bigr\} - 2 (2\ell+3) \bigl\{\integre \calQ^{(-)}_L \bigr\} \right) -\ell (\ell+1) \hat{\partial}_{aL-1} \left( \varepsilon_{iab} \bigl\{\integre \calQ^{(0)}_{bL-1} \bigr\} \right) \right] \biggr\}\,, \\
%%%%%%%%%%%%%%%%%%%%%%%%%%%%%%%%%%%%%%%%%%%%%%%%%%%%%%%%%%%%%%%%%%%%%%%%%%%%%%%%%% 
 (\overline{v}^{ij})' &= \frac{4G}{c^4} \biggl\{ - 3 \, c^2 \delta_{ij} \hat{\partial}_{a}  \bigl\{ \integre\!\!\integre \calQ^{(+)}_{a} \bigr\} + \frac{3}{2} \hat{\partial}_{ij} \left( c^4 \bigl\{ \integre\!\!\integre\!\!\integre \calP \bigr\} + c^2 \bigl\{ \integre \!\! \integre \calQ^{(-)} \bigr\} \right) + 2 \, c^2 \hat{\partial}_{a(i} \left( \varepsilon_{j)ab} \bigl\{\integre\!\!\integre \calQ^{(0)}_b \bigr\} \right) \nn \\ & \qquad \quad + \hat{\partial}_{ija} \left( -\frac{5\, c^4}{2}  \bigl\{ \integre\!\!\integre\!\!\integre \calP_a \bigr\} + 5 \, c^4 \bigl\{\integre\!\!\integre\!\!\integre \!\!\integre \calQ^{(+)}_a \bigr\} - \frac{5 \, c^2}{6} \bigl\{ \integre\!\!\integre \calQ^{(-)}_a \bigr\} \right) \nn \\ & \qquad \quad + \sum_{\ell=2}^{+\infty} \frac{(-)^\ell}{\ell!} \left[ \frac{(2\ell+3) (2\ell+1)}{\ell (\ell-1)} \hat{\partial}_{ijL} \left( c^4 \bigl\{ \integre\!\!\integre \!\!\integre \calP_L \bigr\} + \ell \, c^4 \bigl\{\integre \!\! \integre \!\! \integre \!\! \integre \calQ^{(+)}_L \bigr\} - \frac{(7\ell+3) \, c^2}{(\ell+1) (2\ell+1)} \bigl\{ \integre\!\! \integre \calQ^{(-)}_L \bigr\} \right) \right. \nn \\ & \left. \qquad \qquad \qquad + \delta^{ij} (2\ell+1) \, c^2 \hat{\partial}_L \bigl\{\integre \!\! \integre \calQ^{(+)}_L \bigr\} + \frac{2(2\ell+1) \, c^2}{(\ell-1)} \hat{\partial}_{L-1a(i} \left( \varepsilon_{j)ab} \bigl\{ \integre\!\! \integre \calQ^{(0)}_{bL-1} \bigr\}\right) \right] \biggr\}\,.
\end{align}
\ese
Another convenient choice of linear gauge transformation is obtained by demanding that the $00$ and $0i$ components of the object be as simple as possible. We obtain
\bse \label{eq:vBT2}
\begin{align}
  (\overline{v}^{00})'' &= - \frac{4G}{c^2} \,\bigl\{\integre \calP \bigr\}\,, \\
%%%%%%%%%%%%%%%%%%%%%%%%%%%%%%%%%%%%%%%%%%%%%%%%%%%%%%%%%%%%%%%%%%%%%%%%%%%%%%%%%%
  (\overline{v}^{0i})'' &= \frac{6 G}{c} \,\hat{\partial}_{ia} \bigl\{ \integre\!\!\integre \calP_a \bigr\} \,, \\
%%%%%%%%%%%%%%%%%%%%%%%%%%%%%%%%%%%%%%%%%%%%%%%%%%%%%%%%%%%%%%%%%%%%%%%%%%%%%%%%%%
  (\overline{v}^{ij})'' &= \frac{4G}{c^4} \biggl\{ -\frac{9 \, c^2}{5} \hat{\partial}_{\langle i} \bigl\{\integre \!\! \integre \calQ^{(+)}_{j\rangle} \bigr\} + \frac{3 \, c^2}{2} \hat{\partial}_{ij}  \bigl\{ \integre\!\! \integre \calQ^{(-)} \bigr\} + 2 \, c^2 \hat{\partial}_{a(i} \left( \varepsilon_{j)ab} \bigl\{\integre \!\! \integre \calQ^{(0)}_b \bigr\} \right) \nn \\ & \qquad + \hat{\partial}_{ija} \left(-\frac{5 \, c^4}{2} \bigl\{ \integre\!\! \integre \!\! \integre \calP_a \bigr\} + \frac{c^4}{2} \bigl\{ \integre \!\! \integre \!\! \integre \!\! \integre \calQ^{(+)}_a \bigr\} - \frac{5 \, c^2}{6} \bigl\{ \integre \!\! \integre \calQ^{(-)}_a \bigr\} \right) \nn \\ & \qquad + \sum_{\ell=2}^{+\infty} \frac{(-)^\ell}{\ell!} \left[ \frac{c^4}{2} \hat{\partial}_{ijL} \bigl\{\integre \!\! \integre \!\! \integre  \calP_L \bigr\} - 3 \hat{\partial}_{L-1 \langle i} \left( \frac{(\ell+2) \, c^2}{(2\ell+3)} \bigl\{ \integre \calP_{j\rangle L-1} \bigr\} - \frac{2}{(\ell+1)} \calQ^{(-)}_{j \rangle L-1} \right) \right. \nn \\ &  \qquad \qquad + \hat{\partial}_{L-2} \left(\frac{(\ell+1) (\ell+2)}{2 (2\ell+1) (2\ell-1)} \bigl\{ \dot{\calP}_{ijL-2} \bigr\} + \ell \bigl\{ \calQ^{(+)}_{ijL-2} \bigr\} - \frac{(2\ell+3)}{(2\ell+1)(2\ell-1) \, c^2} \bigl\{ {}^{(-)}\!\ddot{\calQ}_{ijL-2} \bigr\} \right) \nn \\ & \left.  \qquad \qquad -2 \hat{\partial}_{aL-2} \left(\varepsilon_{ab(i} \bigl\{ {}^{(0)}\!\calQ_{j)bL-2} \bigr\} \right) \right] \biggr\}\,.
\end{align}
\ese
We have found that the two previous forms~\eqref{eq:vBT1} and~\eqref{eq:vBT2} drastically simplify the control (made in Sec.~\ref{sec:control}) of all the terms at 4.5PN order with respect to the standard form~\eqref{eq:vab}.

In both cases the linear gauge transformation $\partial\psi\ab\equiv\partial^\alpha\psi^\beta+\partial^\beta\psi^\alpha-\eta\ab\partial_\gamma\psi^\gamma$ is of the form
\begin{align}
	\psi^0 &= \sum_{\ell=0}^{+\infty} \partial_L \bigl\{ \mathcal{A}_L \bigr\} \,,\nonumber \\
	\psi^i &= \sum_{\ell=0}^{+\infty} \partial_{iL} \bigl\{ \mathcal{B}_L \bigr\} + \sum_{\ell=1}^{+\infty} \biggl[ \partial_{L-1} \bigl\{ \mathcal{C}_{iL-1} \bigr\} + \partial_{aL-1} \left(\varepsilon_{iab} \bigl\{ \mathcal{D}_{bL-1} \bigr\} \right) \biggr]\,,
\end{align}
where, for the two cases respectively,
\bse\begin{align}
	\mathcal{A}' & = \frac{G}{c^3} \left[-c^2 \integre \!\! \integre \mathcal{P} + 3 \integre {}^{(-)}\!\mathcal{Q} \right] \,,\nonumber \\
	\mathcal{A}'_i & = \frac{G}{c^3} \left[c^2 \integre \!\! \integre \mathcal{P}_i + 10 \, c^2 \integre \!\! \integre \!\! \integre {}^{(+)}\!\mathcal{Q}_i - \frac{5}{3} \integre {}^{(-)}\!\mathcal{Q}_i \right]\,,\nonumber \\
	\mathcal{A}'_L &= \frac{2 G (-)^\ell (2\ell+1)}{\ell (\ell-1)\ell!\, c^3} \left[- c^2 \integre \!\! \integre \mathcal{P}_L - (2\ell-1) \ell \, c^2 \integre \!\! \integre \!\! \integre {}^{(+)}\!\mathcal{Q}_L + \frac{(2\ell+3)}{(2\ell-1)} \integre {}^{(-)}\!\mathcal{Q}_L \right] \quad \text{for~} \ell \geqslant 2 \,, \nonumber \\
	\mathcal{B}' &= \frac{3G}{c^2} \left[c^2 \integre \!\! \integre \!\! \integre \mathcal{P} +  \integre \!\! \integre {}^{(-)}\!\mathcal{Q} \right] \,, \nonumber \\
	\mathcal{B}'_i &=  \frac{G}{c^2} \left[-5 \, c^2 \integre \!\! \integre \!\! \integre \mathcal{P}_i + 10 \, c^2 \integre \!\! \integre \!\! \integre \!\! \integre {}^{(+)}\!\mathcal{Q}_i - \frac{5}{3} \integre \!\! \integre {}^{(-)}\!\mathcal{Q}_i \right] \,, \nonumber \\
	\mathcal{B}'_L &= \frac{2 G (-)^\ell (2\ell+3)(2\ell+1)}{\ell (\ell-1)\ell!\, c^4} \left[c^4 \integre \!\! \integre \!\! \integre \mathcal{P}_L + \ell \, c^4 \integre \!\! \integre \!\! \integre \!\! \integre {}^{(+)}\!\mathcal{Q}_L - \frac{(7\ell+3) \, c^2}{(2\ell+1)(\ell+1)} \integre \!\! \integre {}^{(-)}\!\mathcal{Q}_L\right] \quad \text{for~} \ell \geqslant 2 \,, \nonumber \\
	\mathcal{C}'_i &= \frac{4G}{c^4} \left[\frac{c^2}{2} \integre \mathcal{P}_i - c^2 \integre \!\! \integre {}^{(+)}\!\mathcal{Q}_i  + \frac{5}{3} {}^{(-)}\!\mathcal{Q}_i\right] \,, \nonumber \\ \mathcal{C}'_L &= \frac{4G (-)^\ell}{(\ell-1)\ell!\, c^4} \left[ - (2\ell+1) \, c^2 \integre \mathcal{P}_L - (2\ell+1) \ell \, c^2 \integre \!\! \integre {}^{(+)}\!\mathcal{Q}_L + \frac{2(2\ell+3)}{(\ell+1)} {}^{(-)}\!\mathcal{Q}_L \right]  \quad \text{for~} \ell \geqslant 2 \,, \nonumber \\
	\mathcal{D}'_i &= \frac{4G}{c^2} \integre \!\! \integre {}^{(0)}\!\mathcal{Q}_i \,, \nonumber \\
	\mathcal{D}'_L &= \frac{4 G (-)^\ell (2\ell+1)}{(\ell-1)\ell!\, c^2} \integre \!\! \integre {}^{(0)}\!\mathcal{Q}_L  \quad \text{for~} \ell \geqslant 2 \,,\\
	\mathcal{A}'' &= -\frac{3 G}{c^3} \integre {}^{(-)}\!\mathcal{Q} \,, \nonumber \\
	\mathcal{A}''_i &= \frac{G}{c^3} \left[c^2 \integre \!\! \integre \mathcal{P}_i + c^2 \integre \!\! \integre \!\! \integre {}^{(+)}\!\mathcal{Q}_i - \frac{5}{3} \integre {}^{(-)}\!\mathcal{Q}_i \right] \,, \nonumber \\
	\mathcal{A}''_L &= \frac{G (-)^\ell}{\ell!\, c} \integre \!\! \integre \mathcal{P}_L \quad \text{for}~\ell \geqslant 2 \,, \nonumber \\
	\mathcal{B}'' &= \frac{3 G}{c^2} \integre \!\! \integre {}^{(-)}\!\mathcal{Q} \,, \nonumber \\
	\mathcal{B}''_i &= \frac{G}{c^4} \left[ -5 c^4 \integre \!\! \integre \!\! \integre \mathcal{P}_i + c^4 \integre \!\! \integre \!\! \integre \!\! \integre {}^{(+)}\!\mathcal{Q}_i - \frac{5c^2}{3} \integre \!\! \integre {}^{(-)}\!\mathcal{Q}_i \right] \,, \nonumber \\
	\mathcal{B}''_L &= \frac{G (-)^\ell}{\ell!} \integre \!\! \integre \!\! \integre \mathcal{P}_L \quad \text{for}~\ell \geqslant 2 \,, \nonumber \\
	\mathcal{C}''_i &= \frac{4G}{c^4} \left[\frac{c^2}{2} \integre \mathcal{P}_i - c^2 \integre \!\! \integre {}^{(+)}\!\mathcal{Q}_i  + \frac{5}{3} {}^{(-)}\!\mathcal{Q}_i\right] \,, \nonumber \\
	\mathcal{C}''_L &= \frac{4G (-)^{\ell+1}}{\ell!\, c^2} \integre \mathcal{P}_L \quad \text{for}~\ell \geqslant 2 \,, \nonumber \\
	\mathcal{D}''_i &= \frac{4G}{c^2} \integre \!\! \integre {}^{(0)}\!\mathcal{Q}_i \,, \nonumber \\
	\mathcal{D}''_L &=0 \quad \text{for}~\ell \geqslant 2 \,.
\end{align}\ese

\subsection{Controlling the harmonicity terms at the 4.5PN order}
\label{sec:control}

The functions $\calB^\alpha_L$ [of which the tensors $\calP_L$, $\calQ^{(+)}_L$, $\calQ^{(0)}_L$, and $\calQ^{(-)}_L$ are the irreducible components] were defined by~\eqref{eq:BL} from the multipole moments $\calA\ab_L$, which are themselves the sum of the $\calF\ab_L$'s and $\calR\ab_L$'s, see~\eqref{eq:AFR}. However, we pointed out that $\calR_L\ab$ represents the tail contribution, whose leading 4PN order has been already computed in previous works \cite{BBFM17}, and can be ignored henceforth. Therefore, we shall consider all the previous formulas with the replacement $\calA_L\ab\longrightarrow\calF_L\ab$ together with $\calB^\alpha_L\longrightarrow\calG^\alpha_L$, where $\calG_L^\alpha$ is related to $\calF_L\ab$ by the equivalent of~\eqref{eq:BL}, \textit{i.e.},
\begin{equation}\label{eq:GL}
	\calG_L^\alpha = \frac{1}{c}\dot{\calF}^{0\alpha}_L - \ell \calF^{\alpha\langle i_\ell}_{L-1\rangle} - \frac{1}{(2\ell+3)c^2} \ddot{\calF}^{i\alpha}_{i L}\,.
\end{equation} 
We shall use a different form of the function $\calG^\alpha_L$ which reads (see Sec.~\ref{sec:notation} for the notation)
\begin{align} \label{eq:GLmu}
	\calG^\alpha_L = \FP \int \dd^3 \mathbf{x}
	\, B\, \widetilde{r}^B r^{-2} x^i \hat{x}_L \,\int^1_{-1} \dd z \,\delta_\ell(z) \,{\overline
		\tau}^{\alpha i}(\mathbf{x}, t+z r/c) \,.
\end{align}
There is an explicit factor $B$, so the range of integration over $\mathbf{x}$ is limited to the neighbourhood of spatial infinity. Hence, we see that this function is the same as the one constructed in the exterior zone in Ref.~\cite{B98mult}. More generally, the above construction is very similar to the one in the exterior zone; simply, all the retarded waves are replaced by their antisymmetric counterparts $\bigl\{\calG^\alpha_L\bigr\}$. Note that from the irreducible decomposition~\eqref{eq:irred1} (with $\calB^\alpha_L\longrightarrow\calG^\alpha_L$), we see that $\calP_L$ is built with the mixed components of the stress-energy tensor, $\Sigma_i=\overline{\tau}^{0i}/c$, while the moments $\calQ^{(+)}_L$, $\calQ^{(0)}_L$, and $\calQ^{(-)}_L$ involve its space components $\Sigma_{ij}=\overline{\tau}^{ij}$.

There remains to compute the function $\calG^\mu_L$ given by Eq.~\eqref{eq:GLmu}. The integration over $z$ is achieved using 
\bse\begin{align}
%\overline{\tau}^{\alpha i}_{[\ell]}(\mathbf{x},t) \equiv  
\int^1_{-1} \dd z \,\delta_\ell(z) \,{\overline \tau}^{\alpha i}(\mathbf{x}, t+z r/c) &= \sum_{j=0}^{+\infty} \alpha^j_\ell \,\frac{r^{2j}}{c^{2j}} \,\partial_t^{2j} \left[\overline{\tau}^{\alpha i}(\mathbf{x},t)\right]\,,
\end{align}
where
\begin{align}
\alpha^j_\ell &= \frac{(2\ell+1)!!}{(2k)!! (2\ell+2k+1)!!}\,.
\end{align}\ese  
When we insert this expression into~\eqref{eq:GLmu} and commute the integration and summation symbols, all the resulting integrals have the form
\begin{align} \label{eq:definition_I}
  \mathcal{I}_{p\,, L}[F^{i}] = \FP B \int \dd^3 \mathbf{x} \, \widetilde{r}^B \frac{x^i}{r^2} \hat{x}^L r^{2j} F^i(\mathbf{x},t)\,,
\end{align}
which may be thus regarded as the master integral of the problem. More precisely, we find
\begin{align} \label{eq:G_moment_formula}
  \bigl(\calP_L, \mathcal{G}^i_L\bigr) = \sum_{j=0}^{+\infty} \alpha^j_\ell \left(\frac{\dd}{c\, \dd t}\right)^{2j} \Bigl[\mathcal{I}_{p, L}[\Sigma_{\alpha a}]\Bigr]\,,
\end{align}
where $\alpha=0$ for the source term of $\calP_L$ and $\alpha=i$ for the source term of $\mathcal{G}^i_L$, with $\Sigma_{0 a}=\Sigma_a$ and we recall that $\Sigma_a=\overline{\tau}^{0a}/c$ and $\Sigma_{ia}=\overline{\tau}^{ia}$.

The master integral~\eqref{eq:definition_I} vanishes when the function $F^i(\mathbf{x},t)$ is locally integrable and has compact support, since the integration volume is limited to spatial infinity and the limit $B \to 0$ and the summation symbol commute in that case. However, the latter condition does not hold for the source functions that arise when solving field equations within the PN iteration scheme. Instead, near $r \to +\infty$, the source in harmonic coordinates, say $F^i$, behaves as
\begin{align} \label{eq:Fi_expansion}
  F^i(\mathbf{x},t) \mathop{\sim}_{r\to +\infty} \sum_{p \leqslant p_{\text{max}},\, q} r^p \ln^q \widetilde{r} \sum_{\ell=0}^{+\infty} f^i_{p, q, L}(t)\, \hat{n}^L\,,
\end{align}
where the $f^i_{p, q, L}(t)$'s are time-dependent STF coefficients, with $p$ and $q$ integers.
%, and $L=i_1 \cdots i_\ell$ multi-indices of length $\ell$.
 The power index $p$ is bounded from above by $p_\text{max}$ and, for given values of $p$ and $\ell$, only a finite number of $f^i_{p, q, L}(t)$ do not vanish. Potential divergences ensuing from the asymptotic expansion~\eqref{eq:Fi_expansion} of the source function $F^i$ of the integral~\eqref{eq:definition_I} are cured by Hadamard's FP regularization. Since those occur at large $r$, we may assume $r$ larger than some cutoff radius $\mathcal{R}$ and restrict the integration domain to $r>\mathcal{R}$ without any incidence on the result.

Inserting the asymptotic expansion~\eqref{eq:Fi_expansion} into the master integral~\eqref{eq:definition_I}, we swap the sum and the integral, and express the latter in spherical coordinates. For a given value of $p$ and $q$, this yields the radial integral
\begin{align} \label{eq:radial_integral}
B \int_\mathcal{R}^{+\infty} \dd r\, \frac{r^{B+\ell+1+2j+p}}{r_0^B} \ln^q \left(\frac{r}{r_0} \right) = - B \left(\frac{\dd}{\dd B} \right)^q\left[ \frac{\mathcal{R}^{B+\ell+2+p+2j}}{r_0^B (B+\ell+2+p+2j)}\right]\,,
\end{align}
where we have used that, for large enough real parts of $B$, the quantity $r^{B+\ell+2+p+2j}$ tends towards zero at infinity. By analytic continuation, we see that this quantity vanishes in the limit $B\to 0$ whenever $p\neq -\ell-2-2j$. If $p = -\ell -2-2j$, only terms for which $q=0$ contain single poles, leading to a contribution $-1$. It remains to perform the angular integration, resorting to the formula
\begin{align} \label{eq:angular_integral}
f^i_{p,q,L'}(t)  \int \frac{\dd^2 \Omega}{4\pi} \hat{n}^L \hat{n}^{L'} = \delta_{\ell,\ell'} \frac{\ell!}{(2\ell+1)!!} f^i_{p,q,L}(t)\,.
\end{align}
Combining Eqs.~\eqref{eq:Fi_expansion}--\eqref{eq:angular_integral}, we obtain an explicit expression for $\mathcal{I}_{p, L}[F^{i}]$ in terms of the coefficients $f^i_{p,q,L}(t)$, namely
\begin{align} \label{eq:I}
  \mathcal{I}_{p, L}[F^{i}] = -\frac{4\pi \ell!}{(2\ell+3)!!} \left[ (\ell+1) f^i_{-\ell-2j-2,0, Li} + (2\ell+3)  f^i_{-\ell-2j-2, 0,\langle L}\, \delta^{i_\ell \rangle i} \right] \,.
\end{align}

Since $\mathcal{I}_{p,L}[F^a]$ vanishes when $F^a$ has compact support, the compact parts of $F^a=\Sigma_a$ and $F^a=\Sigma_{ia}$ do not contribute. As for their non-compact parts, they are non-linear in $G$, and thus of order $\mathcal{O}(2)$ (or higher). Therefore, all moments $\calP_L$, $\calQ^{(+)}_L$, $\calQ^{(0)}_L$, and $\calQ^{(-)}_L$ are 1PN quantities at least. We will now prove that they are in fact of order 2PN at least, by showing that the non-compact support terms entering the 1PN piece of the sources, which are all  quadratic, cannot produce any contribution to the multipole-moment functions $\calG^\alpha_L$. 

To this end, it will be useful to consider, for monomials proportional to $r^p \hat{n}^L$, with $p\in \mathbb{Z}$, such as those entering the asymptotic expansion~\eqref{eq:Fi_expansion}, the ``parity'' of the sum $\ell + p$, and introduce the function $\pi$ defined on the set of such monomials by: $\pi(r^p \hat{n}^L)=0$ if $\ell+p$ is even, $\pi(r^p \hat{n}^L)=1$ if $\ell+p$ is odd. The bit $\pi(r^p \hat{n}^L)$ will be referred to as the $\pi$-value of $r^p \hat{n}^L$. This concept can be extended to more general functions $F(\mathbf{x})$ in the following way. If the asymptotic expansion of $F$ when $r\to +\infty$ is made of terms that all have the same $\pi$-value, this common value is assigned to $\pi(F)$ by definition. For arbitrary functions $F$, no $\pi$-value may be assigned to $F$ in general, but it is often possible to split the asymptotic expansion of $F$ into two pieces, each having a definite $\pi$-value. The main point here is that for any source function $F^i$ of $\pi$-value 0, we have that $\mathcal{I}_{p,L}[F^i]=0$, since all the coefficients entering Eq.~\eqref{eq:I} are themselves equal to zero. 

An important class of functions with definite $\pi$-values are monomials $r^p n^L$, where the angular factor $n^L$ is not STF anymore. Due to the fact that the STF decomposition of $n^L$ only contains terms $\hat{n}^{L-2J}$, with $0\leqslant j \leqslant [\ell/2]$ of the same parity as $\ell$, we see that $\pi(r^p n^L)$ is indeed well defined and equal to $\pi(r^p \hat{n}^L)$. Similarly, the $\pi$-value of the derivative of $r^p \hat{n}^L$, as a sum of monomials $r^{p'} \hat{n}^{L'}$, with $p'=p-1$ and $\text{parity}(\ell')=\text{parity}(\ell)+1$ (mod 2), is $\pi(\partial_i[r^p n]) = \pi(r^p n)$, \textit{i.e.}, space derivatives preserve $\pi$-values. At last, from the trivial identity $(r^{p} n^{L})(r^{p'} n^{L'})=r^{p+p'} n^{LL'}$, it follows that the $\pi$-value of a product of functions is the sum (modulo 2, in the set $\{0,1\}$) of the $\pi$-value of each function whenever it exists.

With those tools in hands, we are in the position to evaluate the $\pi$-value of potentials whose sources have compact support. In the near zone, a symmetric potential $\bar{P}=\widetilde{\Box}\sym^{-1} \bar{S}$ sourced by $\bar{S}$ is given by Eqs.~\eqref{eq:propsym} and~\eqref{eq:genpoissiter}, with $\widetilde{\tau}\ab$ replaced by $\bar{S}$. When $\bar{S}$ is locally integrable with compact support, no regularization is required. Moreover, if we perform the asymptotic expansion of the integrand near $r \to +\infty$, we can commute the sum and integration symbols, so that the multipole expansion of $\bar{P}$ outside the source support is given by
\begin{align}
  \bar{P} = \sum_{k=0}^{+\infty} \frac{1}{(2k)!}  \sum_{\ell=0} \frac{(-)^\ell}{\ell!} \partial_L (r^{2k-1}) \left(\frac{\dd}{c\, \dd t} \right)^{2k} \int \dd^3 \mathbf{x}\, x'^L S(\mathbf{x}',t)\,.
\end{align}
Now, because $\pi(\partial_L r^{2k-1})=\pi(r^{2k-1})=1$, the $\pi$-value of $\bar{P}$ (or its derivatives) exists and is equal to 1. This immediately entails that ``direct'' quadratic terms $\partial \bar{P} \partial \bar{P}'$, precisely those that arise at 1PN in $\Sigma_{\alpha a}$, have a $\pi$-value of zero, which prevents them to contribute to $\calG^\mu_L$ at any PN order.

We have thus shown that $\calP_L=\mathcal{O}(4)$ and also any of the $\calQ^{(+,0,-)}_L=\mathcal{O}(4)$. From this, it is straightforward to list the multipole moments contributing to either of the expressions~\eqref{eq:vBT1} or~\eqref{eq:vBT2} for $\overline{v}_\BT\ab$ in well chosen gauges at the 4.5PN approximation. Those are $\calP$ (\textit{i.e.}, with $\ell=0$) at the 3PN order and $\calP_i$, $\calP_{ij}$, $\calQ_i^{(+)}$, and $\calQ_{ij}^{(+)}$ at the 2PN order. With the help of Eq.~\eqref{eq:G_moment_formula}, we checked (with \textit{Mathematica}) that all those quantities vanish when discarding negligible remainders. This leads us to the important conclusion that $\overline{v}\ab$, in whatever coordinate system (recall from Sec.~\ref{sec:nonlinear} that linear gauge transformations have no incidence on the result), does not contribute to the present work at 4.5PN order.

\section{Balance equations and dimensional identities}
\label{app:id_dim}

The method described in Sec.~\ref{sec:flux_balance} has been supplemented, up to 3.5PN, by an alternative approach that consists in forming, for each of the integrals of motion $H=\{E, J_i, P_i, G_i\}$
%$I = E$, $J^i$, $P^i$, $G^i$ 
expressed in terms of the configuration variables \mbox{$(\bm{y}_1$, $\bm{y}_2$, $\bm{v}_1$, $\bm{v}_2)$}, the most general ansätze for the Schott terms of $(p+1/2)$PN order (with $p=2,3,4$), say $\delta H_{(2p+1)}$, that are compatible with the symmetries of the problem, are polynomial in the two masses and of course have the correct physical dimension. The coefficients of the various monomials therein are initially unknown. However, they must satisfy, for the balance equations to hold, a set of linear equations that result from identifying the left- and right-hand sides of the explicit form of the balance laws. Solving this simple system provides the desired Schott terms $\delta H_{(2p+1)}$.

This approach, whose principle is straightforward, is not as easy as it may seem to implement in practice, for two different reasons. First, the monomials that compose the original ansätze are not \textit{a priori} independent. This can be understood by looking at their structure. They are made of elementary factors, which may be taken to be dimensionless for convenience. The scalar factors are
\begin{align} \label{eq:variable_definition}
& X_1=\frac{G m_1}{r_{12}c^2} \,,  & & X_2=\frac{G m_2}{r_{12}c^2} \,, & & X_3=\frac{(n_{12}v_1)}{c} \,, & & X_4=\frac{(n_{12}v_2)}{c}\,, & & X_5 = \frac{v_1^2}{c^2}\,, \nn \\  & X_6 = \frac{(v_1 v_2)}{c^2}\,, & & X_7 = \frac{v_2^2}{c^2}\,, & & X_8 = \frac{(n_{12} y_1)}{r_{12}}\,, & &X_{9} = \frac{y_1^2}{r_{12}^2}\, & & X_{10} = \frac{(y_1 v_1)}{r_{12}c}\,, & & X_{11} = \frac{(y_1 v_2)}{r_{12}c}\,.
\end{align}
Due to the relation $y_2^i = y_1^i - r_{12} n_{12}^i$, there is no need to include quantities that depend explicitly on the position vector $y_2^i$. For
axial quantities such as the angular momentum, mixed products may appear, but
they can be eliminated by means of the following identity, valid for any four vectors $U_1^i$, $U_2^i$, $U_3^i$, and $U^i$:
\begin{align} \label{eq:mixed_product_identity}
(U_1, U_2, U_3) U^i = (U_3 U) (\bm{U}_1 \times \bm{U}_2)^i + (U_1 U) (\bm{U}_2 \times \bm{U}_3)^i + (U_2 U) (\bm{U}_3 \times \bm{U}_1)^i\,,
\end{align}
where $(U_1, U_2, U_3) = \varepsilon_{jkl} U_1^j U_2^k U_3^l$ and  $(\bm{U}_1 \times \bm{U}_2)^i= \varepsilon^{i~}_{jk} U_1^j U_2^k$ denote the mixed and cross products, respectively.

The monomials appearing in the ansatz for a given Schott term $\delta H_{(2p+1)}$ also contain an isolated vector-like factor that bears the free index, namely, for polar vectors:
\begin{align} \label{eq:polar_variable_definition}
 & X^\text{(p)}_{12} = n_{12}^i\,, & & X^\text{(p)}_{12}=\frac{v_1^i}{c}\,, & & X^\text{(p)}_{14} =\frac{v_2^i}{c}\,, & & X^\text{(p)}_{15}=\frac{y_1^i}{r_{12}}\,,
\end{align}
and for the axial vector $J_i$:
\begin{align} \label{eq:axial_variable_definition}
& X_{12}^\text{(a)} = \frac{(\bm{n}_{12} \times \bm{v}_1)^i}{c}\,, & & X^\text{(a)}_{13} = \frac{(\bm{n}_{12} \times \bm{v}_2)^i}{c}\,, & & X^\text{(a)}_{14} = \frac{(\bm{v}_1 \times \bm{v}_2)^i}{c^2}\,,\nn \\ & X^\text{(a)}_{15} = \frac{(\bm{n}_{12} \times \bm{y}_1)^i}{r_{12}}\,, & &  X^\text{(a)}_{16} = \frac{(\bm{y}_1 \times \bm{v}_1)^i}{r_{12}c}  & & X^\text{(a)}_{17} = \frac{(\bm{y}_1 \times \bm{v}_2)^i}{r_{12}c} \,.
\end{align}
Up to the orders that have been investigated, all Schott terms $\delta H_{(2p+1)}$ are polynomials in the variables $\{X_A\}_{1 \leqslant A \leqslant 11}$. Moreover, $\delta E_{(2p+1)}$, $\delta P^i_{(2p+1)}$, and $\delta G^i_{(2p+1)}$, depend on the set of variables $\{X^\text{(p)}_A\}_{12 \leqslant A \leqslant 15}$, while $\delta  J^i_{(2p+1)}$ depends on $\{X^\text{(a)}_A\}_{12 \leqslant A \leqslant 17}$. In principle, the power of the scalar variables $X_8$, which does not contain any $1/c$ factor, can be arbitrary high at fixed PN order. However, we constrain our ansätze for the $\delta H_{(2p+1)}$'s, regarded again as functions of $(y_1^i$, $n_{12}^i$, $v_1^i$, $v_2^i)$, by requiring that the degree in the variable $y_1^i$ of a given Schott term should never exceed that of the acceleration and the corresponding flux at the same PN order. 

The polynomials defined by the Schott terms are thus made of a finite number of monomials $X_{A_1} \cdots X_{A_r} \times X^\text{(p/a)}_{B_1} \cdots X^\text{(p/a)}_{B_s}$, with $1 \leqslant r \leqslant 11$ and other bounds for $s$ depending on the considered Schott term, which are not all independent from each other. Indeed, in three space dimensions, any of the four vectors of the problems, \mbox{$(n_{12}^i$, $y_1^i$, $v_1^i$}, $v_2^i)$, is linked to the three others by means of the fundamental identity $n_{12}^{[i} y_1^j v_1^k v_2^{l]}=0$, stating that there cannot be antisymmetric $n$-forms for $n > 3$. Contracted with four different vectors, it leads to the scalar dimensional identity
\begin{align} \label{eq:scalar_dimensional_identity}
P_S(X_1, \cdots, X_{11}) \equiv \frac{24}{r_{12}^2 c^4} n_{12}^{[i} y_1^j  v_1^k v_2^{l]} n_{12}^{i} y_1^j v_1^k v_2^{l} =0 \,,
\end{align}
where the prefactor $24/(r_{12}^2 c^4)$ makes the polynomial $P_S(X_1, \cdots, X_{11})$ dimensionless and removes the global antisymmetrization factor $1/4!$. In addition, contracting the fundamental identity with three different vectors yield one vector identity for each of the four choices of vector triplets. The resulting four vector identities $n_{12}^{[i} y_1^j v_1^k v_2^{l]} U_1^j U_2^k U_3^l$, with $U^i_1$, $U^i_2$, $U^i_3\in\{n_{12}^i, y_1^i, v_1^i, v_2^i\}$, define four polynomials, which are zero on the configuration space:
\bse\label{eq:polar_dimensional_identity}\begin{align} 
&  P^\text{(p)}_1(X_1, \cdots, X_{11}, X^\text{(p)}_{12}, \cdots, X^\text{(p)}_{15}) \equiv \frac{24}{r_{12} c^4} n_{12}^{[i} y_1^j v_1^k v_2^{l]} n_{12}^j v_1^k v_2^l = 0\,,\\
  &  P^\text{(p)}_2(X_1, \cdots, X_{11}, X^\text{(p)}_{12}, \cdots, X^\text{(p)}_{15}) \equiv \frac{24}{r_{12}^2 c^4} n_{12}^{[i} y_1^j v_1^k v_2^{l]} v_1^j v_2^k y_1^l = 0 \,,\\
  &  P^\text{(p)}_3(X_1, \cdots, X_{11}, X^\text{(p)}_{12}, \cdots, X^\text{(p)}_{15}) \equiv \frac{24}{r_{12}^2 c^3} n_{12}^{[i} y_1^j v_1^k v_2^{l]} v_2^j y_1^k n_{12}^l = 0 \,,\\
  &  P^\text{(p)}_4(X_1, \cdots, X_{11}, X^\text{(p)}_{12}, \cdots, X^\text{(p)}_{15}) \equiv \frac{24}{r_{12}^2 c^3} n_{12}^{[i} y_1^j v_1^k v_2^{l]} y_1^j n_{12}^k v_1^l = 0\,.
\end{align}\ese
For $\delta J^i_{(2p+1)}$, it is always possible to write the original ansatz in such a way that the free index $i$ is held by a cross product. The dimensional identities involving cross products are found by contracting $n_{12}^{[j} y_1^k v_1^l v_2^{m]}=0$ with the Levi-Civita~%3-form 
$\varepsilon_{ijk}$ ($i$ playing here the role of free index) and any of the six pairs of tensors, $U^l_1 U^m_2$, with $U^l_1\,, U^l_2 \in \{n_{12}^l, y_1^l, v_1^l, v_2^l\}$. The identities $\varepsilon_{ijk} n_{12}^{[j} y_1^k v_1^l v_2^{m]} U_1^l U_2^m=0$ imply the vanishing of six new polynomials $P_{s}^\text{(a)}(X_1, \cdots, X_{11}, X^\text{(a)}_{12}, \cdots, X^\text{(a)}_{15})$. All other dimensional identities relevant for our problem amount to Eq.~\eqref{eq:mixed_product_identity}, which has been already taken into account. This specific relation is actually straightforward to handle thanks to the Hodge dual, defined in our case for some vector $v^i$ by
\be \label{eq:HodgeDual}[{}^* v]_{jk} = \frac{1}{2}\varepsilon_{ijk}v_i \,.\ee
After taking the Hodge dual of both members, 
%\textit{i.e.}, multiplying them by $\varepsilon^{iab}$,
and using the relation $\varepsilon^{iab} \varepsilon_{ijk} = \delta^a_j \delta^b_k - \delta^b_j \delta^a_k$, it becomes a trivial identity. We resorted to this Hodge dual technique, applied to the case where the fundamental objects comprise tensors of higher ranks in addition to vectors, namely the multipole moments, to prove the vanishing of the coefficients entering Eq.~\eqref{eq:CD_vanishing}.

Once those dimensional identities are in hands, an efficient manner to take them into account is to perform all calculations in the quotient space of the relevant ring of polynomials by the ideal $\mathcal{I}$ generated by the corresponding dimensional identities, as shown in table~\ref{tab:rings}.
\begin{table}[!h]
  \begin{center}
    \begin{tabular}{|c||c|c|c|}
\hline Integrals of motion & $E$ & $J^i$ & $P^i$ and $G^i$\\ \hline \hline Ring of polynomials $\mathcal{A}$ & $K[X_1, \cdots,X_{11}]$ & $K[X_1, \cdots,X_{11}, X^\text{(a)}_{12}, \cdots, X^\text{(a)}_{17}]$ & $K[X_1, \cdots, X_{11}, X^\text{(p)}_{12}, \cdots, X^\text{(p)}_{15}]$ \\ \hline Generators of the ideal $\mathcal{I}$ & $P_\text{S}$ & $P_S$, $P^\text{(a)}_{1}$, ..., $P^\text{(a)}_{6}$ & $P_S$, $P^\text{(p)}_{1}$, ..., $P^\text{(a)}_{4}$ \\ \hline
    \end{tabular}
    \caption{For each conserved quantities, we specify the polynomial ring $\mathcal{A}$ it belongs to, where the variables $X_A$, $X_A^{\text{(a)}}$ and $X_A^{\text{(p)}}$ are defined by~\eqref{eq:variable_definition} and~\eqref{eq:polar_variable_definition}--\eqref{eq:axial_variable_definition}. In $J^i$, it is understood that all mixed products are eliminated in favor of cross products by means of~\eqref{eq:mixed_product_identity}. We indicate, in each case, the corresponding dimensional identities defined by~\eqref{eq:scalar_dimensional_identity}--\eqref{eq:polar_dimensional_identity} and in the text.}\label{tab:rings}
  \end{center}
\end{table}

The reduction of a polynomial of $\mathcal{A}$ modulo $\mathcal{I}$, \textit{i.e.}, modulo combinations of dimensional identities, may be achieved efficiently by building an appropriate set of generators for $\mathcal{I}$, referred to as a Gröbner basis (see Ref.~\cite{BK10} for a precise definition). Introducing those particular generators requires the setting of some monomial order. Any polynomial may then be reduced with respect to the chosen basis by subtracting repeatedly multiples of basis elements so as to replace, at each step, a monomial of $P$ by several smaller monomials (in the sense of the considered order relation). As it turns out, the reduction chain $P_0$, $P_1$, \dots, $P_i$, \dots, constructed with this procedure is always finite, due to the particular properties of Gröbner basis and its last element can be regarded as a ``fully reduced'' version of $P$. If $P$ belongs to the ideal $\mathcal{I}$ generated by some dimensional identities, its fully reduced version vanishes. In other words, the reduction algorithm recognizes that $P$ is zero when dimensional identities are taken into account. In practice, to show that two polynomials are equal, under the constraints that elements of $\mathcal{I}$ must vanish, we compute their difference, and check that the reduction procedure yields zero. This reduction, for a given Gröbner basis $B_1$, \dots, $B_n$ of the variables $X_1$, \dots, $X_m$, can be achieved with the help of the command \texttt{PolynomialReduce[\{$\text{B}_1$, $\cdots$, $\text{B}_n$\}, \{$\text{X}_1$, $\cdots$, $\text{X}_m$\}]} within \textit{Mathematica}.

Thanks to this powerful tool, we first reduce the monomials used in our ansätze for the Schott terms. The number of mononials entering the reduced expressions is always lower than the number of original monomials. We then compute the combination $\dd H/\dd t+\mathcal{F}_H$, where $\mathcal{F}_H$ represents the flux of $H$, reduce it through the procedure discussed above, and equate the resulting polynomial to zero. The ensuing system of equations for the ansatz coefficients is however extremely large, unless we use additional physical information (\textit{e.g.}, the minimum power of $G$ that can enter the Schott terms) to forbid certain types of terms in the ansätze. Solving those systems is the second difficulty of this approach. If we insist in being agnostic about the structure of the Schott terms, the computation of $\delta J^i_{(3.5)}$, for instance, implies solving %$49\,673$ equations of $25\,706$ unknowns.
about $5\times10^4$ equations for about $2.5\times10^4$ unknows. The computational time is reasonably long (less than 24 hours in this case), but the required memory is of the order of $\mathcal{O}(1\text{Tb})$.

The quantities $\delta E_{(3.5)}$, $\delta J^i_{(3.5)}$, and $\delta P^i_\text{(3.5)}$ constructed with the previous procedure have been checked to be the same, modulo dimensional identities, as their counterparts obtained in Sec.~\ref{sec:flux_balance}, after replacing the multipole moments by their explicit expressions (we provide the latter in the Supplemental Material \cite{SuppMaterial}). For the CM integral, we found a net difference, but it turns out to be a mere constant, so that only the initial value of the CM vector, namely $G^i_0=(G^i + \Gamma^i) - (P^i + \Pi^i) t$, is modified,
\begin{align}
  G^i_{0\,\text{Sec.~\ref{sec:flux_balance}}} - G^i_{0\,\text{App.~\ref{app:id_dim}}}
  %Z^i_\text{this appendix} 
  = \frac{2 G P^i}{15c^5} \left[ 4 E + \frac{13}{14 m \nu^2 c^2} \left(\left(1+ \frac{380}{13} \nu\right) E - \frac{P^2}{2m} \right) \left( E - \frac{P^2}{2m} \right)  \right] + \calO(9)\,.
\end{align}
Note that the right-hand side vanishes at the considered approximation level when $G^i$ is set to zero, thus the CM~frame is defined unambiguously. At the 4.5PN order, the energy has been computed by means of both methods, leading to identical results. The other Schott terms provided by the algorithm described in Sec.~\ref{sec:flux_balance} have been verified, by checking that, after incorporating them to the corresponding integrals of motion, the latter do satisfy the balance equations~\eqref{eq:fluxbalance} at the 4.5PN order when dimensional identities are properly taken into account.

\section{Results for the Schott terms up to 4.5PN order}
\label{app:Schott}

We have proven in Sec.~\ref{sec:flux_balance} that the balance laws~\eqref{eq:fluxbalance} are satisfied at the required order, where the fluxes~\eqref{eq:fluxes} agree with the known expressions computed at future null infinity. Evidently, the proof must be accompanied by the explicit expressions for the terms in the left-hand sides of the balance laws, and notably the RR contributions or Schott terms \cite{Schott} therein, whose structure was given in~\eqref{eq:Schott}. First, the complete expressions of the Schott terms for the energy, the angular and  linear momenta, and the CM position are provided in extended BT coordinates at the 2.5PN, 3.5PN, and 4.5PN orders in the Supplemental Material~\cite{SuppMaterial}, in the form of  linear expressions in the unreplaced multipole moments, which conform to the procedure described in Eq.~\eqref{eq:split}. Secondly, we provide hereafter the explicit expressions of the Schott term at 2.5PN and 3.5PN orders in a more compact form, which allows for quadratic expressions in the unreplaced multipole moments (the 4.5PN piece was too lengthy to present). Of course, these two forms are strictly equivalent. They read: 
\bse\label{eq:Schott25PN35PN}
\begin{align}
	E_{\text{2.5PN}} &= \frac{G}{c^5} \Bigl(\frac{6}{5} \dM^{(2)}_{ab} \dM^{(3)}_{ab}
	+ \frac{2}{5} \dM^{(1)}_{ab} \dM^{(4)}_{ab}\Bigr)\,,\\
	%%%%%%%%%%%%%%%%%%%%%%%%%%%%%%%%%%%%%%%%%%%%%%%%%%%%%%%%%%%%%%%%%%%%%%%%%%%%%%
	E_{\text{3.5PN}} &= \frac{G}{c^7} \Biggl[\frac{25}{189} \dM^{(3)}_{abi} \dM^{(4)}_{abi}
	+ \frac{16}{189} \dM^{(2)}_{abi} \dM^{(5)}_{abi}
	+ \frac{2}{63} \dM^{(1)}_{abi} \dM^{(6)}_{abi}
	+ \dM^{(6)}_{bi} \biggl [\frac{17}{105} n_{12}^{bi} m_{2} r_{12}^3 (n_{12}{} v_{2}{})
	+ m_{2} r_{12} \Bigl(\frac{17}{105} y_{1}^{bi} (n_{12}{} v_{2}{})\nn\\
	& -  \frac{22}{105} v_{2}^{b} y_{1}^{i} (n_{12}{} y_{1}{})\Bigr)
	-  \frac{17}{105} m_{1} y_{1}^{bi} (v_{1}{} y_{1}{})
	-  \frac{17}{105} m_{2} y_{1}^{bi} (v_{2}{} y_{1}{})
	+ m_{2} r_{12}^2 \Bigl(- \frac{34}{105} n_{12}^{b} y_{1}^{i} (n_{12}{} v_{2}{})\nn\\
	& + \frac{22}{105} n_{12}^{b} v_{2}^{i} (n_{12}{} y_{1}{})
	-  \frac{17}{105} n_{12}^{bi} (v_{2}{} y_{1}{})\Bigr)\biggl]
	+ \dM^{(6)}_{ai} \Bigl(\frac{34}{105} m_{2} n_{12}^{a} r_{12} y_{1}^{i} (v_{2}{} y_{1}{})
	+ \frac{11}{105} m_{1} v_{1}^{a} y_{1}^{i} y_{1}^{2}\nn\\
	& + \frac{11}{105} m_{2} v_{2}^{a} y_{1}^{i} y_{1}^{2}\Bigr)
	+ \Bigl(- \frac{11}{105} m_{2} n_{12}^{a} r_{12} v_{2}^{b} y_{1}^{2}
	-  \frac{11}{105} m_{2} n_{12}^{a} r_{12}^3 v_{2}^{b}
	+ \frac{11}{105} m_{2} r_{12}^2 v_{2}^{a} y_{1}^{b}\Bigr) \dM^{(6)}_{ab}\nn\\
	& + m_{2} r_{12} \dM^{(5)}_{ai} \Bigl(- \frac{4}{35} n_{12}^{a} v_{2}^{i} (v_{2}{} y_{1}{})
	+ \frac{92}{105} n_{12}^{a} y_{1}^{i} v_{2}^{2}\Bigr)
	+ \dM^{(5)}_{bi} \biggl(m_{2} r_{12} \Bigl(- \frac{4}{35} v_{2}^{b} y_{1}^{i} (n_{12}{} v_{2}{})
	+ \frac{22}{105} v_{2}^{bi} (n_{12}{} y_{1}{})\Bigr)\nn\\
	& + m_{2} r_{12}^2 \Bigl(\frac{4}{35} n_{12}^{b} v_{2}^{i} (n_{12}{} v_{2}{})
	-  \frac{46}{105} n_{12}^{bi} v_{2}^{2}\Bigr)
	+ m_{1} \Bigl(\frac{4}{35} v_{1}^{b} y_{1}^{i} (v_{1}{} y_{1}{})
	-  \frac{46}{105} y_{1}^{bi} v_{1}^{2}
	-  \frac{11}{105} v_{1}^{bi} y_{1}^{2}\Bigr)\nn\\
	& + m_{2} \Bigl(\frac{4}{35} v_{2}^{b} y_{1}^{i} (v_{2}{} y_{1}{})
	-  \frac{46}{105} y_{1}^{bi} v_{2}^{2}
	-  \frac{11}{105} v_{2}^{bi} y_{1}^{2}\Bigr)
	+ \frac{G}{r_{12}} \biggl [- \frac{1}{21} n_{12}^{bi} m_{1} m_{2} r_{12} (n_{12}{} y_{1}{})\nn\\
	& + m_{1} m_{2} \Bigl(- \frac{4}{35} n_{12}^{b} y_{1}^{i} (n_{12}{} y_{1}{})
	+ \frac{11}{105} n_{12}^{bi} y_{1}^{2}\Bigr)\biggl]\biggl)
	+ \biggl [- \frac{11}{105} m_{2} r_{12}^2 v_{2}^{ab}
	+ \frac{G}{r_{12}} \Bigl(- \frac{2}{35} m_{1} m_{2} r_{12}^2 n_{12}^{ab}\nn\\
	& + \frac{23}{105} m_{1} m_{2} n_{12}^{a} r_{12} y_{1}^{b}
	-  \frac{17}{105} m_{1} m_{2} y_{1}^{ab}\Bigr)\biggl] \dM^{(5)}_{ab}
	+ \Bigl(\frac{16}{45} n_{12}^{bi} m_{2} r_{12}^2 v_{2}^{a}
	+ \frac{16}{45} m_{1} v_{1}^{a} y_{1}^{bi}
	+ \frac{16}{45} m_{2} v_{2}^{a} y_{1}^{bi}\Bigr) \varepsilon_{aij} \dS^{(5)}_{bj}\nn\\
	& -  \frac{16}{45} m_{2} \varepsilon_{bij} n_{12}^{a} r_{12} v_{2}^{b} y_{1}^{i} \dS^{(5)}_{aj}
	+ \frac{16}{45} m_{2} \varepsilon_{abj} n_{12}^{a} r_{12} v_{2}^{b} y_{1}^{i} \dS^{(5)}_{ij}
	+ \frac{1}{189} \dM_{abi} \dM^{(7)}_{abi}
	+ \frac{32}{15} \dS^{(2)}_{ab} \dS^{(3)}_{ab}
	+ \frac{32}{45} \dS^{(1)}_{ab} \dS^{(4)}_{ab}\Biggl]\,,\\
	%%%%%%%%%%%%%%%%%%%%%%%%%%%%%%%%%%%%%%%%%%%%%%%%%%%%%%%%%%%%%%%%%%%%%%%%%%%%%%
	J^i_{\text{2.5PN}} &= \frac{G}{c^5} \Bigl(- \frac{2}{5} \varepsilon_{ibj} \dM^{(1)}_{ab} \dM^{(3)}_{aj}
	+ \frac{2}{5} \varepsilon_{ibj} \dM^{(4)}_{aj} \dM_{ab}\Bigr)\,,\\
	%%%%%%%%%%%%%%%%%%%%%%%%%%%%%%%%%%%%%%%%%%%%%%%%%%%%%%%%%%%%%%%%%%%%%%%%%%%%%%
	J^i_{\text{3.5PN}} &= \frac{G}{c^7} \Biggl[- \frac{1}{63} \varepsilon_{iab} \dM^{(2)}_{ajk} \dM^{(4)}_{bjk}
	+ \frac{1}{63} \varepsilon_{iab} \dM^{(1)}_{ajk} \dM^{(5)}_{bjk}
	+ \varepsilon_{ibj} \dM^{(6)}_{aj} \Bigl(\frac{11}{105} m_{2} r_{12}^4 n_{12}^{ab}
	-  \frac{11}{105} m_{2} n_{12}^{a} r_{12}^3 y_{1}^{b}\nn\\
	& + \frac{11}{105} m_{2} r_{12}^2 y_{1}^{ab}\Bigr)
	+ \varepsilon_{ijk} \dM^{(6)}_{bk} \biggl [- \frac{22}{105} m_{2} r_{12}^3 n_{12}^{bj} (n_{12}{} y_{1}{})
	+ m_{2} r_{12}^2 \Bigl(\frac{11}{105} n_{12}^{bj} y_{1}^{2}
	+ \frac{22}{105} n_{12}^{b} (n_{12}{} y_{1}{}) y_{1}^{j}\Bigr)\nn\\
	& -  \frac{22}{105} m_{2} r_{12} (n_{12}{} y_{1}{}) y_{1}^{bj}
	+ \frac{11}{105} m_{1} y_{1}^{2} y_{1}^{bj}
	+ \frac{11}{105} m_{2} y_{1}^{2} y_{1}^{bj}\biggl]
	-  \frac{11}{105} m_{2} \varepsilon_{iaj} n_{12}^{a} r_{12}^3 y_{1}^{b} \dM^{(6)}_{bj}\nn\\
	& + \Bigl(- \frac{22}{105} m_{2} \varepsilon_{ikb} n_{12}^{b} r_{12}^2 y_{1}^{j} (n_{12}{} y_{1}{})
	+ \frac{11}{105} m_{2} \varepsilon_{ika} n_{12}^{a} r_{12} y_{1}^{j} y_{1}^{2}\Bigr) \dM^{(6)}_{jk}
	+ \varepsilon_{ibj} \dM^{(5)}_{aj} \Bigl(\frac{11}{105} m_{2} n_{12}^{a} r_{12}^3 v_{2}^{b}\nn\\
	& -  \frac{11}{105} m_{2} r_{12}^2 v_{2}^{a} y_{1}^{b}\Bigr)
	+ \varepsilon_{iaj} \dM^{(5)}_{bj} \Bigl(\frac{11}{105} m_{2} n_{12}^{a} r_{12}^3 v_{2}^{b}
	-  \frac{11}{105} m_{2} r_{12}^2 v_{2}^{a} y_{1}^{b}\Bigr)
	+ \varepsilon_{iab} \dM^{(5)}_{jk} \Bigl(\frac{3}{5} m_{2} r_{12}^2 v_{2}^{a} y_{1}^{b} n_{12}^{jk}\nn\\
	& + \frac{3}{5} m_{2} n_{12}^{a} r_{12} v_{2}^{b} y_{1}^{jk}\Bigr)
	+ \varepsilon_{iak} \dM^{(5)}_{bj} \Bigl(- \frac{3}{5} m_{2} r_{12}^3 v_{2}^{a} n_{12}^{bjk}
	+ \frac{3}{5} m_{1} v_{1}^{a} y_{1}^{bjk}
	+ \frac{3}{5} m_{2} v_{2}^{a} y_{1}^{bjk}\Bigr)\nn\\
	& + \biggl [- \frac{11}{105} m_{2} \varepsilon_{ikb} n_{12}^{a} r_{12} v_{2}^{b} y_{1}^{2}
	+ \varepsilon_{ijk} \Bigl(- \frac{34}{105} m_{2} n_{12}^{a} r_{12} y_{1}^{j} (v_{2}{} y_{1}{})
	-  \frac{11}{105} m_{1} v_{1}^{a} y_{1}^{j} y_{1}^{2}
	-  \frac{11}{105} m_{2} v_{2}^{a} y_{1}^{j} y_{1}^{2}\Bigr)\biggl] \dM^{(5)}_{ak}\nn\\
	& + \biggl(- \frac{11}{105} m_{2} \varepsilon_{ika} n_{12}^{a} r_{12} v_{2}^{b} y_{1}^{2}
	+ \varepsilon_{ijk} \biggl [- \frac{34}{105} m_{2} r_{12}^3 n_{12}^{bj} (n_{12}{} v_{2}{})
	+ m_{2} r_{12}^2 \Bigl(\frac{34}{105} n_{12}^{bj} (v_{2}{} y_{1}{})
	-  \frac{22}{105} n_{12}^{b} (n_{12}{} y_{1}{}) v_{2}^{j}\nn\\
	& + \frac{34}{105} n_{12}^{b} (n_{12}{} v_{2}{}) y_{1}^{j}\Bigr)
	+ \frac{34}{105} m_{1} (v_{1}{} y_{1}{}) y_{1}^{bj}
	+ \frac{34}{105} m_{2} (v_{2}{} y_{1}{}) y_{1}^{bj}
	+ m_{2} r_{12} \Bigl(\frac{22}{105} (n_{12}{} y_{1}{}) v_{2}^{b} y_{1}^{j}\nn\\
	& -  \frac{34}{105} (n_{12}{} v_{2}{}) y_{1}^{bj}\Bigr)\biggl]\biggl) \dM^{(5)}_{bk}
	+ \biggl(\varepsilon_{iak} \Bigl(- \frac{34}{105} m_{2} n_{12}^{a} r_{12} y_{1}^{j} (v_{2}{} y_{1}{})
	-  \frac{11}{105} m_{1} v_{1}^{a} y_{1}^{j} y_{1}^{2}
	-  \frac{11}{105} m_{2} v_{2}^{a} y_{1}^{j} y_{1}^{2}\Bigr)\nn\\
	& + \varepsilon_{ibk} \biggl [\frac{22}{105} m_{2} r_{12} v_{2}^{b} y_{1}^{j} (n_{12}{} y_{1}{})
	+ m_{2} r_{12}^2 \Bigl(- \frac{22}{105} n_{12}^{b} (n_{12}{} y_{1}{}) v_{2}^{j}
	+ \frac{34}{105} n_{12}^{b} (n_{12}{} v_{2}{}) y_{1}^{j}\Bigr)\biggl]\biggl) \dM^{(5)}_{jk}\nn\\
	& -  \frac{32}{45} \varepsilon_{ibj} \dS^{(1)}_{ab} \dS^{(3)}_{aj}
	+ \dS^{(5)}_{ib} \Bigl(- \frac{32}{45} m_{2} n_{12}^{b} r_{12}^2 (n_{12}{} y_{1}{})
	+ \frac{32}{45} m_{2} r_{12} y_{1}^{b} (n_{12}{} y_{1}{})
	-  \frac{16}{45} m_{1} y_{1}^{b} y_{1}^{2}
	-  \frac{16}{45} m_{2} y_{1}^{b} y_{1}^{2}\Bigr)\nn\\
	& + \dS^{(5)}_{ia} \Bigl(\frac{16}{45} m_{2} n_{12}^{a} r_{12}^3
	+ \frac{16}{45} m_{2} n_{12}^{a} r_{12} y_{1}^{2}
	-  \frac{16}{45} m_{2} r_{12}^2 y_{1}^{a}\Bigr)
	+ \biggl [m_{2} r_{12}^2 \Bigl(\frac{16}{45} y_{1}^{i} n_{12}^{ab}
	+ \frac{32}{45} n_{12}^{bi} y_{1}^{a}\Bigr)\nn\\
	& + m_{2} r_{12} \Bigl(- \frac{32}{45} y_{1}^{bi} n_{12}^{a}
	-  \frac{16}{45} n_{12}^{i} y_{1}^{ab}\Bigr)\biggl] \dS^{(5)}_{ab}
	+ \Bigl(- \frac{16}{45} n_{12}^{abi} m_{2} r_{12}^3
	+ \frac{16}{45} m_{1} y_{1}^{abi}
	+ \frac{16}{45} m_{2} y_{1}^{abi}\Bigr) \dS^{(5)}_{ab}\nn\\
	& -  \frac{1}{63} \varepsilon_{iab} \dM^{(6)}_{bjk} \dM_{ajk}
	-  \frac{11}{105} m_{2} \varepsilon_{ijk} n_{12}^{a} r_{12} y_{1}^{j} \dM^{(6)}_{ak} y_{1}^{2}
	+ \frac{6}{5} m_{2} \varepsilon_{iak} r_{12}^2 v_{2}^{a} y_{1}^{b} \dM^{(5)}_{bj} n_{12}^{jk}
	+ \frac{32}{45} \varepsilon_{ibj} \dS^{(4)}_{aj} \dS_{ab}\nn\\
	& -  \frac{6}{5} m_{2} \varepsilon_{ibk} n_{12}^{a} r_{12} v_{2}^{b} \dM^{(5)}_{aj} y_{1}^{jk}\Biggl]\,,\\
	%%%%%%%%%%%%%%%%%%%%%%%%%%%%%%%%%%%%%%%%%%%%%%%%%%%%%%%%%%%%%%%%%%%%%%%%%%%%%%
	P^i_{\text{2.5PN}} &= \frac{G}{c^5} \biggl [\dM^{(4)}_{ia} \Bigl(- \frac{2}{5} m_{2} n_{12}^{a} r_{12}
	+ \frac{2}{5} m_{1} y_{1}^{a}
	+ \frac{2}{5} m_{2} y_{1}^{a}\Bigr)
	+ \dM^{(3)}_{ia} \Bigl(- \frac{2}{5} m_{1} v_{1}^{a}
	-  \frac{2}{5} m_{2} v_{2}^{a}\Bigr)\biggl]\,,\\
	%%%%%%%%%%%%%%%%%%%%%%%%%%%%%%%%%%%%%%%%%%%%%%%%%%%%%%%%%%%%%%%%%%%%%%%%%%%%%%
	P^i_{\text{3.5PN}} &= \frac{G}{c^7} \Biggl[\dM^{(6)}_{ib} \Bigl(\frac{22}{105} m_{2} n_{12}^{b} r_{12}^2 (n_{12}{} y_{1}{})
	-  \frac{22}{105} m_{2} r_{12} y_{1}^{b} (n_{12}{} y_{1}{})
	+ \frac{11}{105} m_{1} y_{1}^{b} y_{1}^{2}
	+ \frac{11}{105} m_{2} y_{1}^{b} y_{1}^{2}\Bigr)\nn\\
	& + \dM^{(6)}_{ia} \Bigl(- \frac{11}{105} m_{2} n_{12}^{a} r_{12}^3
	-  \frac{11}{105} m_{2} n_{12}^{a} r_{12} y_{1}^{2}
	+ \frac{11}{105} m_{2} r_{12}^2 y_{1}^{a}\Bigr)
	+ \biggl [m_{2} r_{12}^2 \Bigl(- \frac{17}{105} y_{1}^{i} n_{12}^{ab}
	-  \frac{34}{105} n_{12}^{bi} y_{1}^{a}\Bigr)\nn\\
	& + m_{2} r_{12} \Bigl(\frac{34}{105} y_{1}^{bi} n_{12}^{a}
	+ \frac{17}{105} n_{12}^{i} y_{1}^{ab}\Bigr)\biggl] \dM^{(6)}_{ab}
	+ \Bigl(\frac{17}{105} n_{12}^{abi} m_{2} r_{12}^3
	-  \frac{17}{105} m_{1} y_{1}^{abi}
	-  \frac{17}{105} m_{2} y_{1}^{abi}\Bigr) \dM^{(6)}_{ab}\nn\\
	& + \dM^{(5)}_{ia} \Bigl(- \frac{34}{105} m_{2} n_{12}^{a} r_{12} (v_{2}{} y_{1}{})
	-  \frac{11}{105} m_{1} v_{1}^{a} y_{1}^{2}
	-  \frac{11}{105} m_{2} v_{2}^{a} y_{1}^{2}
	-  \frac{11}{105} m_{2} r_{12}^2 v_{2}^{a}\Bigr)\nn\\
	& + \dM^{(5)}_{ib} \biggl [\frac{34}{105} m_{2} n_{12}^{b} r_{12}^2 (n_{12}{} v_{2}{})
	+ \frac{34}{105} m_{1} y_{1}^{b} (v_{1}{} y_{1}{})
	+ \frac{34}{105} m_{2} y_{1}^{b} (v_{2}{} y_{1}{})
	+ m_{2} r_{12} \Bigl(\frac{22}{105} (n_{12}{} y_{1}{}) v_{2}^{b}\nn\\
	& -  \frac{34}{105} (n_{12}{} v_{2}{}) y_{1}^{b}\Bigr)\biggl]
	+ \biggl [m_{2} r_{12}^2 \Bigl(- \frac{46}{105} v_{2}^{i} n_{12}^{ab}
	-  \frac{22}{105} n_{12}^{bi} v_{2}^{a}\Bigr)
	+ m_{2} r_{12} \Bigl(\frac{22}{105} n_{12}^{i} v_{2}^{a} y_{1}^{b}
	+ \frac{92}{105} n_{12}^{a} v_{2}^{i} y_{1}^{b}\nn\\
	& + \frac{22}{105} n_{12}^{a} v_{2}^{b} y_{1}^{i}\Bigr)
	+ m_{1} \Bigl(- \frac{22}{105} y_{1}^{bi} v_{1}^{a}
	-  \frac{46}{105} v_{1}^{i} y_{1}^{ab}\Bigr)
	+ m_{2} \Bigl(- \frac{22}{105} y_{1}^{bi} v_{2}^{a}
	-  \frac{46}{105} v_{2}^{i} y_{1}^{ab}\Bigr)\biggl] \dM^{(5)}_{ab}\nn\\
	& + \frac{1}{21} n_{12}^{iab} G m_{1} m_{2} \dM^{(4)}_{ab}
	+ \dM^{(4)}_{ia} \biggl [- \frac{29}{105} m_{2} n_{12}^{a} r_{12} v_{2}^{2}
	+ \frac{G}{r_{12}} \Bigl(\frac{19}{105} m_{1} m_{2} n_{12}^{a} r_{12}
	-  \frac{10}{21} m_{1} m_{2} y_{1}^{a}\Bigr)\biggl]\nn\\
	& + \dM^{(4)}_{ib} \biggl [\frac{4}{35} m_{2} r_{12} v_{2}^{b} (n_{12}{} v_{2}{})
	+ \frac{4}{35} \frac{G m_{1} m_{2} n_{12}^{b}}{r_{12}} (n_{12}{} y_{1}{})
	+ m_{1} \Bigl(- \frac{4}{35} (v_{1}{} y_{1}{}) v_{1}^{b}
	+ \frac{29}{105} v_{1}^{2} y_{1}^{b}\Bigr)\nn\\
	& + m_{2} \Bigl(- \frac{4}{35} (v_{2}{} y_{1}{}) v_{2}^{b}
	+ \frac{29}{105} v_{2}^{2} y_{1}^{b}\Bigr)\biggl]
	+ \biggl [m_{2} r_{12} \Bigl(\frac{4}{35} v_{2}^{bi} n_{12}^{a}
	-  \frac{22}{105} n_{12}^{i} v_{2}^{ab}\Bigr)
	+ \frac{G m_{1} m_{2}}{r_{12}} \Bigl(- \frac{22}{105} y_{1}^{i} n_{12}^{ab}\nn\\
	& + \frac{4}{35} n_{12}^{bi} y_{1}^{a}\Bigr)
	+ m_{1} \Bigl(\frac{22}{105} y_{1}^{i} v_{1}^{ab}
	-  \frac{4}{35} v_{1}^{bi} y_{1}^{a}\Bigr)
	+ m_{2} \Bigl(\frac{22}{105} y_{1}^{i} v_{2}^{ab}
	-  \frac{4}{35} v_{2}^{bi} y_{1}^{a}\Bigr)\biggl] \dM^{(4)}_{ab}\nn\\
	& + \frac{G m_{1} m_{2} \dM^{(3)}_{bj}}{r_{12}} \Bigl(\frac{1}{7} n_{12}^{bij} (n_{12}{} v_{1}{})
	-  \frac{1}{7} n_{12}^{bij} (n_{12}{} v_{2}{})\Bigr)
	+ \frac{G m_{1} m_{2} \dM^{(3)}_{ij}}{r_{12}^2} \Bigl(\frac{12}{35} n_{12}^{j} (n_{12}{} v_{1}{}) (n_{12}{} y_{1}{})\nn\\
	& -  \frac{12}{35} n_{12}^{j} (n_{12}{} v_{2}{}) (n_{12}{} y_{1}{})\Bigr)
	+ \frac{G \dM^{(3)}_{ia}}{r_{12}} \biggl [\frac{m_{1} m_{2}}{r_{12}} \Bigl(- \frac{8}{35} n_{12}^{a} (v_{1}{} y_{1}{})
	+ \frac{8}{35} n_{12}^{a} (v_{2}{} y_{1}{})\Bigr)
	+ m_{1} m_{2} \Bigl(\frac{31}{105} v_{1}^{a}
	+ \frac{1}{15} v_{2}^{a}\Bigr)\biggl]\nn\\
	& + \frac{G m_{1} m_{2} \dM^{(3)}_{bj}}{r_{12}^2} \Bigl(- \frac{22}{35} y_{1}^{i} n_{12}^{bj} (n_{12}{} v_{1}{})
	+ \frac{22}{35} y_{1}^{i} n_{12}^{bj} (n_{12}{} v_{2}{})
	+ \frac{12}{35} n_{12}^{ij} (n_{12}{} v_{1}{}) y_{1}^{b}
	-  \frac{12}{35} n_{12}^{ij} (n_{12}{} v_{2}{}) y_{1}^{b}\Bigr)\nn\\
	& + \dM^{(3)}_{ib} \biggl(- \frac{17}{105} m_{1} v_{1}^{b} v_{1}^{2}
	-  \frac{17}{105} m_{2} v_{2}^{b} v_{2}^{2}
	+ \frac{G}{r_{12}} \biggl [m_{1} m_{2} \Bigl(\frac{1}{15} n_{12}^{b} (n_{12}{} v_{1}{})
	+ \frac{9}{35} n_{12}^{b} (n_{12}{} v_{2}{})\Bigr)\nn\\
	& + \frac{m_{1} m_{2}}{r_{12}} \Bigl(- \frac{8}{35} (n_{12}{} y_{1}{}) v_{1}^{b}
	+ \frac{8}{35} (n_{12}{} y_{1}{}) v_{2}^{b}
	+ \frac{8}{105} (n_{12}{} v_{1}{}) y_{1}^{b}
	-  \frac{8}{105} (n_{12}{} v_{2}{}) y_{1}^{b}\Bigr)\biggl]\biggl)\nn\\
	& + \frac{G \dM^{(3)}_{ab}}{r_{12}} \biggl [m_{1} m_{2} \Bigl(\frac{17}{105} v_{1}^{i} n_{12}^{ab}
	-  \frac{1}{15} v_{2}^{i} n_{12}^{ab}
	-  \frac{22}{105} n_{12}^{bi} v_{1}^{a}
	+ \frac{2}{5} n_{12}^{bi} v_{2}^{a}\Bigr)
	+ \frac{m_{1} m_{2}}{r_{12}} \Bigl(- \frac{8}{35} n_{12}^{i} v_{1}^{a} y_{1}^{b}
	-  \frac{8}{35} n_{12}^{a} v_{1}^{i} y_{1}^{b}\nn\\
	& + \frac{8}{35} n_{12}^{i} v_{2}^{a} y_{1}^{b}
	+ \frac{8}{35} n_{12}^{a} v_{2}^{i} y_{1}^{b}
	+ \frac{88}{105} n_{12}^{a} v_{1}^{b} y_{1}^{i}
	-  \frac{88}{105} n_{12}^{a} v_{2}^{b} y_{1}^{i}\Bigr)\biggl]
	+ \Bigl(- \frac{2}{21} m_{1} v_{1}^{abi}
	-  \frac{2}{21} m_{2} v_{2}^{abi}\Bigr) \dM^{(3)}_{ab}\nn\\
	& + \dM^{(6)}_{iab} \Bigl(- \frac{1}{63} m_{2} r_{12}^2 n_{12}^{ab}
	+ \frac{2}{63} m_{2} n_{12}^{a} r_{12} y_{1}^{b}
	-  \frac{1}{63} m_{1} y_{1}^{ab}
	-  \frac{1}{63} m_{2} y_{1}^{ab}\Bigr)
	+ \dM^{(5)}_{iab} \Bigl(- \frac{2}{63} m_{2} n_{12}^{a} r_{12} v_{2}^{b}\nn\\
	& + \frac{2}{63} m_{1} v_{1}^{a} y_{1}^{b}
	+ \frac{2}{63} m_{2} v_{2}^{a} y_{1}^{b}\Bigr)
	+ \dM^{(4)}_{iab} \Bigl(\frac{2}{63} \frac{G m_{1} m_{2}}{r_{12}} n_{12}^{ab}
	-  \frac{2}{63} m_{1} v_{1}^{ab}
	-  \frac{2}{63} m_{2} v_{2}^{ab}\Bigr)
	+ \varepsilon_{ibj} \dS^{(5)}_{aj} \Bigl(\frac{16}{45} m_{2} r_{12}^2 n_{12}^{ab}\nn\\
	& -  \frac{16}{45} m_{2} n_{12}^{a} r_{12} y_{1}^{b}
	+ \frac{16}{45} m_{1} y_{1}^{ab}
	+ \frac{16}{45} m_{2} y_{1}^{ab}\Bigr)
	-  \frac{16}{45} m_{2} \varepsilon_{iaj} n_{12}^{a} r_{12} y_{1}^{b} \dS^{(5)}_{bj}
	+ \varepsilon_{abj} \dS^{(4)}_{ij} \Bigl(- \frac{16}{45} m_{2} n_{12}^{a} r_{12} v_{2}^{b}\nn\\
	& -  \frac{16}{45} m_{1} v_{1}^{a} y_{1}^{b}
	-  \frac{16}{45} m_{2} v_{2}^{a} y_{1}^{b}\Bigr)
	+ \varepsilon_{iaj} \dS^{(4)}_{bj} \Bigl(\frac{16}{45} m_{1} v_{1}^{a} y_{1}^{b}
	+ \frac{16}{45} m_{2} v_{2}^{a} y_{1}^{b}\Bigr)
	-  \frac{16}{45} m_{2} \varepsilon_{ibj} n_{12}^{a} r_{12} v_{2}^{b} \dS^{(4)}_{aj}\nn\\
	& + \biggl [- \frac{16}{45} \frac{G m_{1} m_{2} \varepsilon_{ijb}}{r_{12}} n_{12}^{ab}
	+ \varepsilon_{ibj} \Bigl(- \frac{16}{45} m_{1} v_{1}^{ab}
	-  \frac{16}{45} m_{2} v_{2}^{ab}\Bigr)\biggl] \dS^{(3)}_{aj}\Biggl]\,,\\
	%%%%%%%%%%%%%%%%%%%%%%%%%%%%%%%%%%%%%%%%%%%%%%%%%%%%%%%%%%%%%%%%%%%%%%%%%%%%%%
	G^i_{\text{2.5PN}} &= \frac{G}{c^5} \biggl [\dM^{(3)}_{ia} \Bigl(- \frac{2}{5} m_{2} n_{12}^{a} r_{12}
	+ \frac{2}{5} m_{1} y_{1}^{a}
	+ \frac{2}{5} m_{2} y_{1}^{a}\Bigr)
	+ \dM^{(2)}_{ia} \Bigl(- \frac{4}{5} m_{1} v_{1}^{a}
	-  \frac{4}{5} m_{2} v_{2}^{a}\Bigr)\biggl]\,,\\
	%%%%%%%%%%%%%%%%%%%%%%%%%%%%%%%%%%%%%%%%%%%%%%%%%%%%%%%%%%%%%%%%%%%%%%%%%%%%%%
	G^i_{\text{3.5PN}} &= \frac{G}{c^7} \biggl(\dM^{(5)}_{ib} \Bigl(\frac{22}{105} m_{2} n_{12}^{b} r_{12}^2 (n_{12}{} y_{1}{})
	-  \frac{22}{105} m_{2} r_{12} y_{1}^{b} (n_{12}{} y_{1}{})
	+ \frac{11}{105} m_{1} y_{1}^{b} y_{1}^{2}
	+ \frac{11}{105} m_{2} y_{1}^{b} y_{1}^{2}\Bigr)\nn\\
	& + \dM^{(5)}_{ia} \Bigl(- \frac{11}{105} m_{2} n_{12}^{a} r_{12}^3
	-  \frac{11}{105} m_{2} n_{12}^{a} r_{12} y_{1}^{2}
	+ \frac{11}{105} m_{2} r_{12}^2 y_{1}^{a}\Bigr)
	+ \biggl [m_{2} r_{12}^2 \Bigl(- \frac{17}{105} y_{1}^{i} n_{12}^{ab}
	-  \frac{34}{105} n_{12}^{bi} y_{1}^{a}\Bigr)\nn\\
	& + m_{2} r_{12} \Bigl(\frac{34}{105} y_{1}^{bi} n_{12}^{a}
	+ \frac{17}{105} n_{12}^{i} y_{1}^{ab}\Bigr)\biggl] \dM^{(5)}_{ab}
	+ \Bigl(\frac{17}{105} n_{12}^{abi} m_{2} r_{12}^3
	-  \frac{17}{105} m_{1} y_{1}^{abi}
	-  \frac{17}{105} m_{2} y_{1}^{abi}\Bigr) \dM^{(5)}_{ab}\nn\\
	& + \dM^{(4)}_{ia} \Bigl(- \frac{4}{35} m_{2} n_{12}^{a} r_{12} (v_{2}{} y_{1}{})
	-  \frac{22}{105} m_{1} v_{1}^{a} y_{1}^{2}
	-  \frac{22}{105} m_{2} v_{2}^{a} y_{1}^{2}
	-  \frac{22}{105} m_{2} r_{12}^2 v_{2}^{a}\Bigr)\nn\\
	& + \dM^{(4)}_{ib} \biggl [\frac{4}{35} m_{2} n_{12}^{b} r_{12}^2 (n_{12}{} v_{2}{})
	+ \frac{4}{35} m_{1} y_{1}^{b} (v_{1}{} y_{1}{})
	+ \frac{4}{35} m_{2} y_{1}^{b} (v_{2}{} y_{1}{})
	+ m_{2} r_{12} \Bigl(\frac{44}{105} (n_{12}{} y_{1}{}) v_{2}^{b}\nn\\
	& -  \frac{4}{35} (n_{12}{} v_{2}{}) y_{1}^{b}\Bigr)\biggl]
	+ \biggl [m_{2} r_{12}^2 \Bigl(- \frac{29}{105} v_{2}^{i} n_{12}^{ab}
	+ \frac{18}{35} n_{12}^{bi} v_{2}^{a}\Bigr)
	+ m_{2} r_{12} \Bigl(- \frac{18}{35} n_{12}^{i} v_{2}^{a} y_{1}^{b}
	+ \frac{58}{105} n_{12}^{a} v_{2}^{i} y_{1}^{b}\nn\\
	& -  \frac{18}{35} n_{12}^{a} v_{2}^{b} y_{1}^{i}\Bigr)
	+ m_{1} \Bigl(\frac{18}{35} y_{1}^{bi} v_{1}^{a}
	-  \frac{29}{105} v_{1}^{i} y_{1}^{ab}\Bigr)
	+ m_{2} \Bigl(\frac{18}{35} y_{1}^{bi} v_{2}^{a}
	-  \frac{29}{105} v_{2}^{i} y_{1}^{ab}\Bigr)\biggl] \dM^{(4)}_{ab}
	-  \frac{4}{21} n_{12}^{iab} G m_{1} m_{2} \dM^{(3)}_{ab}\nn\\
	& + \dM^{(3)}_{ia} \biggl [- \frac{17}{105} m_{2} n_{12}^{a} r_{12} v_{2}^{2}
	+ \frac{G}{r_{12}} \Bigl(\frac{29}{105} m_{1} m_{2} n_{12}^{a} r_{12}
	-  \frac{38}{105} m_{1} m_{2} y_{1}^{a}\Bigr)\biggl]
	+ \dM^{(3)}_{ib} \biggl [- \frac{4}{21} m_{2} r_{12} v_{2}^{b} (n_{12}{} v_{2}{})\nn\\
	& -  \frac{4}{21} \frac{G m_{1} m_{2} n_{12}^{b}}{r_{12}} (n_{12}{} y_{1}{})
	+ m_{1} \Bigl(\frac{4}{21} (v_{1}{} y_{1}{}) v_{1}^{b}
	+ \frac{17}{105} v_{1}^{2} y_{1}^{b}\Bigr)
	+ m_{2} \Bigl(\frac{4}{21} (v_{2}{} y_{1}{}) v_{2}^{b}
	+ \frac{17}{105} v_{2}^{2} y_{1}^{b}\Bigr)\biggl]\nn\\
	& + \biggl [m_{2} r_{12} \Bigl(\frac{8}{105} v_{2}^{bi} n_{12}^{a}
	+ \frac{32}{105} n_{12}^{i} v_{2}^{ab}\Bigr)
	+ \frac{G m_{1} m_{2}}{r_{12}} \Bigl(\frac{32}{105} y_{1}^{i} n_{12}^{ab}
	+ \frac{8}{105} n_{12}^{bi} y_{1}^{a}\Bigr)
	+ m_{1} \Bigl(- \frac{32}{105} y_{1}^{i} v_{1}^{ab}\nn\\
	& -  \frac{8}{105} v_{1}^{bi} y_{1}^{a}\Bigr)
	+ m_{2} \Bigl(- \frac{32}{105} y_{1}^{i} v_{2}^{ab}
	-  \frac{8}{105} v_{2}^{bi} y_{1}^{a}\Bigr)\biggl] \dM^{(3)}_{ab}
	+ \dM^{(2)}_{ib} \biggl [\frac{G m_{1} m_{2}}{r_{12}} \Bigl(\frac{2}{5} n_{12}^{b} (n_{12}{} v_{1}{})
	+ \frac{2}{5} n_{12}^{b} (n_{12}{} v_{2}{})\Bigr)\nn\\
	& -  \frac{2}{5} m_{1} v_{1}^{b} v_{1}^{2}
	-  \frac{2}{5} m_{2} v_{2}^{b} v_{2}^{2}\biggl]
	+ \frac{G m_{1} m_{2} \dM^{(2)}_{ia}}{r_{12}} \Bigl(\frac{2}{5} v_{1}^{a}
	+ \frac{2}{5} v_{2}^{a}\Bigr)
	+ \dM^{(5)}_{iab} \Bigl(- \frac{1}{63} m_{2} r_{12}^2 n_{12}^{ab}
	+ \frac{2}{63} m_{2} n_{12}^{a} r_{12} y_{1}^{b}\nn\\
	& -  \frac{1}{63} m_{1} y_{1}^{ab}
	-  \frac{1}{63} m_{2} y_{1}^{ab}\Bigr)
	+ \dM^{(4)}_{iab} \Bigl(- \frac{4}{63} m_{2} n_{12}^{a} r_{12} v_{2}^{b}
	+ \frac{4}{63} m_{1} v_{1}^{a} y_{1}^{b}
	+ \frac{4}{63} m_{2} v_{2}^{a} y_{1}^{b}\Bigr)
	+ \dM^{(3)}_{iab} \Bigl(\frac{2}{21} \frac{G m_{1} m_{2}}{r_{12}} n_{12}^{ab}\nn\\
	& -  \frac{2}{21} m_{1} v_{1}^{ab}
	-  \frac{2}{21} m_{2} v_{2}^{ab}\Bigr)
	+ \varepsilon_{ibj} \dS^{(4)}_{aj} \Bigl(\frac{16}{45} m_{2} r_{12}^2 n_{12}^{ab}
	-  \frac{16}{45} m_{2} n_{12}^{a} r_{12} y_{1}^{b}
	+ \frac{16}{45} m_{1} y_{1}^{ab}
	+ \frac{16}{45} m_{2} y_{1}^{ab}\Bigr)\nn\\
	& -  \frac{16}{45} m_{2} \varepsilon_{iaj} n_{12}^{a} r_{12} y_{1}^{b} \dS^{(4)}_{bj}
	+ \varepsilon_{ibj} \dS^{(3)}_{aj} \Bigl(- \frac{16}{45} m_{1} v_{1}^{a} y_{1}^{b}
	-  \frac{16}{45} m_{2} v_{2}^{a} y_{1}^{b}\Bigr)
	+ \varepsilon_{abj} \dS^{(3)}_{ij} \Bigl(- \frac{16}{45} m_{2} n_{12}^{a} r_{12} v_{2}^{b}\nn\\
	& -  \frac{16}{45} m_{1} v_{1}^{a} y_{1}^{b}
	-  \frac{16}{45} m_{2} v_{2}^{a} y_{1}^{b}\Bigr)
	+ \frac{16}{45} m_{2} \varepsilon_{iaj} n_{12}^{a} r_{12} v_{2}^{b} \dS^{(3)}_{bj}\biggl)\,.
\end{align}\ese
Finally, we refer to the Supplemental Material~\cite{SuppMaterial} for the RR Schott terms at 4.5PN order in expanded form.

\bibliography{ListeRef_BFT24.bib}

\end{document}